\documentclass[mnsc, nonblindrev]{informs3_WP} 

\OneAndAHalfSpacedXI

\usepackage{natbib}
 \bibpunct[, ]{(}{)}{,}{a}{}{,}%
 \def\bibfont{\small}%

\usepackage{lmodern}
 
\usepackage{tabularx}
\usepackage{booktabs}

\usepackage{bm}
\usepackage[dvipsnames]{xcolor}

\RequirePackage{endnotes}

\usepackage{svg}

 \def\endnotesize{\normalsize}

\newcommand{\comment}[1]{}

\TheoremsNumberedThrough     
\ECRepeatTheorems

\EquationsNumberedThrough    

\MANUSCRIPTNO{MS-RMA-2024-06067.R2}

\usepackage{url}            

\usepackage{algorithm}
\usepackage{algpseudocode}	

\usepackage[hidelinks]{hyperref} 

\usepackage{caption}
\usepackage[labelfont=sf]{subcaption}
\usepackage{mathtools}
\usepackage{comment}

\newcommand{\newmacro}[2]{\newcommand{#1}{\debug{#2}}}		
\newcommand{\newop}[2]{\DeclareMathOperator{#1}{\debug{#2}}}		

\DeclarePairedDelimiterX{\setdef}[2]{\{}{\}}{#1:#2}

\newmacro{\dd}{\:d}		

\newcommand{\col}{C}

\newcommand{\coli}{\col^i}
\newcommand{\colset}{\mathcal{C}}
\newcommand{\Stud}{\mathcal{S}}
\newcommand{\stud}{s}
\newcommand{\capacity}{\alpha}
\newcommand{\capacities}{\boldsymbol{\alpha}}

\newcommand{\capi}{\capacity^{i}}
\newcommand{\gp}{G}
\newcommand{\ngp}{d}
\newcommand{\gpi}{\gp_1}
\newcommand{\gpii}{\gp_2}
\newcommand{\gpj}{\gp_j}
\newcommand{\gpl}{\gp_\ell}
\newcommand{\gpk}{\gp_\ngp}
\newcommand{\prop}{\gamma}

\newcommand{\propj}{\prop_j}
\newcommand{\propl}{\prop_\ell}
\newcommand{\props}{\boldsymbol{\gamma}}
\newcommand{\pref}{\beta}

\newcommand{\prefl}{\pref_\ell}

\newcommand{\prefs}{\boldsymbol{\beta}}
\newcommand{\preflist}{\sigma}
\newcommand{\prefsig}{\pref^{\preflist}}
\newcommand{\prefsj}{\pref_{j}^{\preflist}}

\newcommand{\prefGj}{\prefs_{j}}

\newcommand{\preflistsm}{\Sigma([m])}
\newcommand{\grade}{W}

\newcommand{\gradei}{\grade^{i}}
\newcommand{\gradeis}{\grade^{i}_\stud}
\newcommand{\grades}{\textbf{W}}
\newcommand{\cut}{P}

\newcommand{\cuti}{\cut^i}
\newcommand{\cutoffs}{\textbf{P}}
\newcommand{\cor}{r}

\newcommand{\type}{\mathtt{S}}

\newcommand{\spearman}{\rho}
\newcommand{\kendall}{\tau}

\newcommand{\eff}{E}
\newcommand{\ineq}{L}
\newcommand{\ineqjl}{\ineq_{j, \ell}}

\newcommand{\VE}{V_{\emptyset}}

\newcommand{\Rjsk}{R^{k, \preflist}_j}
\newcommand{\Rjsi}{R^{1, \preflist}_j}
\newcommand{\Rjsm}{R^{m, \preflist}_j}
\newcommand{\RjsE}{R^{\emptyset, \preflist}_j}
\newcommand{\RjE}{R^{\emptyset}_j}

\newcommand{\Rlsk}{R^{k, \preflist}_\ell}
\newcommand{\Rlsi}{R^{1, \preflist}_\ell}

\newcommand{\RlE}{R^{\emptyset}_\ell}

\newcommand{\match}{\mu}
\newcommand{\demand}{D}
\newcommand{\demands}{\textbf{D}}

\newcommand{\param}{\theta}
\newcommand{\tset}{\Theta}

\newcommand{\paramj}{\theta_{j}}
\newcommand{\paraml}{\theta_{\ell}}

\newcommand{\params}{\boldsymbol{\theta}}

\newcommand{\cdf}{F}

\newcommand{\pdfCij}{f_j^{i}}
\newcommand{\cdfCij}{F_j^{i}}

\newcommand{\pdfCil}{f_\ell^{i}}
\newcommand{\cdfCil}{F_\ell^{i}}

\newcommand{\coppdffamily}{\left(\coppdft \right)_{\param \in \tset}}

\newcommand{\copcdffamily}{\left(\copcdft \right)_{\param \in \tset}}

\newcommand{\copcdf}{H}
\newcommand{\coppdft}{h_\param}
\newcommand{\copcdft}{H_\param}

\newcommand{\copcdfj}{H_j}

\newcommand{\pdfj}{f_{j, \paramj}}
\newcommand{\cdfj}{F_{j, \paramj}}

\newcommand{\pdfs}{\boldsymbol{f}}

\newcommand{\xsetj}{I_j}

\newcommand{\xsetCij}{I^i_j}
\newcommand{\xsetCijl}{\underline{I}^i_j}
\newcommand{\xsetCiju}{\Bar{I}^i_j}

\newcommand{\class}{Q}

\usepackage{mathrsfs}

\newcommand{\mass}{\eta}
\newop{\ex}{\mathbb{E}}		
\newcommand{\prob}{\mathbb{P}}		
\newop{\Var}{Var}		
\newop{\simplex}{\hull}		

\newcommand{\R}{\mathbb{R}}		

\usepackage{acronym}		

\newacro{DA}{Deferred Acceptance}
\newacro{CA}{College Admission}
\newacro{LHS}{left-hand side}
\newacro{RHS}{right-hand side}
\newacro{iid}[i.i.d.]{independent and identically distributed}

\begin{document}


 \RUNAUTHOR{Castera, Loiseau, and Pradelski}

\RUNTITLE{Correlation in rankings  in matching markets}

\TITLE{Correlation of rankings in matching markets}
\ARTICLEAUTHORS{%
\AUTHOR{Rémi Castera}
\AFF{MCGT, UM6P, \EMAIL{remi.castera@protonmail.com}}
\AUTHOR{Patrick Loiseau}
\AFF{Inria, Fairplay joint team}
\AUTHOR{Bary S.R.~Pradelski}
\AFF{CNRS, Maison Fran\c{c}aise d'Oxford}
\AFF{Department of Economics, University of Oxford} \vspace{0.3 cm}
\AUTHOR{December 2025 (First version: May 2022)}
} 

\ABSTRACT{
We study the role of correlation in matching markets, where multiple decision-makers simultaneously face selection problems from the same pool of candidates. 
We propose a model in which a candidate's priority scores across different decision-makers exhibit varying levels of correlation dependent on the candidate's sociodemographic group.
Such \emph{differential correlation} can arise in school choice due to the varying prevalence of selection criteria, in college admissions due
to test-optional policies, or due to algorithmic monoculture, that is, when decision-makers rely on the same algorithms and data sets to evaluate candidates.
We show that higher correlation for one of the groups generally improves the outcome for all groups, leading to higher efficiency. 
However, students from a given group are more likely to remain unmatched as their own correlation level increases. 
This implies that it is advantageous to belong to a low-correlation group. Finally, we extend the tie-breaking literature to multiple priority classes and intermediate levels of correlation. Overall, our results point to differential correlation as a previously overlooked systemic source of group inequalities in school, university, and job admissions.
}%



\maketitle

\section{Introduction}
\label{sec:introduction}

Outcome inequalities for different demographic or social groups are ubiquitous, for example, in college admission, job assignment, or investment allocation.  \cite*{arcidiacono_harvard_2022} show that Asian-American applicants have lower admission chances at Harvard than white applicants with similar academic records, \cite{niessen_gender_2019} find significantly lower inflows in female managed mutual funds than in male managed mutual funds, and \cite{bertrand_labor_2004} expose race-based discrimination in callback decisions by job advertisers. Consequently, the sources of observed outcome inequities---as, for example, bias or statistical discrimination---remain the subject of frequent and ongoing controversy and political debate.
A common concern is that the causes of outcome inequalities in matching markets are poorly understood, making them difficult to address \citep{longhofer_rooting_1995}.

We consider a previously overlooked source of outcome inequities.
To this end, 
we study how different correlations of priority scores, and thus rankings, between different sociodemographic groups---\emph{differential correlation}---affect outcome inequalities and efficiency in matching markets. Our findings point to a source of inequity between different groups that is specific to matching markets. 
This source may have been overlooked, as most policies, including those aiming to correct for biases or implement affirmative action, are designed with one decision-maker in mind. 
In particular, we find that differential correlation across groups leads to outcome inequities even when the rankings by each college are ``fair'', i.e., all groups are represented at all levels of each college's ranking in the same proportion as in the applicant population.
Furthermore, we identify the impact of variations of the correlation of rankings on the efficiency of the resulting matching.

\begin{figure}[ht]
    \FIGURE{
    \includegraphics[width=0.3\linewidth]{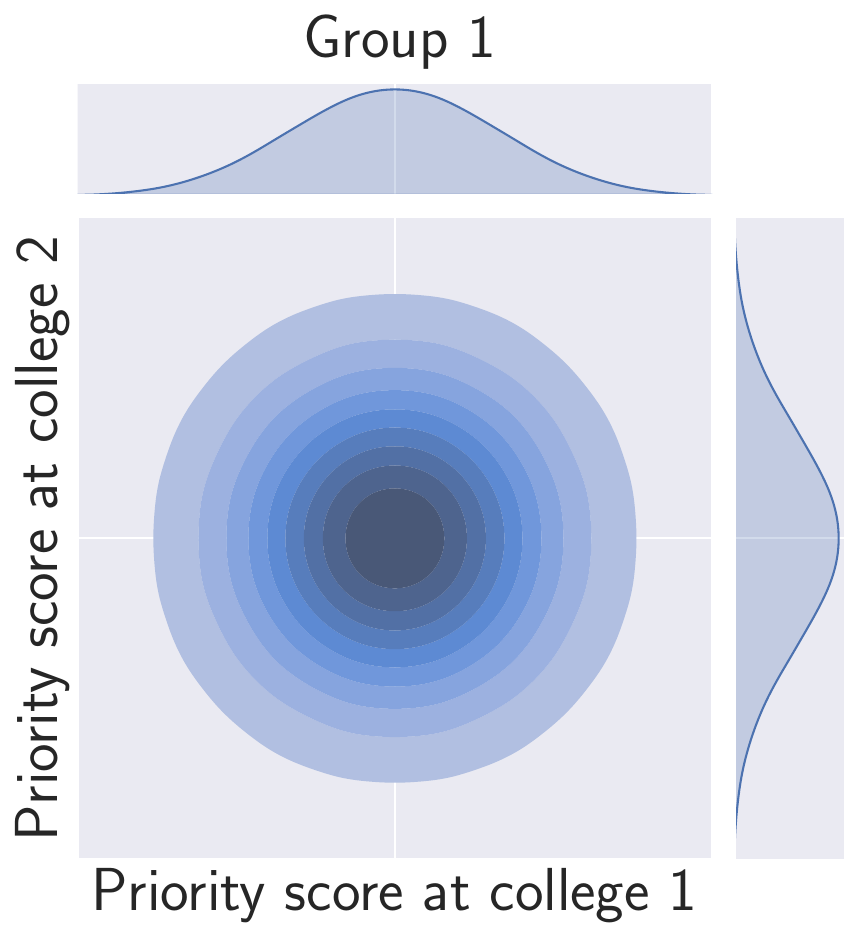} \hspace{1 cm}
    \includegraphics[width=0.3\linewidth]{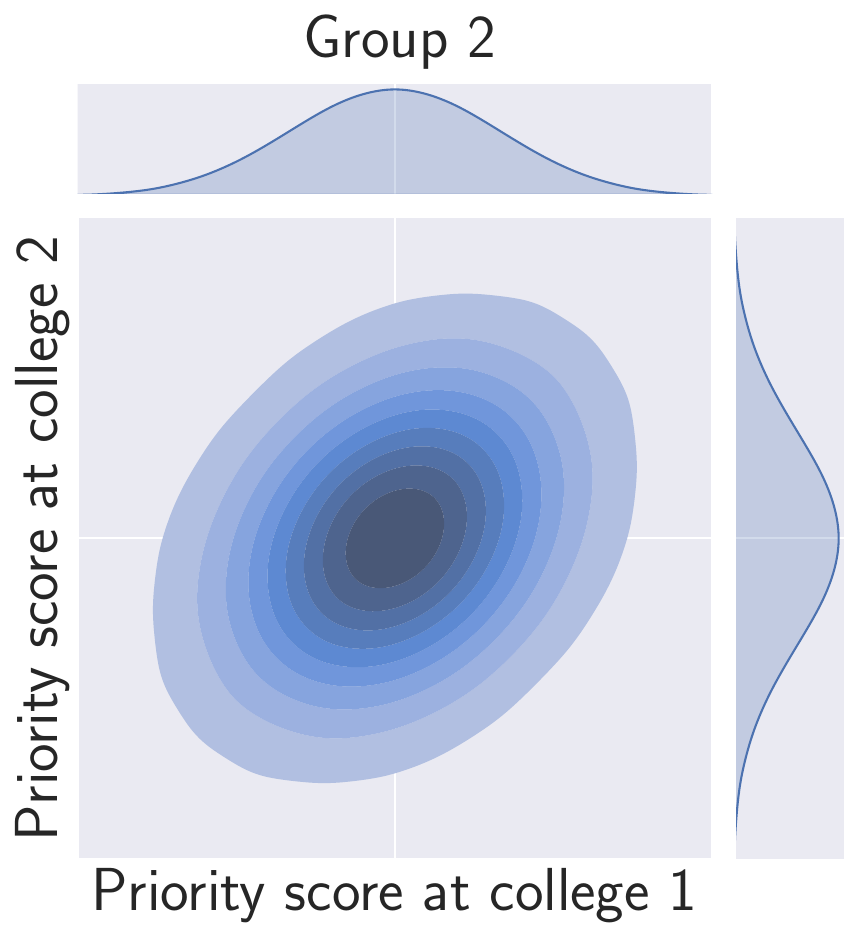}}
    {Differential correlation between two groups at two colleges. \label{fig:example_into}\\ \vspace{-0.3cm}}
    {The left and right panels represent the priority score distributions for Groups 1 and 2, respectively. 
    The color shading indicates density, with darker shades representing higher probability density. The marginal distributions of priority scores at each college are shown along the top and right edges of each plot. While both groups share the same marginal distributions, their joint distributions differ due to differential correlation, with Group 2 exhibiting higher correlation than Group 1.}
\end{figure}

Figure \ref{fig:example_into} illustrates differential correlation for two groups at two colleges, showing distributions of priority scores, from which ordinal rankings can be derived. The marginal distributions are the same for all students (Gaussian), however, the joint distributions differ due to differential correlation. Looking ahead, we will show that, ceteris paribus, the group with lower correlation is better off.

Differential correlation arises when different decision-makers, such as colleges, use different information on candidates from different (sociodemographic) groups, such as students, when assigning priority scores to rank and admit them. 
Decision-makers hereby have to rely on observable, partial information, as candidates' latent qualities are unknown to the decision-maker.\footnote{The idea to study a model of latent quality plus noise can be traced to \cite{phelps72}. More recently, this idea has been used by \cite*{chade_student_2014,garg_dropping_2021,emelianov22}. }
Differences in information between groups may be due to the varying cost of information acquisition, the availability of information, or varying prevalence of attributes between different groups.

For concreteness, consider school choice and selection criteria that are more or less prevalent between different groups, e.g., diversity with respect to the current student body, proximity, or sibling priority. 
The latter is, for example, used in the centralized school choice mechanism in Chile \citep{correa_chile_2022}, where, in 2023,
 21.3\% students among the economically disadvantaged received sibling priority, while only 15.6\% of the remaining students received  sibling priority.
As a result, a group for which the criterion is more prevalent will, ceteris paribus, exhibit lower correlation than a group for which the criterion is less prevalent (assuming that a student only receives sibling priority at one school). Thus, the effect of  differential correlation is conjugated with the advantage from receiving sibling priority.

 Differential correlation may also result from test-optional policies for standardized tests, such as the SAT used in the United States for college admissions \citep[cf., e.g.,][]{Leon24}. Standardized testing increases the correlation of assessment results between different colleges. Moreover, if different demographic groups exhibit different participation patterns in standardized tests, then differential correlation arises.

Finally, correlation of rankings can also change due to the use of algorithms to support decisions. Algorithms are increasingly used in selection tasks, ranging from bail decisions
\citep*[cf., e.g.,][]{Ang16}
to screening and selection of candidates for university admission, employment, tenancy, or mortgages \citep[cf., e.g.,][]{citron2014scored}.
The increased use of the same or similar machines, algorithms, or data sets has been termed ``algorithmic monoculture’’ \citep{goth2003addressing}.\footnote{The terminology is borrowed from its use in agriculture.}
While various risks have been associated with algorithmic monoculture, e.g., vulnerability to flaws, hacker attacks, or favoring persistent biases \citep*[cf., e.g.,][]{ goth2003addressing, citron2014scored, kleinberg_algorithmic_2021,peng_monoculture_2023}, algorithmic monoculture also leads to increased correlation between decision-makers' assessments of candidates and can lead to differential correlation (for instance if the data does not have the same informative value between different populations).

Our results identify differential correlation as an overlooked source of inefficiency and inequity in matching markets. 
Policy-makers and managers who orchestrate matching markets should be aware that only considering and enforcing the \emph{fairness} of each decision-maker separately is insufficient to ensure overall equity (and efficiency). In practice, school choice mechanisms have been frequently redesigned over the past several decades and researchers have actively contributed to these efforts (cf. e.g.,  \citealt*{abdulkadiroglu_school_2003,abdulkadiroglu_college_2005,abdulkadiroglu_strategy-proofness_2009}, \citealt{correa_chile_2022},  \citealt*{kamada_fair_2018}). According to our results, differential correlation should be accounted for when (re)designing these and other markets. The increasing use of algorithms to facilitate screening and selection tasks may lead to changes in (differential) correlation that should be taken into account; optimistically, 
new interventions and regulations may be available, for example, against monopolies by software and data providers.

\subsection{Our contribution}

 We study the college admissions problem, where multiple decision-makers select a subset of applicants from an applicant pool with stability as the solution concept \citep{gale_college_1962, azevedo_supply_2016}. Specifically, we suppose that an infinite population of students divided into groups $\gpi, ~\dots, ~\gpk$  applies to colleges $\col^1, ~\dots, ~\col^m$. The groups represent, for example, protected attributes such as gender or race. 
Each college assigns a priority score to each student and thus  
each student receives a priority score at each college. We propose an original model for the distributions of these priority scores to study the correlation between the rankings produced by different colleges. Our model allows for any number of groups, different marginals, and student preferences that depend on the priority scores.

To formalize correlation and thus capture the vague notion of ``a connection between two things in which one thing changes as the other does",\footnote{Oxford Advanced Learner's Dictionary, 2023.} we
leverage prior work on copulas and their relation with classical notions of correlation via \emph{coherence}. This allows us to model correlation without a specific functional form and, in particular, nest classical notions as special cases, e.g.,  
Spearman's and Kendall's correlation indices. With this, we assume that the correlation between the priority scores at different colleges depends on 
a candidate's group; we call this feature \emph{differential correlation}. 

We investigate two main questions:
\begin{itemize}
    \item \emph{How does correlation impact efficiency, i.e., the number of students who obtain their first choice (or one of their top $k$ choices)?} 
    \item \emph{How does differential correlation impact  inequality, i.e., the difference between groups in the number of students remaining unmatched?}
\end{itemize}

\begin{figure}[ht]
    \FIGURE
    {\shortstack{
    {
    \centering\textbf{\footnotesize{Baseline.}}}\\
    \includegraphics[width =  0.35 \textwidth]{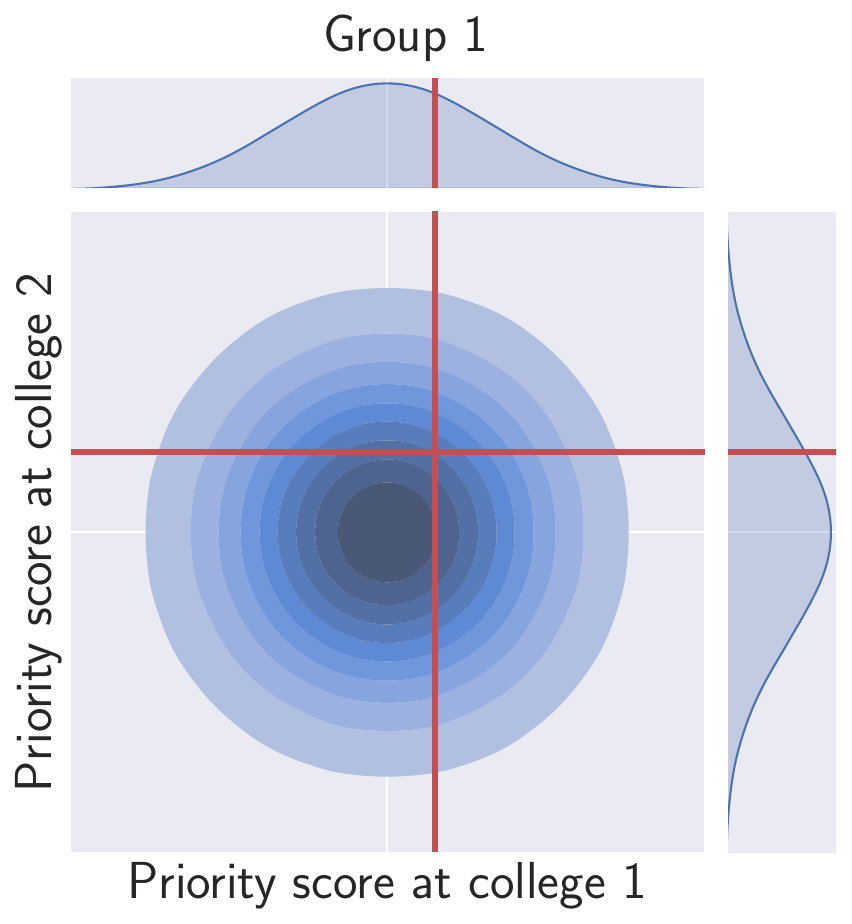}
    \hspace{1.12cm}\includegraphics[width =  0.35 \textwidth]{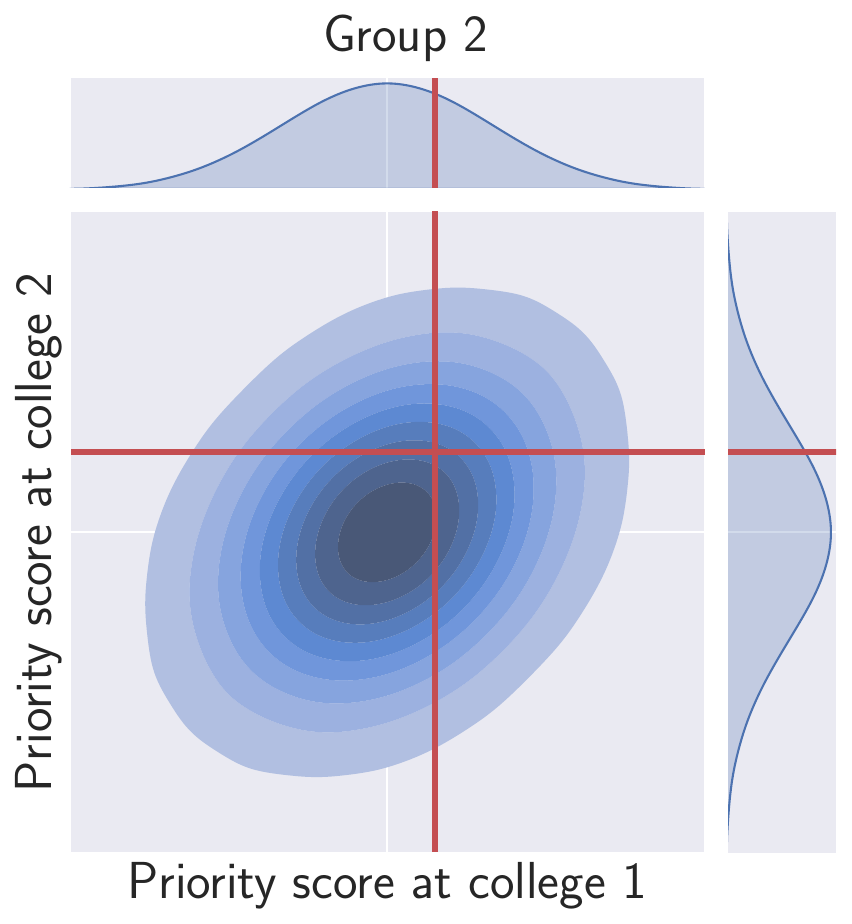} \\  \vspace{0.1cm} \\
     \textbf{\endnotesize{After increase of the correlation between priority scores of Group $1$.}}\\
    \includegraphics[width =  0.35 \textwidth]{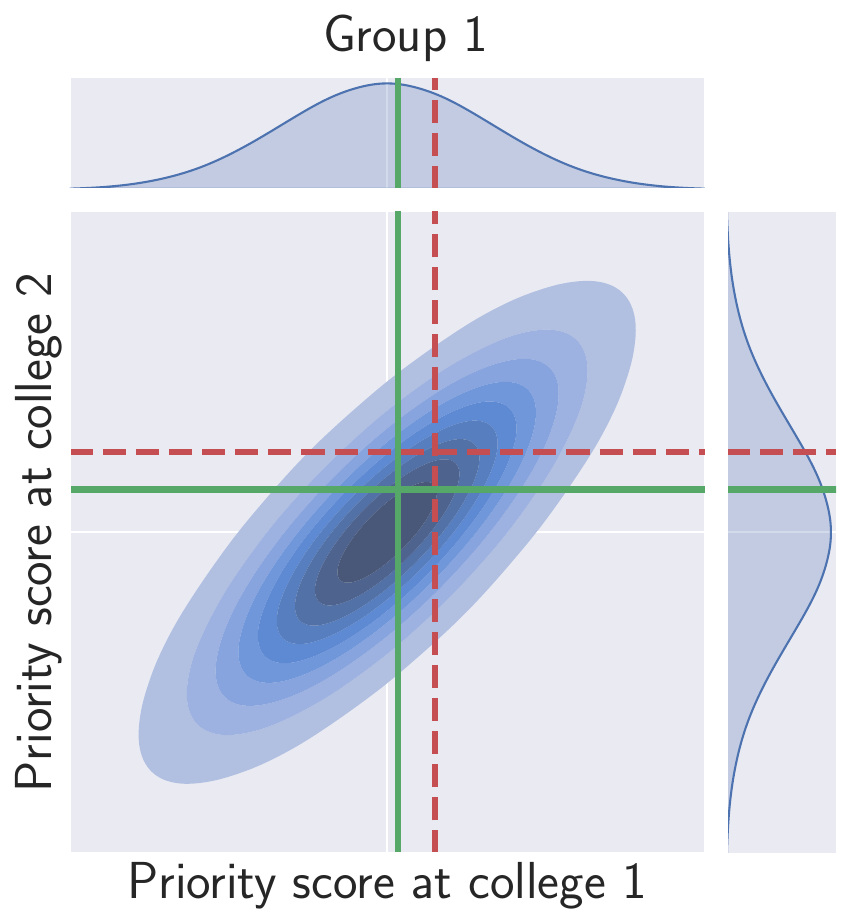}
    \hspace{1cm}
    \includegraphics[width =  0.35 \textwidth]{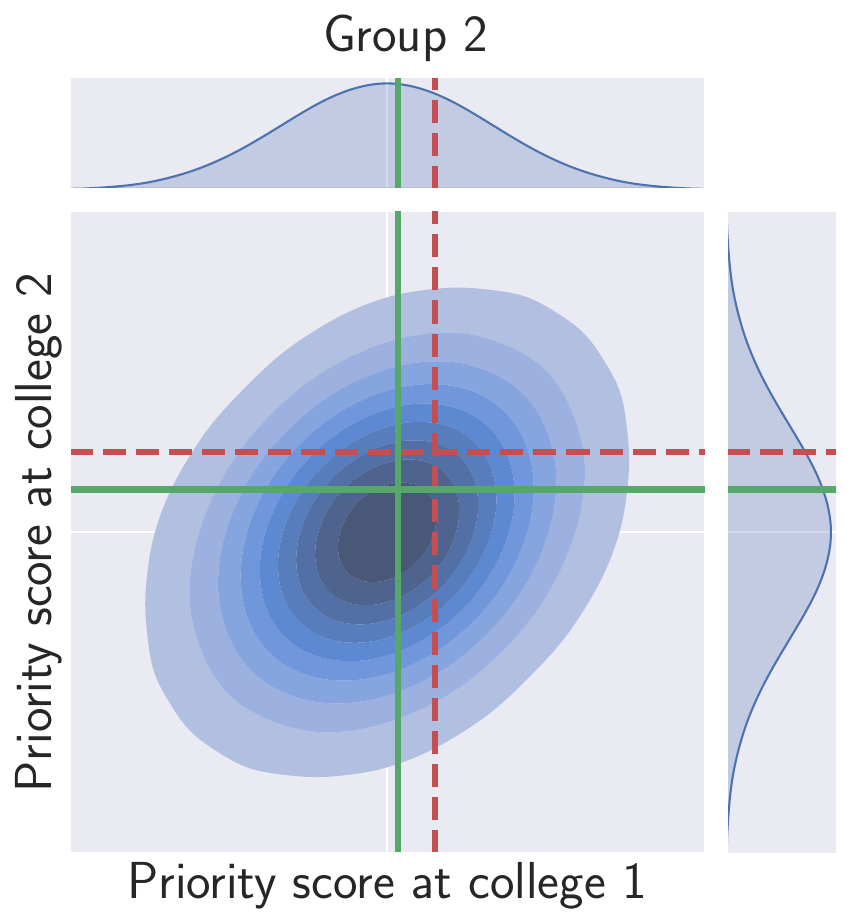}}}
    {Illustration of the effect of correlation increase on cutoffs and efficiency. \vspace{0.45cm} \label{fig:cor-cutoff_intro} }
    {The left and right columns represent the priority score distributions for Groups $1$ and $2$, respectively. The top row shows the baseline scenario, while the bottom row illustrates the effect of increasing correlation  for Group $1$. The color shading indicates density, with darker shades representing higher probability density. The  red lines indicate the initial cutoffs (shown dashed for reference in the bottom row), while the green lines show the updated cutoffs after increasing correlation. As correlation increases, cutoffs decrease, allowing more students to be matched to their first choice, as can be seen from the plots of the marginal distributions (top and right of each plot). The impact differs between groups since Group $1$'s priority score distribution changed while Group $2$'s remained the same; for the former less students are above either cutoff, while for the latter the overall change is entirely beneficial.}   
\end{figure}

First, we show that efficiency increases with each group's correlation level; i.e., 
increasing the correlation level of any group increases the number of students who obtain their first choice in all groups, and also increases the number of students who obtain one of their top $k$ choices, for all $k$, in all groups except the one whose correlation increased (Theorem \ref{thm:V1-inc}). Intuitively, this is the case because increasing correlation leads to a decrease in admission cutoffs, otherwise the number of unmatched students would increase. As a consequence, it becomes easier for students to be assigned to one of their preferred colleges. However, the benefit is countered for the group whose correlation increased, since their probability of being below the cutoff of several colleges also increases.
Figure \ref{fig:cor-cutoff_intro}  illustrates the impact of increasing correlation on the cutoffs for two colleges and two groups.

Considering inequality, we show that the proportion of students in a given group who remain unmatched increases in its own correlation level and decreases in the correlation level of all other groups (Theorem \ref{thm:unmatched}).
This implies that it is advantageous to belong to a low-correlation group.  
Intuitively, this is the case because an ``independent second chance'' is preferable to ``carrying over the bad signal from previous rejection''. However, as stated before, low correlation also leads to higher admission cutoffs. 
From an individual's viewpoint, to benefit from both effects, it is best to have a low correlation in their own priority scores, while everyone else's priority scores exhibit high correlation.

Next, we show that a given efficiency level can be reached by a continuum of different correlation vectors, generating different levels of inequality; this shows that
differential correlation has a distinct impact on
efficiency on one hand and on inequality on the
other (Proposition \ref{prop:efficiency}).

Finally, our results imply extensions of known results on tie-breaking (cf. \citealt*{ashlagi_assigning_2019-1}, \citealt{ashlagi_competition_2020}, and  \citealt{arnosti_lottery_2022}), particularly for multiple priority classes and intermediate levels of correlation (Proposition \ref{prop:TB}).

All the aforementioned results hold under the assumption that admission cutoffs decrease in the correlation parameters. We show that the assumption always holds for two colleges (Theorem \ref{thm:cutoff_inc}). Moreover, under standard symmetry conditions from the literature, the assumption holds for any number of colleges (Proposition \ref{prop:symmetry}). Finally, we show through simulations that our results remain qualitatively valid across the parameter space.

\subsection{Outline}

The remainder of the article is organized as follows. Section \ref{sec:model} introduces the model and the concept of differential correlation. Section \ref{sec:metrics} introduces our welfare metrics and presents preliminary results. Our main results are presented in Section \ref{sec:main_results}: we first derive comparative statics regarding the impact of correlation on our metrics, under the assumption that admission cutoffs are decreasing in the correlation parameters (Section \ref{subsec:compstat}), then apply them to the tie-breaking problem in Section \ref{subsec:TB}, and, in Section \ref{sec:dec_cutoffs},  study the validity of the decreasing cutoffs assumption. Finally, in Section \ref{sec:litearture}, we provide an extended overview of the related work and Section \ref{sec:conclusion} concludes.

\section{Setup}
\label{sec:model}

We introduce the college admissions problem with a \textit{continuum} of students and \textit{correlated priority scores}, formalize the notion of correlation, and introduce the supply and demand framework to identify stable matchings. 
A table of notation is provided in Appendix \ref{app:notation} for convenience.

\subsection{Model} \label{subsec:model}

A finite set of $m$ colleges $\colset = \{ \col^1,\dots,\col^m\}$ and a continuous set of students $\Stud$ with unit mass, endowed with a measure $\eta$, need to be matched.\footnote{The formal definition of the measure is deferred to Online Appendix A.1.2.} Colleges have respective capacities $(\capacity^1, \dots, \capacity^m) \coloneqq \capacities \in (0, 1)^m$, which represent the maximal mass of students each can admit, and we assume that capacity is constrained, i.e., $\sum\limits_{i \in [m]} \capi < 1$,  using the notation $[k]:=\{1,2,\ldots,k\}$ for $k\in\mathbb N$. The set of students $\Stud$ is divided into $\ngp$ groups $\gpi, \dots, \gpk$, with a mass $\propj \in [0, 1]$ of students belonging to $\gpj$, such that $\sum_{j=1}^\ngp \propj=1$ and each student belongs to exactly one group.\footnote{Note that while this assumption is common, a recent literature has shown the importance of considering intersectional groups, e.g., the intersection of ethnic and sexual identity \citep*[cf., e.g.,][]{carvalho2022identity,molina2022bounding,carvalho2024intersectionality}. We consider the extension to intersectional groups an interesting avenue of future research in the matching setting.}
Define the vector $\props \coloneqq (\propj)_{j \in [\ngp]}$. Let $\gp (s)$ be the group student $\stud \in \Stud$ belongs to.

Each student has strict preferences over colleges: when student $\stud$ prefers college $\col$ to college $\col'$, we write $\col \succ_\stud \col'$. Preferences can also be represented by a permutation $\preflist \in \preflistsm$, where $\preflistsm$ is the set of all permutations of $m$ elements, and $\preflist(1)$ represents the favorite college and $\preflist(m)$ the least favorite college. 
In group $\gpj$ a share $\prefsj \in (0, 1)$ of students has preference list $\preflist$ (with the share thus depending on the group).\footnote{
Notice that $\prefsj$ is a share that is conditional on the group, and not a mass: $\eta (\setdef {\stud \in \gpj }{ \preflist_{\stud} = \preflist }) = \propj\prefsj$.} We assume that each preference list is used by a non-zero mass of students, and that all students prefer attending some college to remaining unmatched. Let $\prefGj=(\prefsj)_{\preflist \in \preflistsm}$ and  $\prefs=(\prefGj)_{j \in [\ngp]}$.

Each college assigns a priority score to each student, and the higher the priority score, the better the student's evaluation. Thus, each student $\stud$ is assigned a vector of priority scores $(\grade^1_s, \dots, \grade^m_s)$. College $\coli$ prefers $\stud \in \Stud$ to $\stud'$ if and only if $\gradeis > \gradei_{s'}$. 
The (marginal) distribution of priority scores $\gradei$ given by college $\coli$ to students in $\gpj$ is described by a probability density function\footnote{This implies that ties happen with probability zero.} (pdf) $\pdfCij$ defined on the support $\xsetCij\subseteq \mathbb{R}$, assumed to be an interval $(\xsetCijl, \xsetCiju)$.\footnote{The bounds of this interval can be finite or not.} Let $\xsetj = \prod\limits_{i = 1}^{m} \xsetCij$. Denote by $\cdfCij$ the cumulative distribution function (cdf) associated with $\pdfCij$. Let $\pdfs \coloneqq (\pdfCij)^{i \in [m]}_{j \in [\ngp]}$ be the set that contains all the information on the marginal priority score distributions.

\subsubsection*{Differential correlation.}
 
Consider the joint distribution of the vectors $(\grade^1_\stud, \dots, \grade^m_\stud)$. A joint distribution can be characterized by its marginals, i.e., the distribution of each component of the vector, and the shape of the joint distribution, captured by a coupling function, called \emph{copula}. A copula is a cdf over $[0, 1]^m$ with uniform marginals. 
\cite{sklar_copula_1959} shows that any joint distribution can be decomposed into (independent) marginals and a unique copula.

\textbf{\citet[Theorems 1, 2, and 3]{sklar_copula_1959}}
\emph{Let $F$ be an m-dimensional cdf with marginal cdfs $F^1, \dots, F^m$. Then there exists a unique m-dimensional copula $H: [0, 1]^m \to [0, 1]$ such that 
$$ F(x^1, \dots, x^m) = H(F^1(x^1), \dots, F^m(x^m)).$$
Conversely, for any m-dimensional copula $H$ and any set of 1-dimensional cdfs $F^1, \dots, F^m$, $F(x^1, \dots, x^m) \coloneqq H(F^1(x^1), \dots, F^m(x^m)) $ is an m-dimensional cdf with marginals $F^1, \dots, F^m$.
}

The priority score vector of each group $\gpj$ can then be represented by its marginals and a unique copula $\copcdf_{j}$.  
We consider a family of $m$-dimensional copulas $\copcdffamily$ ($\tset$ being an interval of $\mathbb{R}$) such that, for all $j \in [\ngp]$, there exists a parameter $\param_j \in \tset$ such that $\copcdf_{\theta_j}$ is the copula associated with $\gpj$'s distribution; i.e., $\copcdf_{\theta_j}=\copcdf_{j}$ (such a family always exists by \citealt{sklar_copula_1959}). This feature of the model allows us to vary the shape of the joint distribution without changing the marginals; this is crucial to model correlation and isolate its effects. We see this as our principal modeling contribution.

\begin{figure}[ht]
    \FIGURE
  {\shortstack{
    \textbf{\endnotesize{Gaussian copula family.}}\\ 
    \includegraphics[width = 0.19 \textwidth]{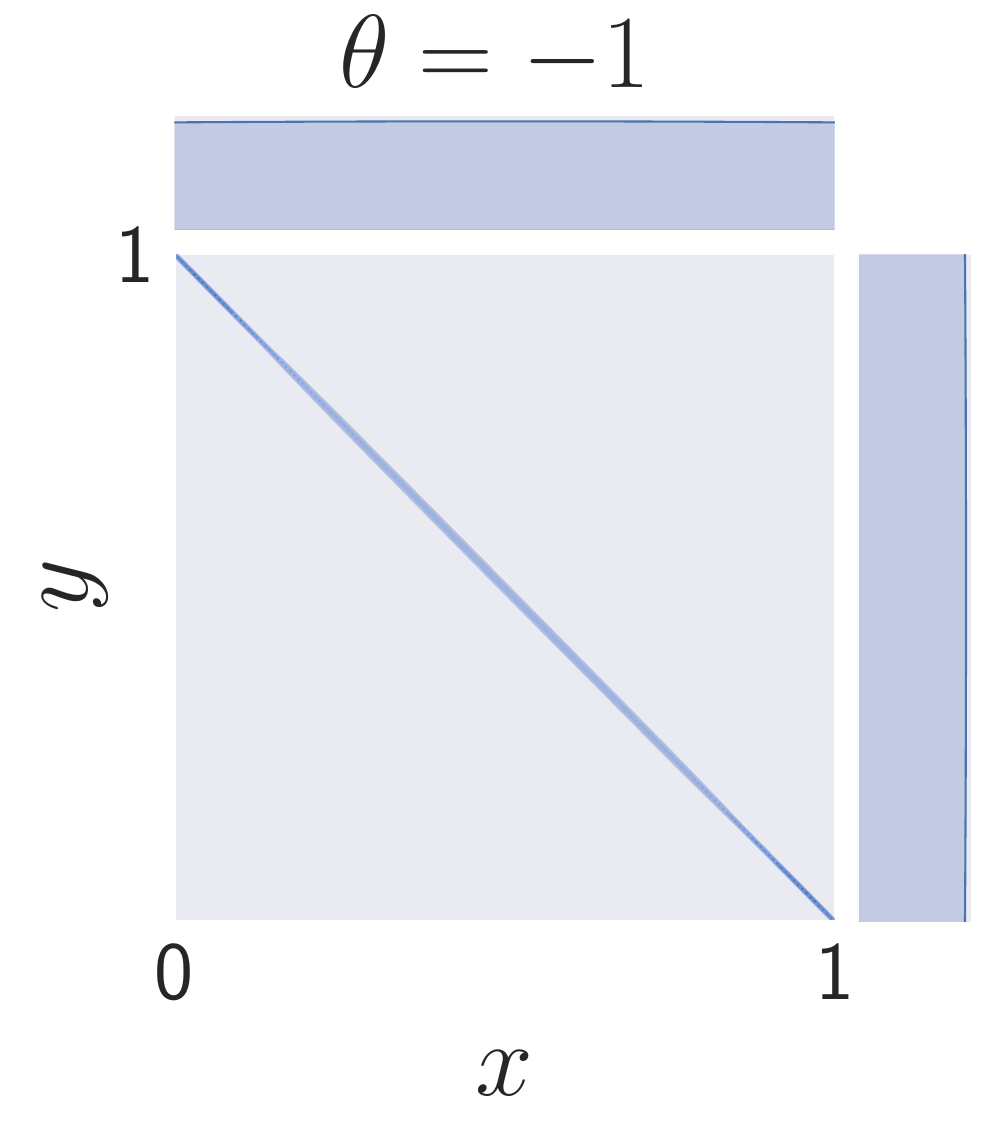}
    \includegraphics[width = 0.19 \textwidth]{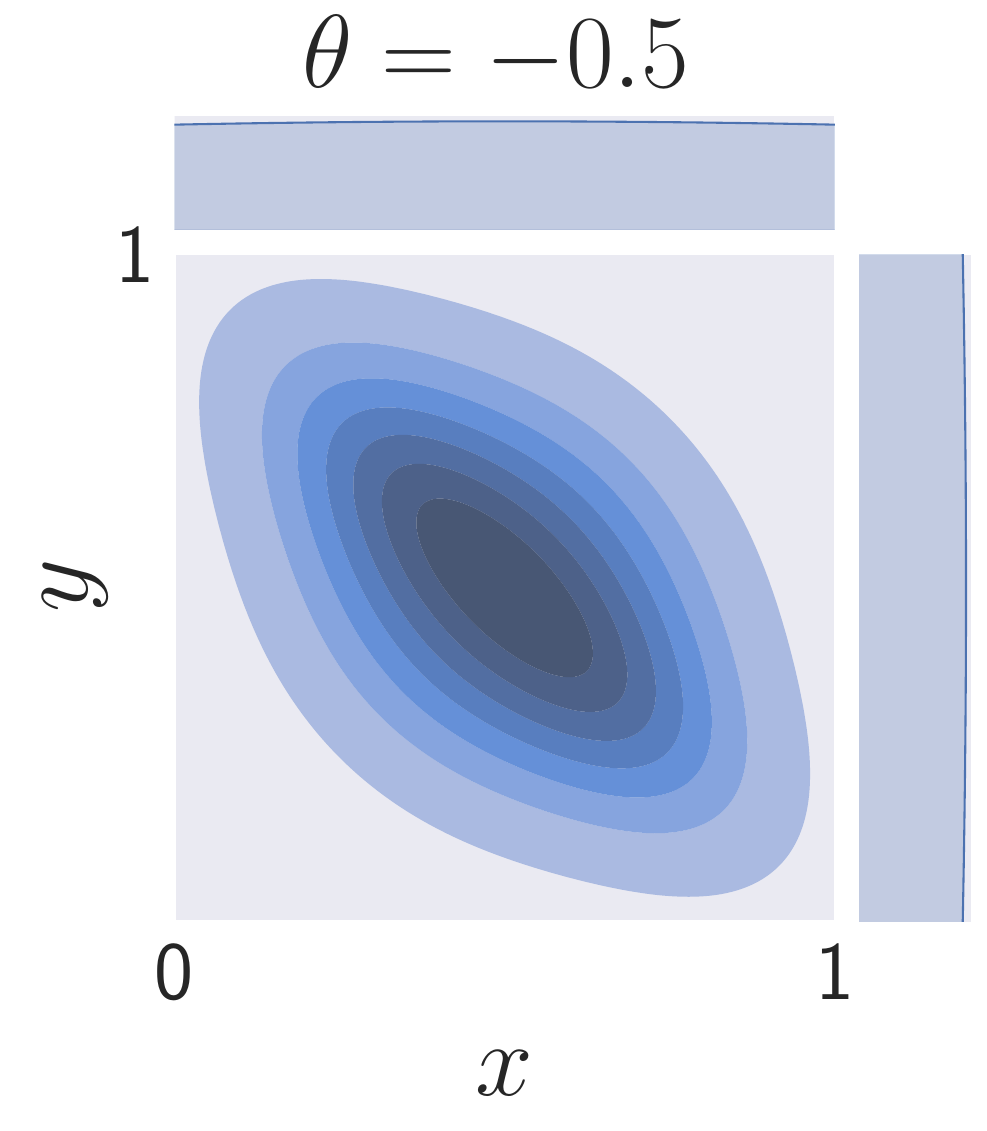}
    \includegraphics[width = 0.19 \textwidth]{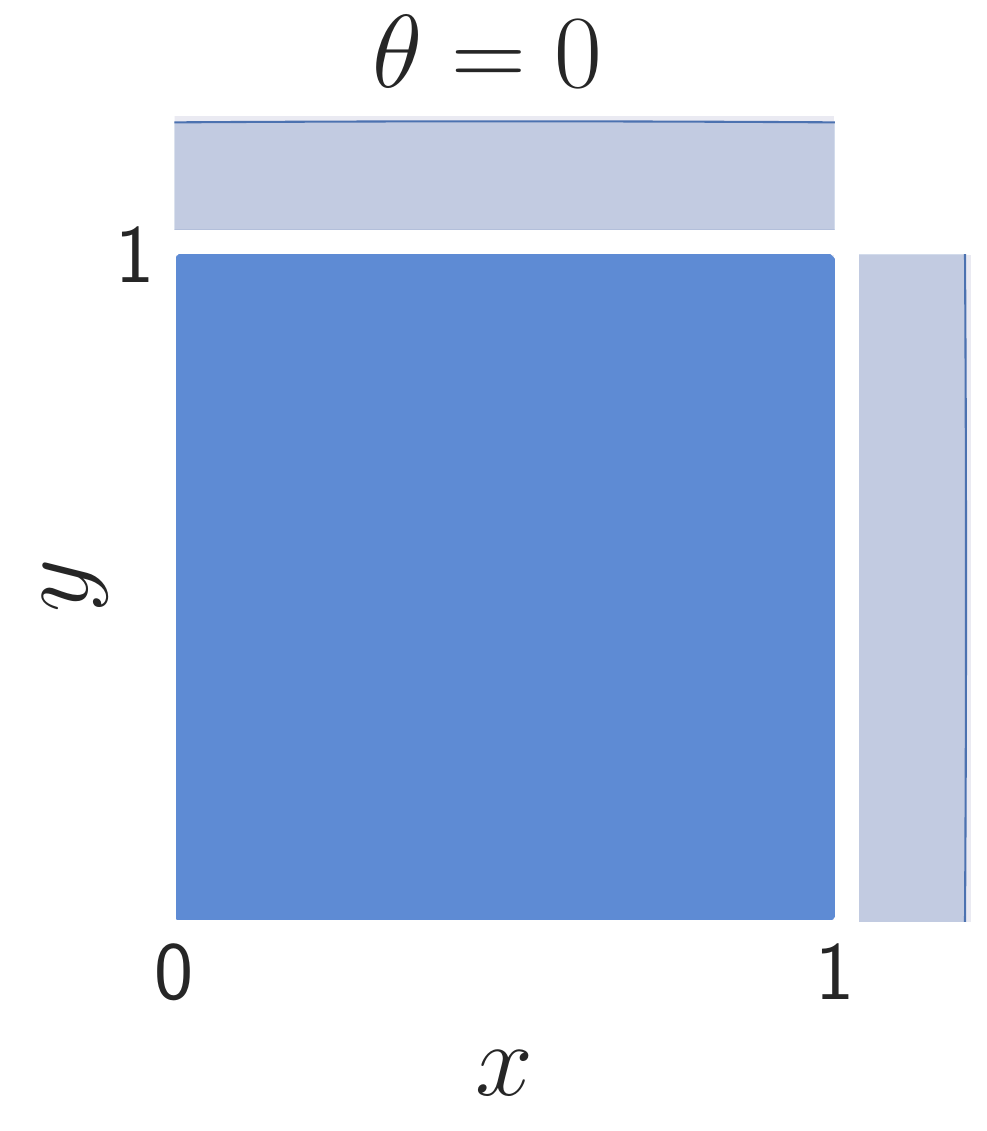}
    \includegraphics[width = 0.19 \textwidth]{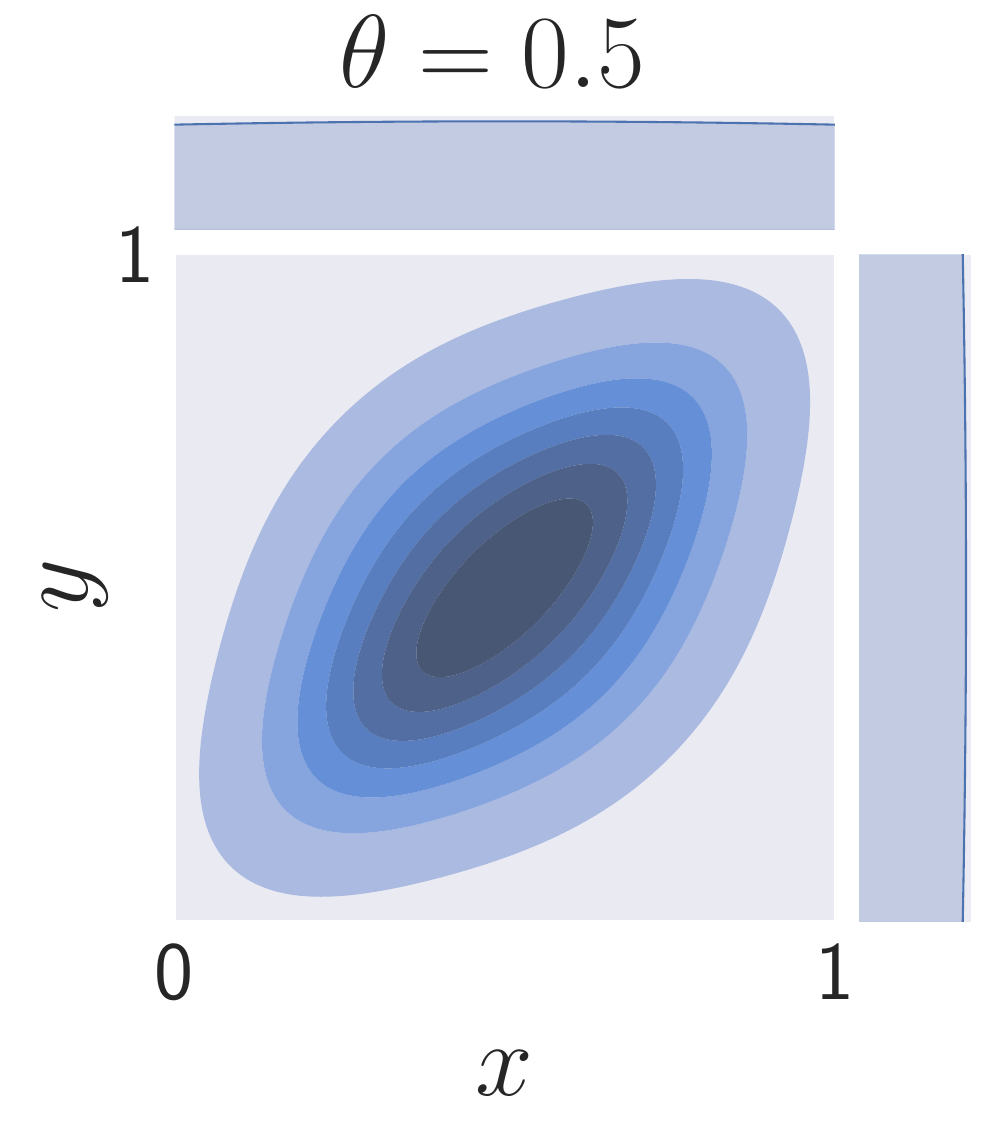}
    \includegraphics[width = 0.19 \textwidth]{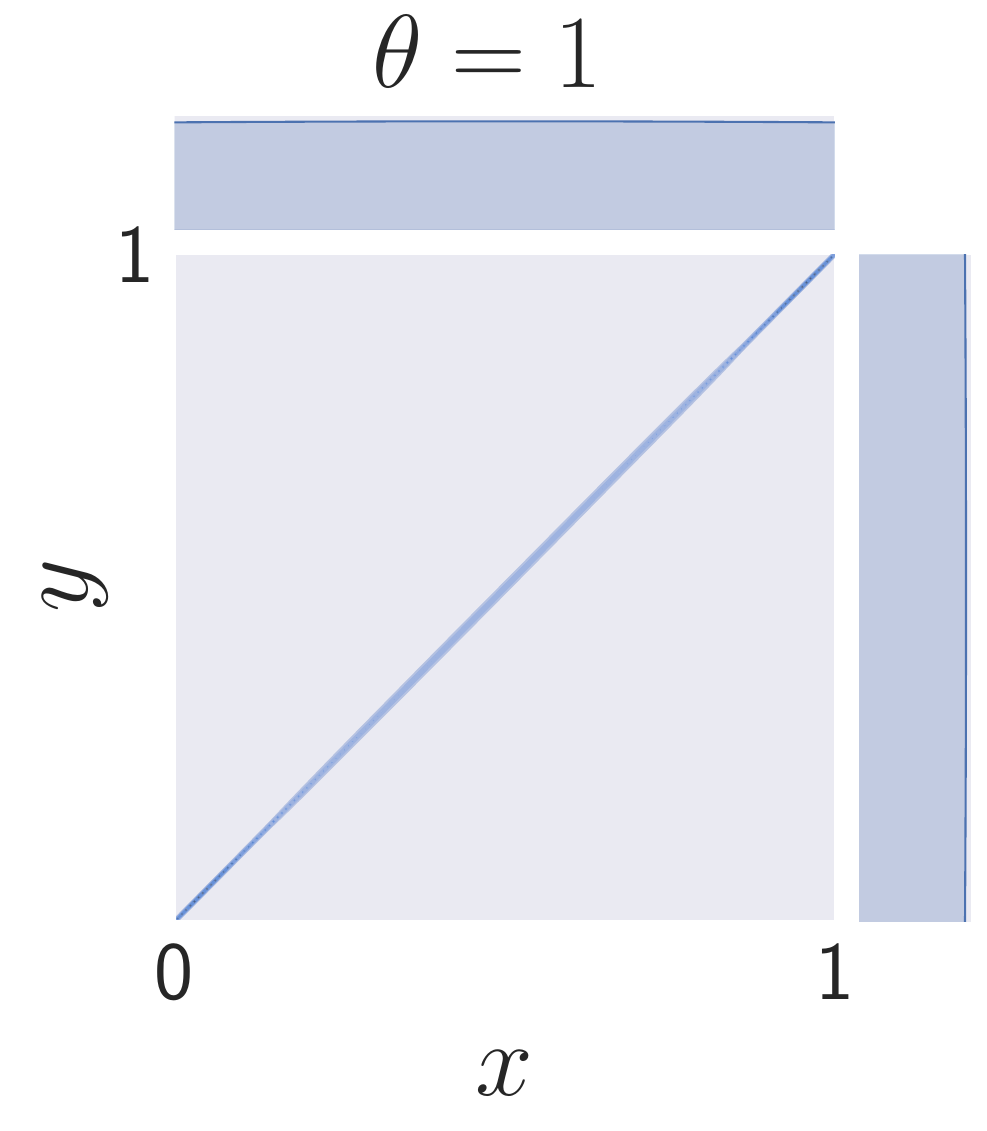} \\ \vspace{0.1cm} \\
    \textbf{\endnotesize{Joint distribution of priority scores.}}\\
    \includegraphics[width = 0.19 \textwidth]{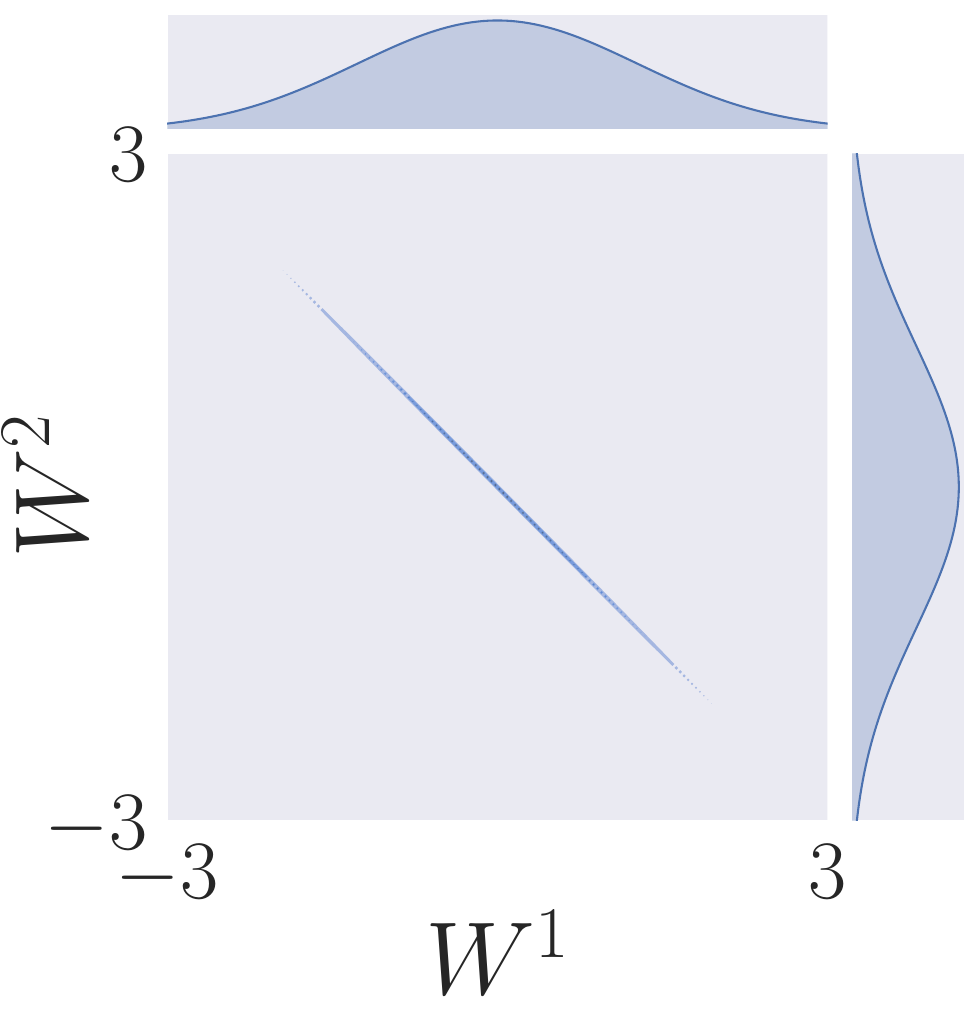}
    \includegraphics[width = 0.19 \textwidth]{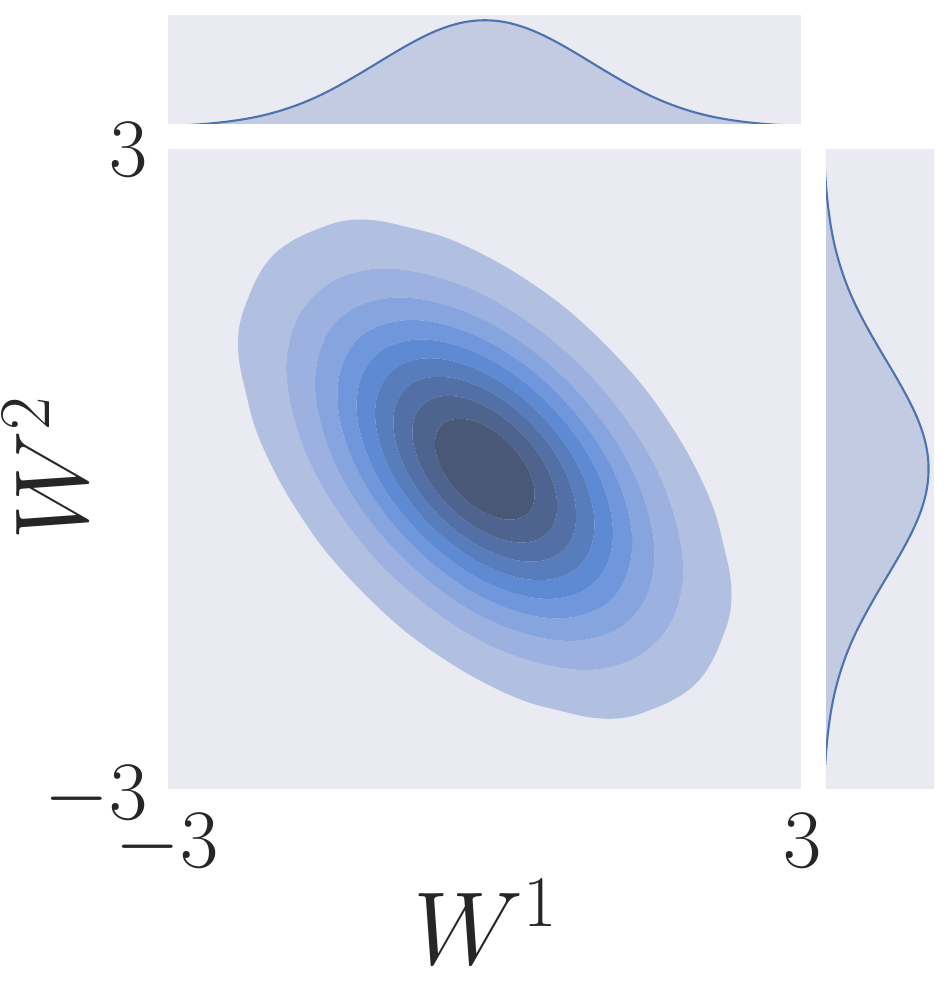}
    \includegraphics[width = 0.19 \textwidth]{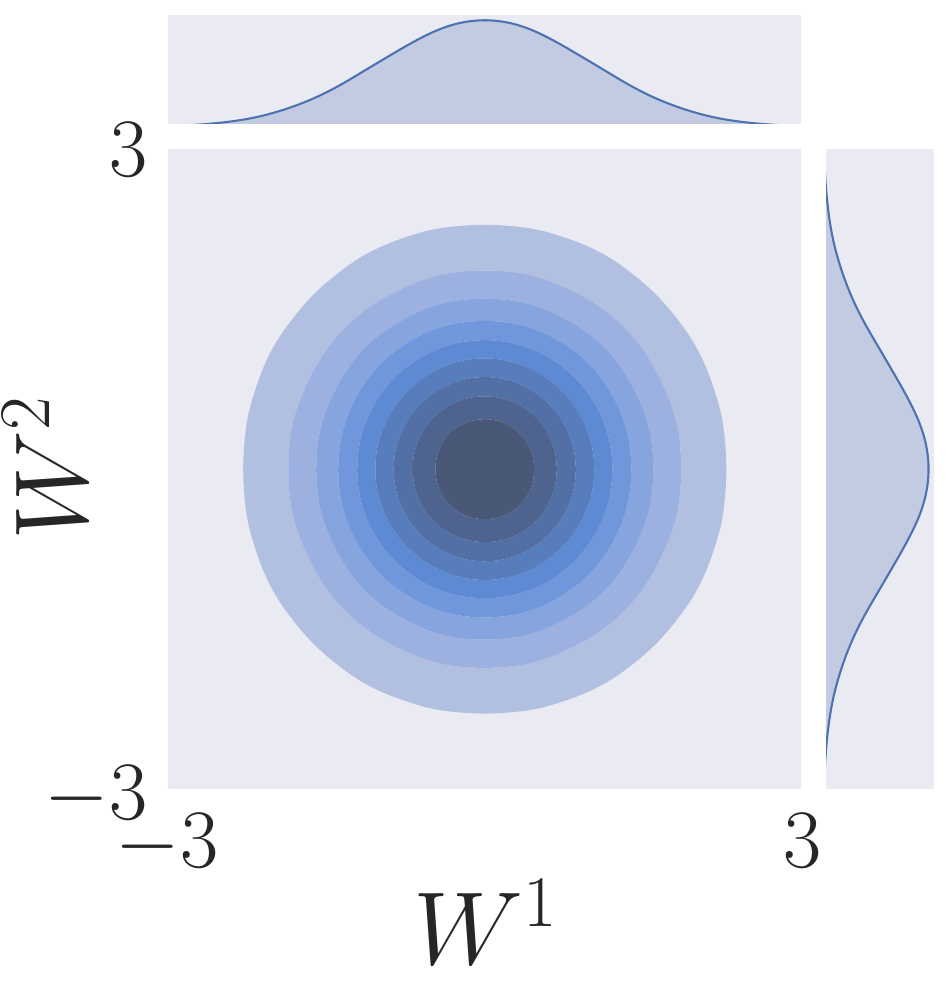}
    \includegraphics[width = 0.19 \textwidth]{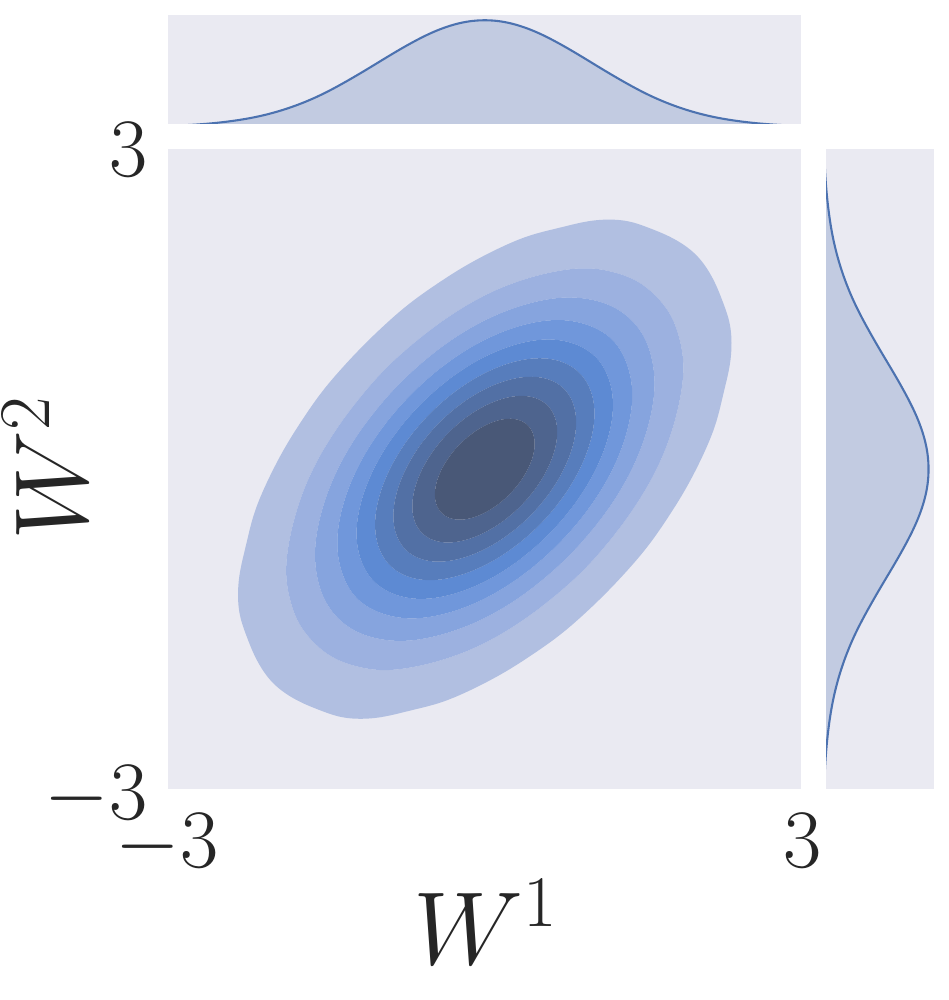}
    \includegraphics[width = 0.19 \textwidth]{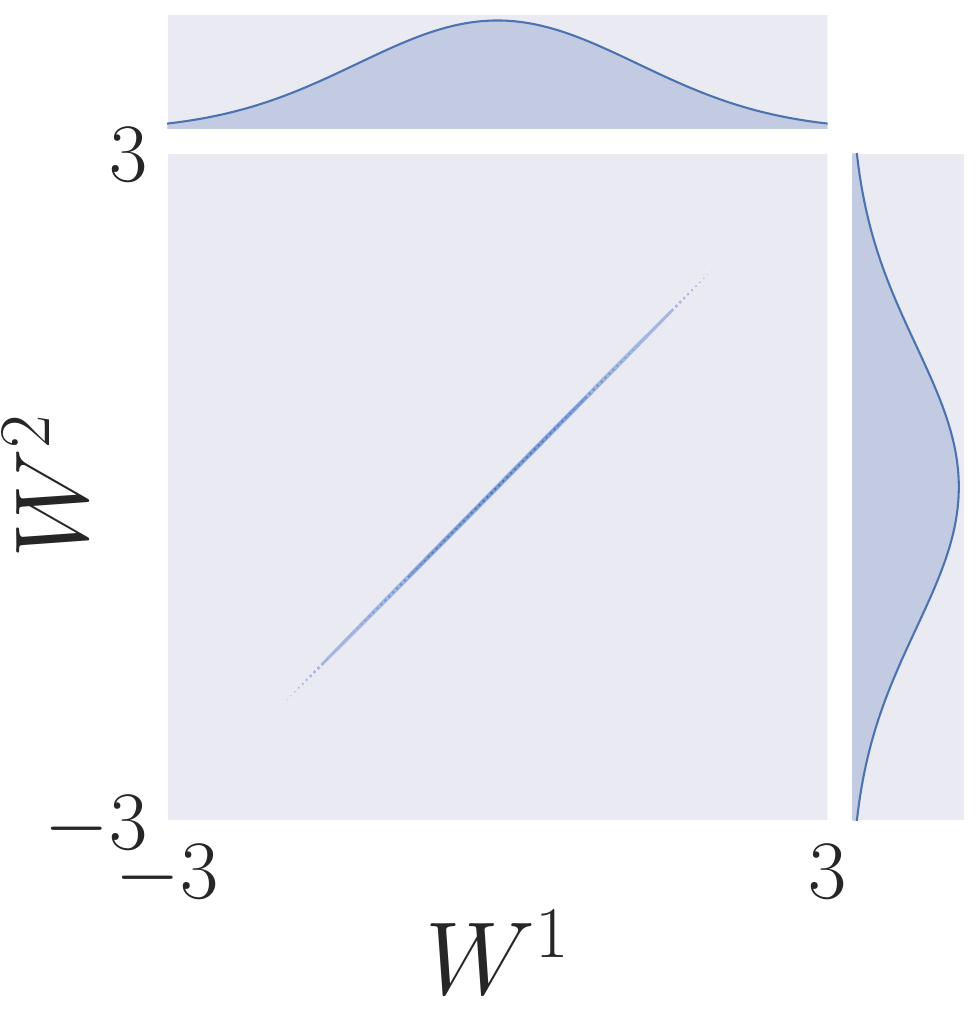}}}
    {Gaussian copula and joint  distributions for two colleges.
    \label{fig:distrib}\vspace{0.45cm}}
    {The columns represent the priority score distributions for different correlation levels. The $x$-axis represents the priority scores assigned by college $\col^1$, and the $y$-axis represents the priority scores assigned by college $\col^2$. The top row illustrates the Gaussian copula for five different correlation levels $\param$, demonstrating how correlation influences the dependence structure between priority scores. The bottom row presents the corresponding joint distributions of priority scores, assuming standard Gaussian marginals for both colleges. 
    The color shading indicates density, with darker shades representing higher probability density. 
    As correlation increases, the distribution shifts toward alignment along the diagonal (higher correlation) or anti-diagonal (negative correlation), meaning that a student’s ranking at one college becomes more predictive of their ranking at another. The marginal distributions of priority scores at each college are shown along the top and right edges of each plot.} 
\end{figure}

Figure \ref{fig:distrib} provides an example, showing the Gaussian copula family with covariance used as parameter $\param$. 

Denote by $\params \coloneqq (\paramj)_{j \in [\ngp]}$ the vector that contains the parameter for each group. Note that each group has a different $\paramj$ and thus a different joint distribution, in addition to possibly different marginals. With some foresight to the explanations provided in Section \ref{sec:correlation}, we call this feature of the model \emph{differential correlation}.\footnote{This is in the spirit of the notion of differential variance studied in \cite*{emelianov22} and \cite*{garg_dropping_2021}.} Finally, we assume that all copulas in $\copcdffamily$ have full support over $[0, 1]^m$. We write $\pdfj$ and $\cdfj$ for the joint multivariate pdf and cdf of group $\gpj$.

Given a family of copulas $\copcdffamily$, we refer to the tuple $(\props, \prefs, \capacities, \pdfs, \params)$ as a \emph{college admissions problem}. Notice that so far, we only assume that distributions admit a density and have full support.
This model allows each group to have different marginal distributions at each college. However, we implicitly assumed that priority scores were independent of preferences within each group; this is a loss of generality, but it can often be overcome.\footnote{
If the independence assumption is not verified, it might be recovered in some cases by subdividing the groups. For example, if the two groups are men and women, and there are two tiers of students, good and bad, with the preferences distribution depending on the tier, then we can divide the existing groups by tiers and obtain four groups within which priorities are independent of preferences. More generally, the independence assumption can be recovered as long as 
there exists a finite subdivision of the initial
groups such that preferences and priority scores are independent within each newly defined subgroup.}

\subsection{Correlation and coherence}\label{sec:correlation}

The purpose of our model is to use $\param$ as a proxy for correlation of rankings, rather than some specific functional form. However, all classical measures of correlation are defined for couples of variables, and there is no standard generalization of the notion of correlation to vectors of higher dimension. We employ the notion of \emph{coherence}, that ensures that $\param$ is aligned with classical measures of correlation for two colleges ($m=2$), and still encompasses the intuitive idea of correlation for more than two colleges ($m>2$).

\begin{assumption}[Coherence]\label{assumption:coherence}
We say that the family of copulas $\left(\copcdft \right)_{\param \in \tset}$ is \emph{coherent} if
for all $x \in (0, 1)^m$, $\copcdft(x)$ is increasing in $\param \mbox{ on } \tset$.
\end{assumption}

When $m=2$, coherence induces an equivalence between $\param$ and two classical ordinal correlation measures; using \cite{scarsini_concordance_1984}, we can show that Spearman's and Kendall's correlation coefficients are increasing functions of $\param$ (cf. Online Appendix A.2 for a further discussion). The Gaussian copula depicted in Figure $\ref{fig:distrib}$ is an example of a coherent family. When $\param = 0$, the variables are independent; when $\param$ is positive, the joint distribution gets closer to the diagonal $x = y$; and when $\param$ is negative, the joint distribution gets closer to the diagonal $y = -x$.

Finally, we introduce a technical assumption that will be required for some of our results, especially when considering comparative statics in $\param$. Denote by $\mathring{\tset}$ the interior of $\tset$.

\begin{assumption}[Differentiability]\label{assumption:differntiability}
We say that the family of copulas $\left(\copcdft \right)_{\param \in \tset}$ is \emph{differentiable} if for all $x \in (0, 1)^m$ and for all $\param \in \mathring{\tset}$, $\copcdft(x)$ is differentiable in the variable $\param$.
\end{assumption}  \vspace{0.3 cm}

The coherence and differentiability assumptions are not particularly restrictive; for instance, the Gaussian copula (with covariance chosen as $\param$) verifies them, and so do other commonly occurring copulas (cf. discussion on distributional assumptions in Online Appendix A.3). Moreover, our model nests the setting where students are assumed to have an---unobserved---latent quality and decision-makers only observe a noisy measurement (cf. Appendix \ref{app:latentnoise}).

\subsection{Stable matching and the supply and demand framework}
\label{sec:matching}

We now introduce the key elements of matching theory used throughout the paper, following \cite{azevedo_supply_2016}.\footnote{For further details on matching theory and the deferred acceptance algorithm, see Online Appendix A.5.}

\begin{definition}[Matchings, cutoffs, and stability]
\phantom{a}
    \begin{itemize}
    \item A matching $\mu:\Stud \to \colset \cup \emptyset$ is a mapping associating each student to a college, or to the empty set if they are unmatched, such that $\mass(\match^{-1}(\coli)) \leq \capi$ for all $i \in [m]$.
    \item For a matching $\match$, for all $i \in [m]$, let $\cuti \coloneqq \inf \setdef{\gradeis}{\match(s)  = \coli}$, be the \emph{cutoff} of college $\coli$.
    \item A matching $\match$ is \emph{stable} if each student $s$ is matched to their favorite college among those for which their priority score is above the respective cutoff, i.e.,  $\grade^i_s \geq P^i$.
    \end{itemize}
\end{definition}

Given a cutoff vector $\cutoffs = (\cut^1, \dots, \cut^m)$, 
let  $\demand_\stud(\cutoffs) \in \colset \cup \emptyset$ be  the \emph{demand} of student $\stud$, i.e., 
their favorite college among those where their priority score is above the respective cutoff (or the empty set if there is none). The \emph{aggregate demand} at college $\coli$ is the mass of students demanding it: $\demand^{i}(\cutoffs) = \mass(\setdef{s}{\demand_\stud(\cutoffs) = \coli})$.
Finding a stable matching then amounts to finding cutoffs such that demand is equal to supply, the latter being the capacity of each college.

\begin{definition}[Market-clearing]
The cutoff vector $\cutoffs$ is \emph{market-clearing} if for $i \in [m]$, $\demand^i(\cutoffs) \leq \capacity^i$, with equality if $\cuti > \min\limits_{j \in [\ngp]} \xsetCijl$. 
\end{definition}

A cutoff vector is therefore market-clearing if it induces a demand that is equal to colleges' capacities when they reach their capacity constraint and lower for colleges that are not full ($\cuti = \min\limits_{j \in [\ngp]} \xsetCijl$ means that $\coli$ rejects no one, and therefore is not full). When the constraint is reached in all colleges, i.e., when $\sum\limits_{i \in [m]} \capi < 1$, the system 
\begin{equation}\label{eq:market-clearing}
    \demands(\cutoffs)  = \capacities
\end{equation}
is called the \emph{market-clearing equation}, and the corresponding \emph{market-clearing cutoffs} can be computed by solving the equation.

The following results from \cite{azevedo_supply_2016} establish the link between market-clearing cutoffs and stable matchings
(for an illustration, see Online Appendix A.4).

\vspace{0.1 cm}

\textbf{\citet[Lemma 1, Theorem 1]{azevedo_supply_2016}.}\footnote{Note that the original theorem specifies conditions on the distribution of students' types, such as being continuous and having full support, which hold in our definition of a college admissions problem. The result stated here is thus a special case of the theorem as stated in \cite{azevedo_supply_2016}.}\hfill
\emph{For any college admissions problem $(\props, \prefs, \capacities, \pdfs, \params)$:
\begin{enumerate}
\item A matching $\match$ is stable if and only if the associated cutoff vector $\cutoffs$ is market-clearing.
\item There exists a unique stable matching. \label{lemma:unique-matching}
\end{enumerate} }

\noindent This result allows us to study stable matchings in the college admissions problem through the unique cutoff vector $\cutoffs(\params)$ associated with it.\footnote{\cite{azevedo_supply_2016} further show that the stable matching varies continuously in the parameters of the problem and that the set of stable matchings from a college admissions problem with a finite number of students converges to the unique stable matching of the continuum problem with the same parameters. The latter result justifies the approximation of large finite instances by their limit. For a better approximation of markets with a small number of students, \cite{arnosti_continuum_2022} proposes a related framework.} We study $\cutoffs$ as a function of $\params$ since we assume all other parameters to be fixed.

\section{Welfare metrics and preliminary results} \label{sec:metrics}

 In selection problems, inequalities between groups are measured by the proportion of admitted candidates in each group. In a matching setting, the situation is more complex; the proportion of students who get one of their top $k$ choices is a relevant metric for all $k \in [m]$, and some group might be advantaged compared to another for some $k$, while being disadvantaged for another $k'$.

\begin{definition}[Rank functions]
Let $\params$ be a correlation vector. In the matching induced by $\params$, let $\text{Rank}(\stud)$ be the rank of the match of student $\stud$ in their preference list, e.g., $\text{Rank}(\stud) = 1$ means that $s$ got their first choice. For any $k \in [m]$, let  
\begin{equation}
\Rjsk(\params) \coloneqq \mathbb{P}(\text{Rank}(\stud)\leq k \vert \stud \in \gpj, \preflist_\stud = \preflist) = \frac{\mass(\{\stud \in \Stud :\text{Rank}(\stud)\leq k, \stud \in \gpj, \preflist_\stud = \preflist\})}{\propj \prefsj}\end{equation}
be the proportion of students from group $\gpj$ with preferences $\preflist$ who get one of their top $k$ choices.
\end{definition}

For example, $\Rjsi( \params)$ is the relative mass of students who obtain their first choice among those in group $\gpj$ who have preferences $\preflist$, or equivalently, it is the probability of a randomly drawn student obtaining their first choice conditionally on belonging to $\gpj$ and having preferences $\preflist$.
Note that we condition on preferences to isolate the effect of differential correlation from differences in preferences. 
The rank metrics can be easily computed from the cutoff vector, as shown in Online Appendix A.5.

\begin{proposition}\label{prop:1st-choice}
If two groups, $\gpj, \gpl$, have the same marginal distribution at college $\coli$, then for students whose first choice is college $\coli$, the probability of obtaining this college is the same for both groups, i.e., $\Rjsi = \Rlsi$. 
\end{proposition}

The proof is provided in Online Appendix C.1.\footnote{Note that Proposition \ref{prop:1st-choice} remains valid when capacity is not constrained, i.e., $\sum\limits_{i \in [m]}\capi \geq 1$.
} 
Proposition \ref{prop:1st-choice}, though simple, is an important property of the model. If two students prefer the same college, then their probabilities of obtaining it only depend on their respective groups' marginals and not on their correlations; thus, differential correlation has no effect on this metric.

Another metric of interest is the proportion of students who remain unmatched, which is the complement of the proportion of students getting one of their top $m$ choices:
\begin{equation} \label{eq:Rjsk}
    \RjsE(\params) \coloneqq \mathbb{P}(\match(s) = \emptyset \vert \stud \in \gpj, \preflist_\stud = \preflist) = 1 - \Rjsm(\params)
\end{equation}
We can derive a simple yet important result. 

\begin{proposition}\label{lemma:unmatched}
The probability that a student remains unmatched depends only on their group and is independent of their preferences, i.e., for all $j,\preflist,\preflist'$, $\RjsE(\params) = R_j^{\emptyset, \preflist'}( \params)$.
Moreover, the total mass of unmatched students is constant in the correlation of any group and is equal to $1 - \sum\limits_{i \in [m]}\capi$.
\end{proposition}

The proof is provided in Online Appendix C.2.\footnote{Note that the statement of Proposition \ref{lemma:unmatched} is only true because we have assumed throughout that students prefer any college over staying unmatched. Otherwise, students with a short preference list would have a higher probability of staying unmatched than those with a long preference list.}
With Proposition \ref{lemma:unmatched} at hand, we use the notation $\RjE$ since these quantities do not depend on the preferences of the students. For all the metrics we defined, when there is no ambiguity, we also omit the dependence on $\params$ and write $\Rjsk$ and $\RjE$ instead.

We next define two global metrics, i.e., metrics that are not conditioned on the groups and preferences of students.

\begin{definition}[Efficiency and Inequality] \label{def:eff_ineq}
For $\params \in \tset^\ngp$ and $\match_{\params}$ the induced stable matching, define the \emph{efficiency} $\eff(\params)$ of a matching as the proportion of students obtaining their first choice, and the \emph{inequality} $\ineqjl(\params)$ between two groups, $j,\ell \in [\ngp]$ as the difference in the probability of remaining unmatched between those two groups:
\begin{align}
\eff(\params)  & = \mass\left(\{\stud \in \Stud : \match_{\params}(\stud) = \col^{\preflist_s(1)} \} \right)  \\
\ineqjl(\params)  & = \vert \RjE(\params) - \RlE(\params)  \vert.
\end{align}

\end{definition}

We sometimes focus on these metrics as they represent
the two extremes, getting one's first choice or not getting admitted at any college, and are thus arguably of particular importance. Also, note that when there are only two colleges, efficiency and inequality fully capture all rank metrics defined above.

Before we turn to our main results concerning the impact of correlation, we characterize the impact of marginal distributions on our metrics.

\begin{proposition} \label{prop:FOSD}
Suppose the priority score distributions of group $\gpj$ first-order stochastically dominate those of group $\gpl$ at all colleges\footnote{This means that $\cdfCij < \cdfCil$ for all $i \in [m]$.} and correlations are equal ($\paramj = \paraml$). Then, group $\gpj$ is favored, that is,
 $\Rjsk > \Rlsk$ for all $k \in [m]$ and all $\preflist \in \preflistsm$ and there is positive inequality,
$\ineqjl > 0$. 
\end{proposition}

The proof is provided in Online Appendix C.3. 
This proposition shows that differences in the marginal distributions create inequalities between students in the probability of being matched, but also in the probability of getting a top $k$ choice, for all $k$. Marginal distributions impose a baseline level of inequality that will be increased or mitigated by differential correlation, as we shall see in the following section.

\section{Main results} \label{sec:main_results}

We consider college admissions problems where 
the
 group sizes 
($\props$), the colleges' capacities ($\capacities$), 
the students' preferences ($\prefs$), and the marginal priority score distributions ($\pdfs$) are fixed and study the impact of the correlation, $\params$, on the stable matching. Section \ref{subsec:compstat} contains general comparative statics, and Section \ref{subsec:TB}
examines tie-breaking.
We assume that Assumptions \ref{assumption:coherence} (coherence) and \ref{assumption:differntiability} (differentiability) hold. Finally, the results presented here rely on an additional, crucial assumption.\footnote{Recall also that we assume that capacity is constrained, which is needed for our results. See Appendix \ref{app:excess} for a result on unconstrained capacity.}

\begin{assumption}[Decreasing cutoffs]\label{assumption:dec_cutoffs}
For fixed parameters $\props, \prefs, \capacities, \pdfs$, we say that \emph{decreasing cutoffs} holds if
for all $\params \in \tset^\ngp$, for all $i \in [m]$, for all $j \in [\ngp]$, the cutoff $\cuti(\params)$ decreases when $\paramj$ increases.
\end{assumption}

We show in Section \ref{sec:dec_cutoffs} that this assumption always holds for two colleges and holds for more than two colleges under common symmetry assumptions on preferences and capacities
from the literature. Moreover, we show through simulations that our results remain qualitatively valid across the parameter space.

\subsection{Comparative statics}\label{subsec:compstat}

We first consider how the efficiency of the matching varies when changing the correlation for one group.

\begin{theorem}[Efficiency is increasing in correlation]\label{thm:V1-inc} Suppose Assumptions \ref{assumption:coherence}, \ref{assumption:differntiability} and \ref{assumption:dec_cutoffs} hold. For all $k \in [m]$ and all preferences $\preflist$, for $j \in [\ngp]$, 
 \begin{enumerate}
     \item the proportion of students in group $\gpj$ with preferences $\preflist$ who obtain one of their top $k$ choices, $\Rjsk$, is increasing
in all the other groups' correlations $\paraml$ for $\ell \neq j$, and
\item the proportion of students in group $\gpj$ with preferences $\preflist$ who obtain their first choice, $\Rjsi$, is also increasing in $\gpj$'s own correlation $\paramj$.
\end{enumerate}
Consequently, the global efficiency $\eff(\params)$ is increasing in all components of $\params$. 
\end{theorem}

\proof{Proof.}
Lemma 4 in Online Appendix A.6 states that $\cutoffs(\params)$ is of class $\mathcal{C}^1$.\footnote{This is not necessarily the case if $\params$ is not in the interior of $\tset^\ngp$. However, since the $\Rjsk$ are continuous, they are still increasing over the whole interval $\tset$.} Moreover, under Assumption \ref{assumption:dec_cutoffs} the cutoffs are decreasing in all $\paramj$. Since $\Rjsk = \mathbb{P}_{j, \paramj}(\grade^{\preflist(1)} \geq \cut^{\preflist(1)} \cup \dots \cup \grade^{\preflist(k)} \geq \cut^{\preflist(k)}) $\footnote{The notation $\mathbb{P}_{j, \paramj}$ denotes the probability distribution of  priority scores of group $\gpj$, with correlation $\paramj$.}, if $\paraml$ increases for $\ell \neq j$, the distribution $\mathbb{P}_{j, \paramj}$ does not change but all cutoffs decrease, therefore $\Rjsk$ increases. If $\paramj$ increases, since $\Rjsi = \mathbb{P}_{j}(\grade^{\preflist(1)} \geq \cut^{\preflist(1)})$, we can conclude with the same argument. Moreover, $\eff$ is a convex combination of all the $\Rjsi$ and therefore is also increasing in all $\param_\ell$.
\hfill \Halmos
\endproof

\
 Theorem~\ref{thm:V1-inc} implies that generally the outcome improves as correlation increases. In particular, students from a given group always benefit from an increase in correlation in other groups. However, if one's own group's correlation increases, this is only guaranteed to be beneficial in terms of admittance at one's favorite college, 
 because the mass of its priority score distribution accumulates below all cutoffs. This can be observed by going back to Figure \ref{fig:cor-cutoff_intro}. Finally, note that Theorem~\ref{thm:V1-inc} also provides insight into the impact of correlation on the efficiency of a matching without groups (i.e., only one group).

We next consider how the inequality of a matching varies when changing the correlation for one group. Recall from Proposition \ref{lemma:unmatched} that the probability that a student remains unmatched is independent of their preferences, and thus we consider $\RjE$.

\begin{theorem}[Low-correlation groups are advantaged]\label{thm:unmatched}
Suppose Assumptions \ref{assumption:coherence}, \ref{assumption:differntiability} and \ref{assumption:dec_cutoffs} hold. For $j \in [\ngp]$, the proportion of students in group $\gpj$ remaining unmatched, $\RjE$, 
\begin{enumerate}
     \item is decreasing in  $\paraml$, for all $\ell \neq j$, 
     and
     \item  is increasing in $\gpj$'s own correlation, $\paramj$
     .
\end{enumerate}
Consequently, the inequality between any two groups $\ell\neq j$, $\ineqjl(\params)$, decreases in the correlation  of the group with the lower rate of unmatched students and increases in the correlation  of the other group. 
\end{theorem}

\proof{Proof sketch.}
Fix $\ell \in [\ngp]$. We show that the $\RjE$ are increasing in $\paraml$ for $j \neq \ell$ using Assumption \ref{assumption:dec_cutoffs} (decreasing cutoffs). We then use the fact that total capacity is constant to deduce that $\RlE$ has to be decreasing in $\paraml$ to compensate for the others. The complete proof is provided in Online Appendix C.4.
\hfill 
\endproof

Theorem \ref{thm:unmatched} implies that, ceteris paribus, a group's  students are more likely to be matched when having low correlation. This is the case because an ``independent second chance'' is preferable to ``carrying over the bad signal from previous rejections''.
Since a given group's students are more likely to be matched as their correlation decreases, and capacity is limited, it follows that all other students become less likely to be matched. 
 Consequently, inequality decreases when the correlation of the group with the lower proportion of unassigned students (that is, the better-off group) increases or when the correlation of the group with the higher proportion of unassigned students (that is, the worse-off group) decreases.

Note that the effect of one's own group's correlation on obtaining a match other than one's favorite college (but not remaining unmatched) is neither covered by Theorems \ref{thm:V1-inc} and \ref{thm:unmatched}; the effect is not monotone in correlation and depends on the other parameters of the college admissions problem.

Theorem \ref{thm:unmatched} also allows us to formulate a corollary
relating the inequality from different marginal distributions to the inequality from differential correlation.

\begin{corollary}\label{cor:unmatched} 
Suppose Assumptions \ref{assumption:coherence}, \ref{assumption:differntiability} and \ref{assumption:dec_cutoffs} hold. 
    If groups $\gpj$ and $\gpl$ have the same marginals, $\RjE < \RlE$ if and only if $\paramj < \paraml$.  
   If $\gpj$ first-order stochastically dominates $\gpl$ at all colleges, for any fixed $\paraml$, there exists $\hat{\param} > \paraml$ such that $\RjE < \RlE$ if and only if $\paramj < \hat{\param}$, i.e.,  the inequalities from different marginal distributions and from differential correlation are cumulative. 
\end{corollary}

\proof{Proof.} 
If marginal distributions are equal, and $\paraml = \paramj$ there is no inequality. As $\paraml$ increases the proportion of unmatched students increases for $\gpl$ and decreases for $\gpj$ by Theorem \ref{thm:unmatched}, proving the first part. If $\gpj$ first-order stochastically dominates $\gpl$ at all colleges, and $\paramj = \paraml$, Proposition \ref{prop:FOSD} states that the inequality is in favor of $\gpj$. By Theorem \ref{thm:unmatched}, a decrease in $\paramj$  further increases inequality, while an increase in $\paramj$ decreases the inequality until it reaches zero (if it does), then the inequality is reverted.\hfill \Halmos
\endproof

Corollary \ref{cor:unmatched} shows that the proportion of unmatched students is higher in the higher-correlation group, even with identical marginals. This proves the existence of inequalities specific to matching problems. Indeed, even when the rankings of each college are ``fair'', i.e., all groups are represented at all levels of each college's ranking in the same proportion as in the applicant population (which is equivalent to saying that all groups have the same marginals), inequality exists due to differential correlation.
This finding contrasts with Proposition \ref{prop:1st-choice}, which states that with identical marginals, the proportion of students who get their first choice is the same in all groups (conditional on preferences). 
The second part of Corollary \ref{cor:unmatched} considers the case when marginals are different; hence, there is some ``baseline" inequality when both groups have the same correlation. When the group with the higher marginal distribution also exhibits lower correlation, inequality increases further. Conversely, if it exhibits higher correlation, inequality is reduced and might even be reversed when the correlation gap becomes large.

So far, we examined efficiency and inequality separately. The following proposition describes their interaction. Specifically, different correlation vectors $\params$ can lead to the same efficiency, while inducing different levels of inequality between groups. This shows that differential correlation has a distinct impact on efficiency on one hand and on inequality on the other.

\begin{proposition} \label{prop:efficiency} 
Suppose Assumptions \ref{assumption:coherence}, \ref{assumption:differntiability} and \ref{assumption:dec_cutoffs} hold, and let $\params \in \tset^\ngp$. 
\begin{enumerate}
    \item
    There exists a continuum\footnote{Unless $\params = (\inf \tset, \dots, \inf \tset) \text{ or } (\sup \tset, \dots, \sup \tset)$, in which case it is a singleton.} of correlation vectors that achieve a given efficiency. Formally, the set of vectors $\params'$ such that $\eff(\params') = \eff(\params)$ is a connected hypersurface of dimension $\ngp-1$.\footnote{In this context, a connected hypersurface is defined as the graph of a continuous function from $\tset^{\ngp-1}$ to $\tset$.}
    \item  Fixing efficiency, correlations are substitutes, i.e., for any two groups $\gpj, \gpl$, there exists an interval $U \coloneqq [\underline{\param}, \Bar{\param}]$ and a decreasing function $\phi:U \to \tset$ such that $\paraml = \phi(\paramj)$ for all $\paramj \in U$. 
    \item Over the graph of $\phi$, $\RjE$ is minimized at $(\underline{\param}, \phi(\underline{\param}))$ and maximized at $(\Bar{\param}, \phi(\Bar{\param}))$. Moreover, there is a unique $\hat{\param}\in U$ such that $(\hat{\param},\phi(\hat{\param}))$ minimizes inequality $\ineqjl(\params)$.
\end{enumerate}
\end{proposition}

\proof{Proof sketch.}
We combine Theorem \ref{thm:V1-inc} with the implicit function theorem. For point 1, we fix the efficiency, arbitrarily choose some $\paramj$ and express $(\paraml)_{\ell \neq j}$ as a function of $\paramj$, the graph of this function is a connected hypersurface. For point 2, we do the same, but this time we only express one specific $\paraml$ as a function of $\paramj$. Finally, point 3 is obtained by applying Theorem \ref{thm:unmatched} to this function.
The complete proof is provided in Online Appendix C.5. 
\endproof

Beyond the intuition that correlation favors efficiency, Proposition \ref{prop:efficiency} provides a precise insight into the relation between efficiency and inequality and the trade-off between the two. The first part states that, in general, there are infinitely many correlation vectors achieving the same level of efficiency.
The second part considers the comparative statics between two groups. Fixing the efficiency, the correlation parameters behave as rival goods. As the correlation increases for one group, it necessarily decreases for the other group. Using Theorem \ref{thm:unmatched}, the third part of Proposition \ref{prop:efficiency} states that this dynamic also implies that the proportion of unmatched students increases in the first group and decreases in the second; therefore there is a unique point minimizing inequality. 

\begin{figure}[ht]
    \FIGURE
    {\includegraphics[width = \textwidth]{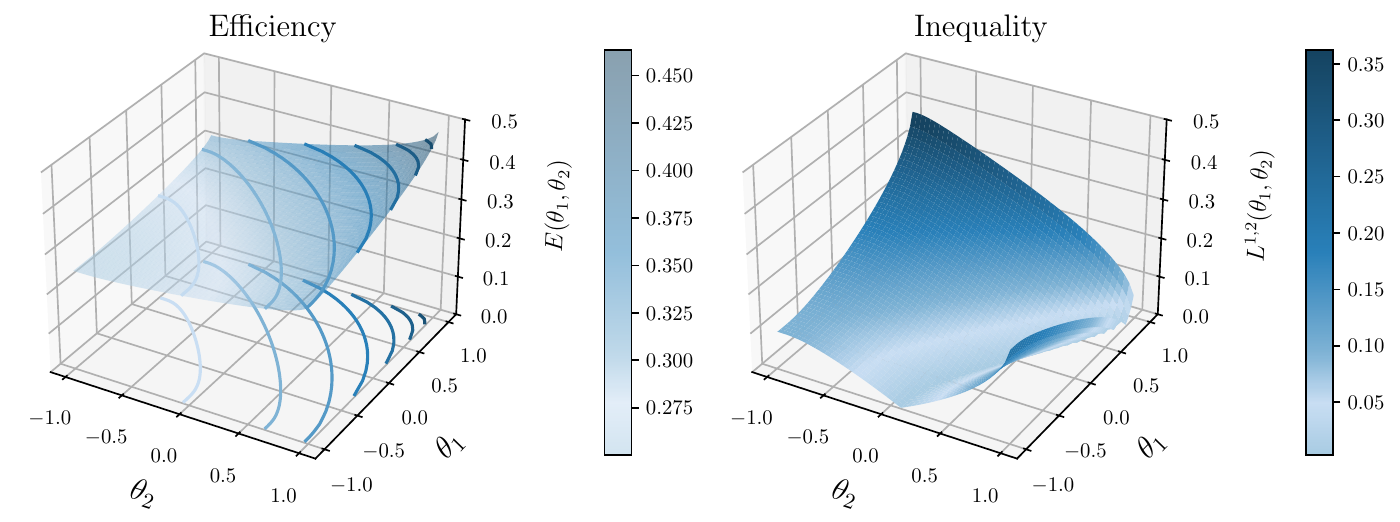}}
    {Variations of efficiency ($\eff$) and inequality ($\ineq$). \label{fig:prop_efficiency}
    }
    {Illustration for two colleges and two groups, with bivariate Gaussian distributions (mean = (0, 0) for $\gp_1$, (0.2, 0.2) for $\gp_2$, variance = 1 for both) and $\param_1, \param_2$ chosen as each group's covariance. Other parameters: $\capacity^1 = \capacity^2 = 0.25$, $\prop_1 = \prop_2 = 0.5$, $\pref_{12} = \pref_{21} = 0.5$. Left: The surface shows the efficiency and the level lines indicate constant efficiency (also projected to the bottom of the figure). Right: The surface shows the inequality.}
    
\end{figure}

Figure \ref{fig:prop_efficiency} shows the levels of efficiency (left panel) and inequality (right panel) for two groups and two colleges with Gaussian distributions as functions of the correlation parameters $\param_1,\param_2$.
On the left panel, the level lines show the decreasing relation between $\param_1$ and $\param_2$ when $\eff$ is kept constant. On the right panel, following these level lines, observe that inequality is maximized when the gap between $\param_1$ and $\param_2$ is maximal, and minimized on a line located on the right of the diagonal $\param_1 = \param_2$ due to the fact that group $\gp_2$ first-order stochastically dominates group $\gp_1$, as predicted by Corollary \ref{cor:unmatched}.

\subsection{Tie-breaking} \label{subsec:TB}

A recent literature has studied the impact of tie-breaking rules on school choice problems, which is a special case of our model where colleges do not have preferences over students (or, equivalently, are indifferent). In this section, we extend some of the prior results and discuss the relation to the literature.

In a \emph{tie-breaking problem}, colleges partition students into priority classes. Students belonging to the same priority class at some college are assumed to have the same priority at that college; however, due to limited capacity, the college might need to choose between them. To achieve this, colleges use a random ranking of students to which they refer each time they need to choose between students from the same priority class; this random ranking is called \emph{tie-breaker}. 

A natural question that has been actively studied in recent years is whether there is a difference in students' welfare if colleges all use the same tie-breaker (called a single tie-breaker, or STB) instead of each producing an independent tie-breaker (called multiple tie-breakers, or MTB). \cite*{ashlagi_assigning_2019-1, ashlagi_competition_2020, arnosti_lottery_2022} show---with slightly different models and assumptions (and among other results)---that when the total capacity of colleges is lower than the number of students, then students are better off under STB than under MTB. To ease the comparison, we restate their results in a simplified form.

\begin{itemize}
    \item[\emph{{\citealt*[Main Theorem]{ashlagi_assigning_2019-1}}}:] Suppose that students' preferences are drawn uniformly at random,  there is only one priority class (the whole ranking is random), and the number of students is close to the number of seats; then, for any fixed $k$, the fraction of students matched to one of their top $k$ choices approaches a positive constant under STB, but
    approaches $0$  under MTB  as the market gets large.
    \item[\emph{{\citealt[Theorem 3.1]{ashlagi_competition_2020}}}:] Suppose there is one slot per college, students' preferences are drawn uniformly at random, there is only one priority class, and there is capacity shortage; then, with high probability, the fraction of students matched to one of their top $k$ choices is higher under STB than MTB for all $k$.
    \item[\emph{{\citealt[Theorem 2]{arnosti_lottery_2022}}}:] Suppose there is only one priority class, students' preferences are drawn uniformly at random, and students do not have to list all colleges. Then, there exists a threshold $\ell$ such that for $k \leq \ell$, students who list $k$ colleges or more have a higher chance of getting a top $k$ choice under STB than MTB, and for $k > l$ it is the opposite. In particular, the number of students getting their first choice is higher under STB.\footnote{Note that this also implies that there are more unmatched students overall under STB than MTB, which means that STB does not dominate MTB. However this is due to students not listing all schools. In our model, we assume that students list all schools and therefore the global mass of matched students does not depend on the correlation or choice of tie-breaker.}
\end{itemize}

Our model, compared to prior work on tie-breaking, allows for any number of priority classes, intermediate levels of correlation or even negative correlation, and several groups of students with differently correlated tie-breaking rules. Intermediate correlations can arise in tie-breaking if, for example, student characteristics are introduced into rankings to break ties, e.g., sibling priority \citep{correa_chile_2022}. 
This is commonly done to
render algorithms more deterministic, and thus explainable. 
Consider sibling priority; ceteris paribus, a student who enjoys sibling priority at one school
exhibits lower correlation between the priority scores at this and any other school than without sibling priority. 

To this end, let $\copcdffamily$ be a family of copulas such that $\param = 0$ gives independent random variables and $\param=1$ gives fully correlated variables. Define the $\param$-TB as the tie-breaker drawn according to $\copcdft$. Thus, MTB corresponds to $\param = 0$, and STB corresponds to $\param = 1$. Moreover, we can assume the existence of several groups with different values of $\param$.

Let there be a continuum mass of students and assume that students prefer any college over being unmatched.
Suppose that each college $\coli$ has $\tau^i$ priority classes $\{\class^i_{1}, \dots, \class^i_{\tau^i}\}$, each student belongs to a product of priority classes (one for each college), and all products of priority classes contain a positive mass of students from each group. Furthermore, suppose that the students are divided into $\ngp$ groups, such that the $\paramj$-TB is used for group $\gpj$. Define 
the group sizes 
($\props$), the colleges' capacities ($\capacities$), and the students' preferences ($\prefs$) as before. Finally, assume that Assumption \ref{assumption:dec_cutoffs} (decreasing cutoffs) holds on the interior of each product of priority classes. 
\begin{proposition} \label{prop:TB}
 Suppose Assumptions \ref{assumption:coherence}, \ref{assumption:differntiability} and \ref{assumption:dec_cutoffs} hold. Then, in the school choice problem described above, for any $\ell \in [\ngp]$, the metrics $\Rjsk$ (for all $j\neq \ell$, $k\in [m]$), $\Rlsi$, and the efficiency $\eff$, are almost surely  increasing in $\paraml$ (and otherwise constant),\footnote{More precisely, the set of vectors $(\props, \prefs, \capacities)$ such that those metrics are constant in some $\paramj$ has Lebesgue measure 0.} and always increasing if there is only one priority class.

    \noindent Inequality between two groups, $\ineqjl(\params)$, is non-decreasing in the tie-breaker correlation of the group with the higher proportion of unassigned students and non-increasing in the other group's tie-breaker correlation.
\end{proposition}

\proof{Proof sketch.}
We build a distribution family that encompasses the priorities of students at each college, accounting for priority classes and tie-breakers, such that MTB and STB correspond to $\param = 0$ and $\param = 1$, respectively. The distribution obtained, while complex, still satisfies most of the assumptions required by our model; with some adjustments, we are able to apply Theorems \ref{thm:V1-inc} and \ref{thm:unmatched} and conclude. The proof is  provided in Online Appendix C.6.
\endproof

This result shows that increasing the correlation of tie-breakers, for one or several groups, increases the proportion of students who obtain their first choice in all groups, as well as the proportion who obtain one of their top $k$  choices in all groups except the one whose correlation increased. Moreover, it also shows that a policy-maker who is able to change the correlation of tie-breakers for some groups can use it to mitigate inequalities between groups.

Proposition \ref{prop:TB} is in some regards more restrictive than the results from the literature presented above, because it requires decreasing cutoffs and that students prefer any college over being unmatched. It also does not prove that STB matches more students to one of their top $k$ choices than MTB for $k>1$, but it is worth noting that the aforementioned results rely on stronger assumptions to do so. On the other hand, it is more general in other regards, as already discussed above: Proposition \ref{prop:TB} allows for several priority classes, does not require students' preferences to be uniformly distributed, accommodates for several groups with different tie-breaking rules, and allows for intermediate tie-breaking rules that interpolate between MTB and STB, as well as for negatively correlated tie-breaking rules.

\subsection{The decreasing cutoffs assumption}

\label{sec:dec_cutoffs}

We explore the conditions under which Assumption \ref{assumption:dec_cutoffs} (decreasing cutoffs) holds, since it is required for all the results presented in Sections \ref{subsec:compstat} and \ref{subsec:TB}. We first provide theoretical guarantees for specific assumptions on the parameters, and then we show numerical experiments  examining the results across the parameter space.

\subsubsection{Theoretical guarantees.} \label{subsec:theor_guar}

\begin{theorem} \label{thm:cutoff_inc}
   Suppose there are two colleges and  the family of copulas $\copcdffamily$ is coherent (Assumption \ref{assumption:coherence}), then decreasing cutoffs always holds.
\end{theorem}

\proof{Proof sketch.}
By coherence, at least one of the cutoffs has to be decreasing. By contradiction, we suppose that the other is not, and we show that the demand at some college is decreasing while it is supposed to be constant and equal to the capacity of this college.
The key element that is only true for two colleges is the fact that $\mathbb{P}_{j, \paramj}(\grade^1 < \cut^1, \grade^2 \geq \cut^2)$ is decreasing in $\paramj$ due to the coherence assumption. The proof is provided in Online Appendix C.7.
\endproof

From Theorem \ref{thm:cutoff_inc}, it follows that all the results presented in Sections \ref{subsec:compstat} and \ref{subsec:TB} are always true for two colleges.
This observation allows to build intuition, since our results are easier to understand and to illustrate for two colleges.
In addition,
it is straightforward to show that in a model where all students prefer two colleges over all others, decreasing cutoffs holds for the two preferred colleges (see Appendix \ref{app:2_cols} for a formal statement). Such preferences appear in practice; consider, for example, the universities of Oxford and Cambridge in the United Kingdom, École polytechnique and École Normale Supérieure in France, or, departing from colleges, McKinsey and BCG among consulting firms.

When there are more than two colleges, Assumption \ref{assumption:dec_cutoffs} holds under additional assumptions that are common in the literature (cf., e.g., \citealt*{ashlagi_assigning_2019-1, ashlagi_competition_2020, peng_monoculture_2023}), namely that colleges have identical capacities and students' preferences are uniformly distributed (i.e., $\forall j\in [\ngp], \forall \preflist \in \preflistsm$, $\prefsj = 1/m!$).

\begin{proposition} \label{prop:symmetry}
    Suppose Assumption \ref{assumption:coherence} holds, all colleges have the same capacity, and student preferences are uniformly distributed. Then decreasing cutoffs holds for any $\props$ and $\pdfs$.
\end{proposition}

\proof{Proof.}
Under the mentioned assumptions all colleges are identical; then, by a symmetry argument, in the unique stable matching all cutoffs are equal. Since the coherence assumption implies that at least one cutoff has to decrease, they are in fact all decreasing. \hfill \Halmos
\endproof

By a continuity argument decreasing cutoffs holds when appropriately relaxing the assumptions of Proposition \ref{prop:symmetry}.

\begin{corollary}\label{cor:continuity}
   Suppose Assumptions \ref{assumption:coherence} and \ref{assumption:differntiability} hold. Then, decreasing cutoffs holds on a subset of positive measure of the space of parameters $\props, \capacities, \prefs$.
\end{corollary}

\proof{Proof.}
By Lemma 4 (c.f. Online Appendix A.6), cutoffs are $\mathcal{C}^1$ functions of $\params$, and by \citet[][Theorem 2]{azevedo_supply_2016} they are continuous in all parameters $\props, \capacities, \prefs$. Therefore, for all $i \in [m], j \in [\ngp], d \cuti/d\paramj$ is continuous in $(\props, \capacities, \prefs)$, so it is negative in a neighborhood of the set of points that verify the assumptions of Proposition \ref{prop:symmetry}. \hfill \Halmos
\endproof

\subsubsection{Numerical Experiments.} \label{subsec:numerical}

We perform numerical experiments to examine whether
Assumption \ref{assumption:dec_cutoffs} (decreasing cutoffs) and the results presented in Sections \ref{subsec:compstat} and \ref{subsec:TB}
hold beyond the settings covered by Theorem \ref{thm:cutoff_inc},  Proposition \ref{prop:symmetry}, and Corollary \ref{cor:continuity}. 
Our experiments show that, while the results sometimes fail, they remain qualitatively valid. 

We analyze a broad range of parameter values for three colleges and two groups (and four colleges and two groups in Online Appendix D.2). We model priorities via Gaussian copulas, with covariance matrix
$\Gamma_\param = \begin{pmatrix}
    1 & \param & \param \\
    \param & 1  & \param \\
     \param & \param & 1
\end{pmatrix}$
and marginals distributed uniformly over $[0, 1]$.
We perform a grid search, varying the group sizes 
($\props$), the colleges' capacities ($\capacities$), the students' preferences ($\prefs$), and the correlation of group $\gpi$ ($\param_1$). Overall, we consider 324 different parameter combinations. For each parameter combination we compute all cutoffs, $\cuti$,
and ranks, $\Rjsk(\params)$, as functions of $\param_2$. The list of parameters used and the detailed results of the experiments for three colleges are provided in Online Appendix D.1.

\begin{figure}[ht]
    \FIGURE{\includegraphics[width=0.49\linewidth]{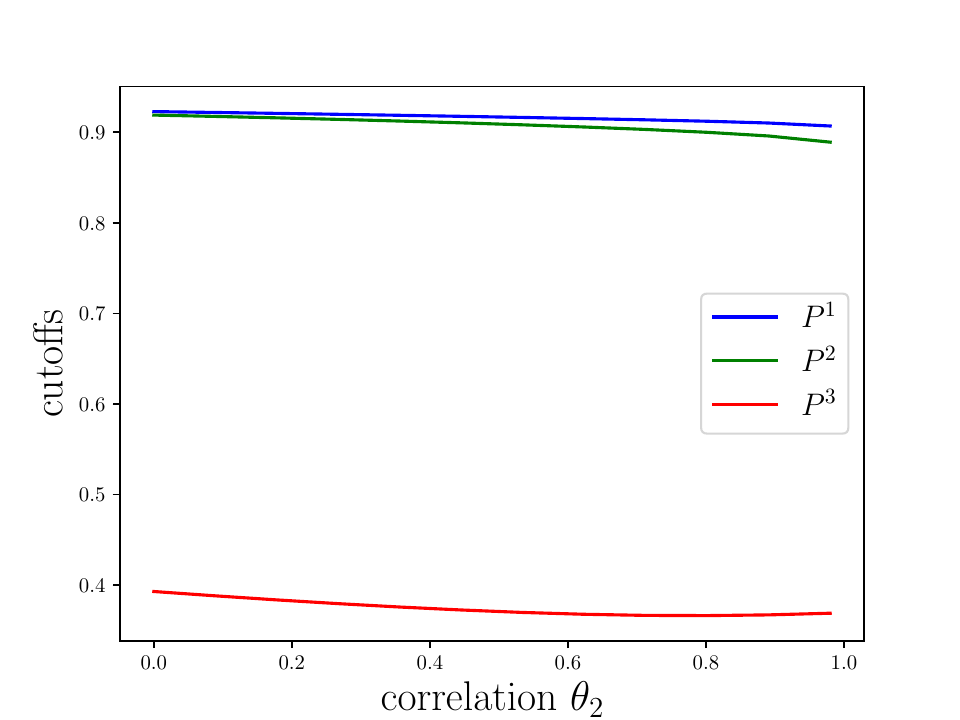}
    \includegraphics[width=0.49\linewidth]{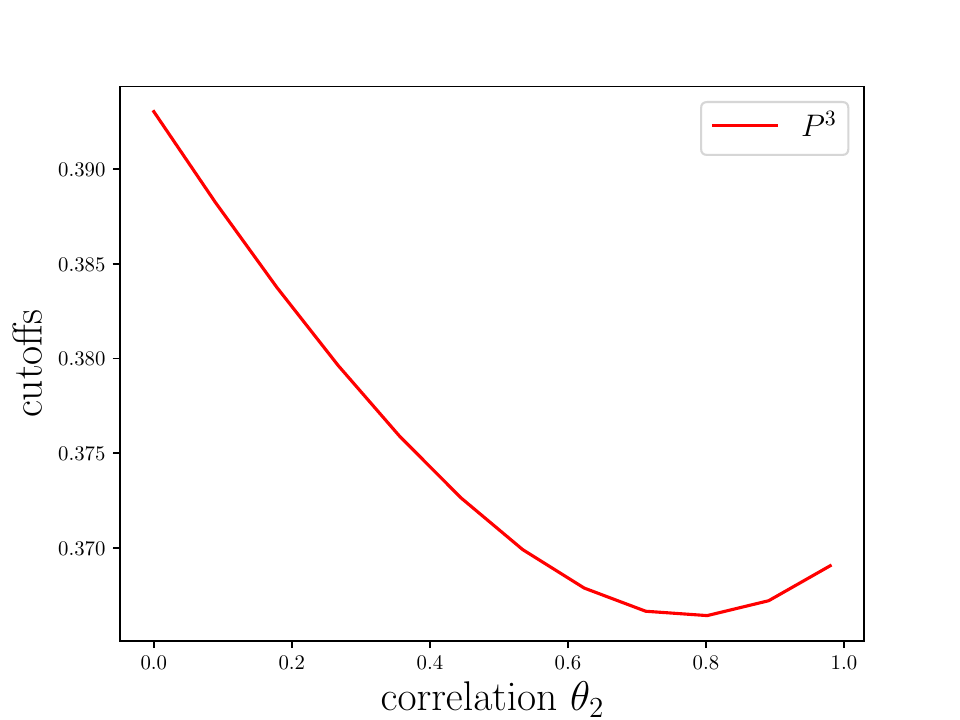}}
    {Cutoffs as functions of groups 2's correlation $\theta_2$.  \label{fig:sim_cutoffs}}
    {On the left are the three cutoffs, and  the right shows only cutoff $\cut^3$, zoomed in. Parameters: three colleges, two groups, each of size $1/2$, preferences $(1/2, 1/16, 1/4, 1/32, 1/8, 1/32)$, $ \theta_1 = 1/3$, and capacity $(1/15, 1/15, 8/15)$.  The preference vector is to be read along the following ordering of permutations: $(123, 132, 213, 231, 312, 321)$. }
\end{figure}

\paragraph{Cutoffs.} We find that ca. 12\% (38 of 324) of the scenarios considered do not exhibit decreasing cutoffs throughout the range of $\param_2$.
This shows that Theorem \ref{thm:cutoff_inc} does not generalize beyond two colleges. Counterexamples occur when one college has significantly higher capacity and lower demand than the others; then the cutoff of this college slightly increases when $\param_2$ moves towards $1$. The increase for high levels of $\param_2$ is orders of magnitude smaller than the overall decrease across the whole range. This suggests that even when the assumption of decreasing cutoffs fails, the impact of differential correlation will qualitatively be mostly aligned with our theoretical results.
 Figure \ref{fig:sim_cutoffs} displays one of those counterexamples, the shape of the cutoff that is not always decreasing is representative of all counterexamples.

\begin{figure}[ht]
    \FIGURE{
    \includegraphics[width=0.49\linewidth]{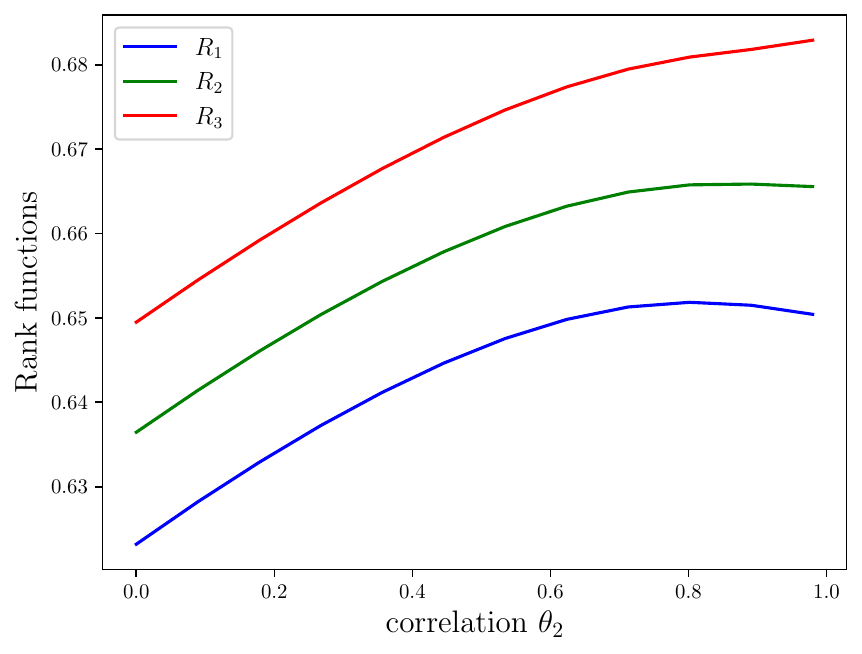}}
    {Rank functions for students of $\gp_1$ with preferences $(312)$, as functions of groups 2's correlation $\theta_2$. \label{fig:sim_rk_gp_pref}}
    {Parameters: three colleges, two groups of size $1/2$, capacities $(1/15, 1/15, 8/15)$, $ \theta_1 = 1/3$, and preference vector $(1/2, 1/16, 1/4, 1/32, 1/8, 1/32)$. The preferences vectors are to be read along the following ordering of permutations: $(123, 132, 213, 231, 312, 321)$.}  
\end{figure}

\paragraph{Ranks.} The counterexamples that violate Assumption \ref{assumption:dec_cutoffs} also imply that our results do not generally hold beyond two colleges. This is the case because when some cutoff $\cuti$ increases with $\param_2$, the same reasoning used to prove Theorem \ref{thm:V1-inc} shows that the proportion of students in group $\gp_1$ who prefer college $\col^i$, and are assigned to it, decreases ($R_1^{1, \preflist}$, 
for preferences $\preflist$ such that $\preflist(1) = i$).
However, our results do not have implications for the variations of lower ranks. We consider the parameters for which we found non-decreasing cutoffs, i.e., where our theoretical results do not hold, and compute $R_1^{k, \preflist}$ for $k \in \{1, 2, 3\}$ and $\preflist = (312)$ (a preference profile for which we expect $R_1^{1, \preflist}$ to not be increasing). As expected, we find that $R_1^{1, \preflist}$ decreases exactly where $\cut^3$ increases. We also find that
out of the 38 counterexamples,  $R_1^{2, \preflist}$   decreases slightly in only six instances, whereas it increases in the other scenarios. One of those instances is shown in Figure \ref{fig:sim_rk_gp_pref}. The amount of students who receive their first choice  decreases when $\param_2$ is high, as predicted, and the amount of students getting one of their top two choices also decreases for even higher values of $\param_2$.

Finally, note that, building on above counterexamples, it is straightforward to construct counterexamples where our results do not hold for any number of colleges. To this end, it suffices to take a counterexample with three colleges and add any number of additional colleges such that all students prefer the original three colleges over the added colleges.

\section{Related Literature}
\label{sec:litearture}

\paragraph{Matching.} \label{para:matching}
The college admissions problem, i.e., how to assign prospective students to colleges given each student's preferences and colleges' priorities over students and capacities such that the outcome is stable, was introduced by \citet{gale_college_1962}. The variant of this model where colleges do not have priorities over students is  called the school choice problem (cf., e.g., \citealt*{balinski_tale_1999},  \citealt*{abdulkadiroglu_school_2003}, \citealt*{abdulkadiroglu_college_2005}, \citealt*{ergin_games_2006}, \citealt*{yenmez_incentive-compatible_2013}). The idea of considering a continuum of students and a finite number of colleges, as we do in our model, has previously been exploited for various purposes due to its analytical tractability. \cite{chade_student_2014} use it to compute students' optimal application strategies when applications are costly, and \cite{abdulkadiroglu_expanding_2015} to study a new and more complex tie-breaking rule for school choice. \cite{azevedo_supply_2016} provide the framework to compute admission thresholds that we used in this paper to derive most of our results, and \cite{arnosti_continuum_2022} proposes a framework for finite markets using continuum techniques, that outputs a probability distribution for admission thresholds and coincide with \cite{azevedo_supply_2016} when the number of students grows to infinity.

\paragraph{Matching with correlated types.}

We study matching in the presence of correlation between the priority scores given by each college to a given student. A special case of this problem has been studied for school choice problems, where many students have the same priority and ties are broken at random. The implications of this feature for students' welfare have been studied by \cite*{erdil_tie_breaking_2008, abdulkadiroglu_strategy-proofness_2009} and \cite*{abdulkadiroglu_expanding_2015}. 
Most closely related 
and as discussed in Section \ref{subsec:TB}, \cite*{ashlagi_assigning_2019-1, ashlagi_competition_2020}, and  \cite{arnosti_lottery_2022} compare the welfare of students in two settings, i.e., that in which either one common lottery is used by all colleges or all colleges draw independent lotteries. 
In our model, this corresponds to a correlation of either 1 or 0, and our results nest some findings of these prior papers. Another line of work has considered correlation between other features, e.g., between students' preferences and colleges' rankings (cf., \citealt{brilliantova_fair_2022}, \citealt{che_efficiency_2019} and \citealt{leshno_cutoff_2020}). 
 Considering the use of copulas to model correlation, \cite{gola_supply_2021} leverages this idea to study how workers sort into two competing sectors in a market with transferable utility.

\paragraph{Fairness.} \label{para:fairness}

The analytical study of fairness in selection problems goes back to at least \citep{phelps72, arrow73}. In a recent influential study, \citet{kleinberg_selection_2018} study the effect of bias and the efficiency of affirmative action policies.\footnote{Recently, reducing outcome inequalities in ranking rather than in final selection has been actively studied, cf. \cite*{celis_interventions_2020}, \cite*{yang_causal_2021}, \cite*{mehrotra2022selection}, and \cite*{zehlike_fairness_2021}.} \citet*{emelianov_fair_2020,emelianov22} and \citet*{garg_dropping_2021} study statistical discrimination. Candidates have a latent quality, and the colleges or companies they apply to have access to only a biased  or noisy estimator of this quality that varies depending on their group, which is called \emph{differential variance}. 
We depart from those models by considering several decision-makers instead of one; that is, we consider the matching problem instead of the selection problem. 
Studies on fairness in matching have considered various affirmative action policies, including upper and lower quotas, to reduce discrimination (\citealt*{abdulkadiroglu_college_2005, kamada_efficient_2015, kamada_fair_2018, delacretaz_refugee_2020, krishnaa_envy-freeness_2020, dur_targetting_2020}). However, these works focus on finding stable matchings under some constraints, accounting for different fairness notions. In contrast, we aim to explain outcome inequalities that naturally occur in stable matchings without constraints. 
Finally, the inequity we identify is a source of systemic discrimination, i.e., discrimination that arises only through the interaction of decision-makers---via the matching mechanism---and is not due to intentional or non-intentional discrimination by single decision-makers (for references on systemic discrimination see \citealt*{Pin96,Fea13,Boh22}).

\paragraph{Individual fairness.}

All of the aforementioned studies consider what is generally termed \emph{group fairness}, that is, the idea that
all (relevant) groups should be treated similarly. In contrast, \emph{individual fairness} posits that individuals with similar characteristics should be treated similarly. In this latter spirit \cite*{karni_fairness_2021} broke new ground, showing that an individually fair ranking does not necessarily lead to an individually fair matching. This conclusion can also be drawn from our results in the context of group fairness. \cite*{devic_uncertainty_2023} also consider individual fairness and adapt the classical notion to incorporate agents' preferences; i.e., they require that similar agents be matched to a college in a similar position on their respective preference list.
Our work is concerned with group fairness; however, we note that if the quality of applicants is similar in each group and the notion of individual fairness put forth by \cite*{devic_uncertainty_2023} is fulfilled, then group inequalities are theoretically mitigated, which would make \cite*{devic_uncertainty_2023}'s notion stricter. However, as noted by \cite{fleisher_individual_2021}, individual fairness is very sensitive to several biases that may make it insufficient to prevent inequalities.

\paragraph{Algorithmic monoculture.}
Our work also contributes to the recent literature on algorithmic monoculture, i.e., the fact that recommendations, choices, and preferences become homogeneous with the rise of algorithmic curation and analysis.  
\cite{kleinberg_algorithmic_2021} study the utility of multiple decision-makers who use algorithms to evaluate candidates. They show that decision-makers are sometimes better off using different low-precision algorithms than when using the same high-precision algorithm. In contrast, we focus on the impact of correlation on applicant welfare and, moreover, introduce differential correlation between groups.
\cite{peng_monoculture_2023,peng2024} recently study the impact of correlation in rankings in a model in which the number of decision-makers grows large. 
In contrast, our results hold in particular  for a small number of decision-makers. Their findings on the impact of correlation with many decision-makers show qualitative effects similar to ours. In further contrast to \cite{peng_monoculture_2023,peng2024}, the main focus of our analysis is on the impact of differential correlation between groups.
Finally, in empirical work,  \cite*{bommasani_picking_2022} find that outcomes are more homogeneous when models and training data sets are shared between decision-makers. Through our theoretical analysis, we thus elucidate the impact of algorithmic monoculture from the candidate's viewpoint: while increasing correlation increases the efficiency of the outcome, differential correlation may lead to increased inequality between different groups.

\section{Conclusion}
\label{sec:conclusion}

We have introduced a tractable model to study the impact of differential ranking correlation between different groups and studied its effect on outcome inequality and efficiency in matching markets.
We have shown that efficiency generally improves as correlation increases; however, low-correlation groups are advantaged with regard to not remaining unmatched.
Differential correlation leads to inequalities between groups, even when each college has a perfectly fair ranking. When there is only one group, our results quantify the role of correlation in student satisfaction and efficiency. Our framework can accommodate almost any priority score distribution, any number of groups with different distributions and different student preferences, and any number of colleges of any capacity. Although our results are most general when we limit ourselves to two colleges, they remain true with additional assumptions for any number of colleges, and simulations suggest that they quantitatively hold across the parameter space. 

While our model can accommodate many extensions, our results required additional assumptions. First, all results rely on the cutoffs being decreasing in the correlation, and we only analytically proved this to be true in limited settings, i.e., for two colleges or for specific values of capacities and preference distributions. Although our numerical computations indicate that this assumption holds in the majority of cases, and our counterexample shows that it cannot be proven to always hold, stronger theoretical guarantees could help to understand the domain of validity of our results. 

In addition to further exploring the results within our model, there is ample scope to extend our model, and thus explore features that are often relevant in practice. For example, candidates may not list all colleges and have incomplete preferences, possibly due to application costs. This is known to impact some results on stable matchings, and  \cite{arnosti_lottery_2022} shows that it breaks the dominance of single tie-breaker over multiple tie-breakers, which implies that comparative statics involving correlation of priorities are affected. Further, when decision-makers use noisy estimates of applicants' latent quality, interesting directions of future research include allowing applicants to invest in accurate assessment, e.g., by acquiring certifications or participating in in-person interviews, or considering the effects of risk aversion. Finally, we have not explored the effect of correlation on the utility of colleges.

In conclusion, there is ample scope to study themes that have already been considered in single decision-maker settings in the matching context. Our analysis suggests that in matching new phenomena arise, which necessitate a deeper understanding.

\ACKNOWLEDGMENT{We thank the reviewers and editors at Management Science for their work that has helped us to greatly improve this article. We also would like to thank Nick Arnosti, Itai Ashlagi, Jean-Paul Carvalho, Julien Combe, Vitalii Emelianov, Simon Finster, Ravi Jagadeesan, Simon Jantschgi, Negar Matoorian, Meg Meyer, Faidra Monachou, Marek Pycia, and Jakob Weissteiner, as well as the participants of the 23rd ACM Conference on Economics and Computation, the 12th Conference on Economic Design,  the 33rd Stony Brook International Conference on Game Theory, the Nuffield Economic Theory workshop, the Alpine Game Theory Symposium (Grenoble, 2023), the From Matching to Markets workshop (CIRM, Marseille, 2024), and the 19th Matching in Practice Workshop (Zurich, 2024) for their feedback and advice, all of which greatly helped us improving this work. All remaining errors are ours. This work was partially supported by MIAI @ Grenoble Alpes (ANR-19-P3IA-0003), by the
French National Research Agency (ANR) through grants ANR-19-CE48-0018 and ANR-20-CE23-0007, and by the
National Science Foundation under Grant No.~DMS-1928930 and by the Alfred P. Sloan Foundation under grant G-2021-16778 that funded Bary Pradelski's residency at the Simons Laufer Mathematical Sciences Institute (formerly MSRI) in Berkeley, California, during the Fall 2023 semester.}

\bibliographystyle{informs2014} %
\bibliography{biblio}

\renewcommand{\theHchapter}{A\arabic{chapter}}
\begin{APPENDICES}
 
\label{sec:appendix}
\newpage

Appendix \ref{app:def_and_tech} provides definitions and technical results, Appendix \ref{app:additional} additional results omitted in the main body, Appendix \ref{app:proof} omitted proofs, and Appendix \ref{app:numerical} details on our numerical experiments.

\section{Definitions and technical details} \label{app:def_and_tech}

\subsection{Definitions} \label{app:defs}

\subsubsection{Notation} \label{app:notation}
Table \ref{table:notation} provides a summary of the notation used throughout the paper. In addition, we use the following convention: bold letters are used for vectors, superscripts represent colleges' indices or preference lists, and subscripts represent groups' or students' indices.

\begin{table}[ht]
    \caption{Notation}
    \label{table:notation}
    {\small
    \begin{tabularx}{\textwidth}{p{0.22\textwidth}X}
    \toprule
      {\underline{Agents}: } \\
      $\col^1, \dots, \col^m$, $\colset$ & Colleges and set of all colleges \\ 
      $\stud$, $\Stud$ & An arbitrary student and the set of all students\\
      $\gp_1, \dots \gp_\ngp$ & Groups of students, partition of $\Stud$ \\
      $\eta$ & Measure over $\Stud$ \\ \vspace{0.03 cm}
     
      {\underline{Agents' features}: } \\
      $\capi, \capacities$ & College $\coli$ 's capacity $\in (0, 1)$ and vector of all capacities\\ 
      $\propj, \props$ & Mass of students in group $\gpj$ $\in [0, 1]$ and vector of all groups' masses\\
      $\preflist$ & Permutation of colleges (represents preferences), $\in \preflistsm$ \\
      $\prefsj, \prefs$ & Share of $\gpj$ students with preferences $\preflist$ $\in (0, 1)$ and vector of all the $\prefsj$\\ \vspace{0.03 cm}
      
      {\underline{Priority scores}: } \\
      $\gradeis$ & Priority score at $\coli$ of student $\stud$ \\
      $\pdfCij, \cdfCij, \pdfs$ & Marginal pdf and cdf of college $\coli$ for group $\gpj$ and vector of all marginal pdfs \\
      $\copcdffamily, \coppdffamily$ & Copula family and associated pdfs, indexed by $\param$ \\
      $\tset$ & Set of possible values for $\param$ \\
      $\paramj$, $\params$ & Correlation level of group $\gpj$ and vector of all correlations \\
      $\pdfj, \cdfj$ & Group $\gpj$'s priority score vectors' joint pdf and cdf, $\cdfj = \copcdf_{\paramj}(\cdf^1_j, \dots, \cdf^m_j)$\\
      $\xsetCij, \xsetj$ & Support of $\pdfCij$ and $\pdfj$ respectively. $\xsetj = \prod\limits_{i \in [m]} \xsetCij$ \\
      $\xsetCijl, \xsetCiju$ & Lower and upper bounds of $\xsetCij$ \\ \vspace{0.03 cm}

      {\underline{Matching and metrics}: } \\
      $\match$ & Matching \\ 
      $\cuti$, $\cutoffs$ & Cutoff of college $\coli$ and vector of all cutoffs \\
      $\Rjsk$ & Share of students of group $\gpj$ with preferences $\preflist$ who get a top $k$ choice \\
      $\RjE$ & Share of students of group $\gpj$ who remain unassigned \\
      $\eff$ & Total mass of students getting their first choice \\
      $\ineqjl$ & Inequality between $\gpj$ and $\gpl$, equal to $\vert \RjE - \RlE \vert$ \\
      \bottomrule
     \end{tabularx}
     }
    \end{table}

\subsubsection{Definition of the mass $\mass$} \label{app:mass_def}

Here we formally define the notion of mass for a subset of students. This section is self-contained and is not necessary to understand the results of the paper; the notations introduced here are not used elsewhere. We identify $\Stud$ to $\type \coloneqq \R^m \times \{ \gp_1 , \dots , \gp_\ngp \} \times \preflistsm$. We partition $\type$ into several subsets: $\type_{j , \preflist} \coloneqq \setdef{(x, \gpj, \preflist) \in \type}{ x \in \R^m}$ is the subset of students belonging to group $\gpj$ with preferences $\preflist$. We say that a subset $J \subseteq \type$ is measurable if and only if for all $j, \preflist$, $J_{j, \preflist} \coloneqq \setdef{\grades}{(\grades, \gpj, \preflist) \in J}$ is Borel-measurable in $\R^m$. Let $\mathcal{B}(\type)$ be the set of measurable subsets of $\type$. We assume that to each group $\gpj$, there is an associated probability measure $\mathbb{P}_j: \mathbb{R}^m \to \mathbb{R}$. For all $j, \preflist$ we define a measure $\mass_{j, \preflist}$ as follows: for $J \subseteq \type$ measurable, 
\begin{equation}
\begin{array}{l}
\mass_{j, \preflist}(J) = \propj \prefsj \prob_{j}(\grades \in J_{j, \preflist}).
\end{array}
\end{equation}
We define over $\mathcal{B}(\type)$ the probability measure $\mass: \mathcal{B}(\type) \to [0, 1]$ such that for any  $J \in \mathcal{B}(\type)$, 
\begin{equation}
\mass(J) = \sum\limits_{j \in [\ngp]} \sum\limits_{\preflist \in \preflistsm} \mass_{j, \preflist}(J).
\end{equation}
This definition is consistent with the notations introduced in the model, as it verifies $ \mass (\gpj) = \propj$, $\mass (\setdef{s \in \gpj}{\preflist_s =  \preflist} ) = \propj \prefsj$ and so on.

\subsection{Elements of correlation theory} \label{app:correlation}

In this section, we present common measures of correlation used in the literature, and some of their properties.

\begin{definition}[Common measures of correlation]
Let $(X,Y)$ be two random variables with respective cdfs $\cdf_X$ and $\cdf_Y$. Define:
\begin{enumerate}
    \item Pearson's correlation: assume $X$ and $Y$ have finite standard deviations $\sigma_X$ and $\sigma_Y$. Then $\cor_{X, Y} = \frac{\mathrm{Cov}(X, Y)}{\sigma_X \sigma_Y}$. 
    \item Spearman's correlation: let $rk_X = \cdf_X(X)$ and $rk_Y = \cdf_Y(Y)$ be the quantile variables of $X$ and $Y$. Then Spearman's correlation is $\rho_{X, Y} = \cor_{rk_X, rk_Y}$. 
    \item Kendall's correlation: let $(X_1, Y_1)$ and $(X_2, Y_2)$ be two independent pairs of random variables with the same joint distribution as $(X, Y)$. Then Kendall's correlation is $$
    \begin{array}{ll}
        \tau_{X, Y} = & \prob\left[ (X_1>X_2 \cap Y_1 > Y_2) \cup (X_1<X_2 \cap Y_1 < Y_2) \right] - \\
         & \prob\left[ (X_1>X_2 \cap Y_1 < Y_2) \cup (X_1<X_2 \cap Y_1 > Y_2) \right].
    \end{array} $$ 
\end{enumerate}
\end{definition} \vspace{0.3 cm}

The covariance of the standard bivariate Gaussian is equal to Pearson's correlation $\cor$. Moreover, for this distribution simple expressions exist for the two other correlation coefficients:
$$ \spearman = \frac{6}{\pi} \arcsin(\cor/2), \kendall = \frac{2}{\pi} \arcsin(\cor).$$

A correlation measure should be zero when variables are independent, and reach its maximum when the variables are totally dependent on each other. The following lemma provides these properties for the measures we just introduced.

\begin{lemma}[{\citealt[Theorems 1, 4, and 5]{scarsini_concordance_1984}}] \label{lemma:corr_mono_rel}
Let $X, Y$ be two real random variables. 
\begin{enumerate}
    \item $\cor_{X, Y}, \spearman_{X, Y}, \kendall_{X, Y} \in [-1, 1]$.
    \item $\spearman_{X, Y} = 1$ if and only if $Y = g(X)$ with $g:\mathbb{R}\to \mathbb{R}$ increasing. The same holds for $\kendall_{X, Y}$. $\cor_{X, Y} = 1$ if and only if the relation is affine.
    \item If $X$ and $Y$ are independent, then $\cor_{X, Y} = \spearman_{X, Y} = \kendall_{X, Y} = 0$.
\end{enumerate}
\end{lemma} 

\begin{lemma}[{\citealt[Theorems 4 and 5]{scarsini_concordance_1984}}] \label{lemma:scarsini_2}
\emph{Suppose $m = 2$, the family of copulas $\left(\copcdft \right)_{\param \in \tset}$ is coherent, and $(X_\param, Y_\param)$ is a random vector drawn according to $\copcdft$, then Spearman's and Kendall's correlation coefficients $\rho(X_\param, Y_\param)$ and $\tau(X_\param, Y_\param)$ are  increasing functions of $\param$.
}
\end{lemma} 

Lemma \ref{lemma:scarsini_2} originally applies to all functions that verify a set of assumptions and that the authors call \emph{measures of concordance}. Assuming that $\copcdffamily$ is coherent makes $\param$ a measure of concordance, which gives the lemma as stated above.

\subsection{Discussion on distributional assumptions}\label{app: restrictive}

We assume that priority score distributions admit a density and have full support over the product space of the marginal distributions' supports, and that they can be represented using a copula family that is coherent (Assumption 1) and differentiable (Assumption 2). We show here that these assumptions are not very restrictive by presenting canonical examples of classical copulas satisfying our assumptions (they are given for $m=2$ for simplicity but extend naturally to any $m$).
\begin{enumerate}
    \item \textit{Gaussian copula}: The Gaussian copula is obtained by composing the cdf $\Phi_\param$ of a bivariate Gaussian with covariance matrix $\begin{pmatrix}{cc} 1 & \param \\ \param & 1 \end{pmatrix}$ and the univariate cdf $\phi$ of the standard Gaussian: $\copcdft(x, y) = \Phi_\param(\phi(x), \phi(y))$. Here, the parameter $\param$ controls the covariance. Notice that when $m \geq 3$, many choices are possible. We could set the covariance of every pair of colleges as $\param$, or only use $\param$ for one pair and keep a constant covariance for all other pairs, or in between those two extremes have the covariance of each pair of colleges equal to a (different) non decreasing function of $\param$, such that for any $\param$ at least one of those functions is increasing. All those options give coherent and differentiable copula families.
    \item \textit{Archimedean copulas}: Archimedean copulas are a broad class of copulas, each member of this class being itself a parametric family of copulas with a real parameter $\param$. The general formula is 
    $$ \copcdft(x, y) = \psi_\param^{-1}\left( \psi_\param(x) + \psi_\param(y) \right) $$
    where $\psi_\param:[0, 1] \to \mathbb{R}_+$ is a continuous  decreasing and convex function such that $\psi_\param(1) = 0$. Examples include:
    \begin{itemize}
        \item Clayton: $\copcdft(x, y) = \left( \max\{x^{-\param} + y^{-\param} -1; 0 \}\right)^{-1/\param}$
        \item Frank: $\copcdft(x, y) = -\frac{1}{\param} \log\left( 1 + \frac{(\exp(-\param x) - 1) (\exp(-\param y) - 1)}{\exp(-\param) - 1} \right)$
        \item Gumbel: $\copcdft(x, y) = \exp\left( -((-\log(x))^\param + (-\log(y))^\param )^{1/\param} \right)$
    \end{itemize}
\end{enumerate}

Clayton's, Frank's, Gumbel's and some other Archimedean copulas all satisfy our coherence and differentiability assumptions.

The only assumption our model makes on the marginals is that they are continuous. This is not particularly restrictive as long as there are no ties (see Section 4.2 for a treatment of ties).

\hypertarget{hyper:app:matching}{}
\subsection{Stable matching} \label{app:matching}

We introduce  elements of matching theory used throughout the paper. 
To define matching in a continuum context, we follow \cite{azevedo_supply_2016}. 
\begin{definition}
A \emph{matching} is an assignment of students to colleges, described by a mapping $\match : \Stud \cup \colset \to 2^\Stud \cup \colset$, with the following properties:
\begin{enumerate}
\item for all $s \in \Stud$, $\match (\stud) \in \colset \cup \{\emptyset\}$;
\item for $i \in [m]$, $\match (\coli ) \subseteq \Stud$ is measurable and $\mass (\match^{-1} (\coli )) \leq \capi$;
\item $\coli = \match (\stud)$ if and only if $s \in \match (\coli )$;
\item for $i \in [m]$, the set $\setdef{s \in \Stud}{ \match(\stud) \preceq_s \coli }$ is open.
\end{enumerate}
\end{definition}

The first three conditions are common to almost all definitions of matching in discrete or continuous models. Condition (1) ensures that a student is either matched to a college or to the empty set, which means that they remain unmatched. Condition (2) ensures that colleges are assigned to a subset of students that respects the capacity constraints. Condition (3) ensures that the matching is consistent, i.e., if a student is matched to a college, then this college is also matched to the student. Condition (4) was introduced by \citet{azevedo_supply_2016} and is necessary to ensure that there do not exist several stable matchings that only differ by a set of students of measure 0.

We next define the notions of blocking and stability. 
\begin{definition}[Stability]
The pair $(s, \coli)$ \emph{blocks} a matching $\match$ if $s$ would prefer $\coli$ to her current match, and either $\coli$ has remaining capacity or it admitted a student with a lower priority score than $s$; formally, if $\match(\stud) \prec_s \coli$ and either $\mass(\match(\coli)) < \capi$ or $\exists s' \in \match(\coli)$ such that $\gradei_{s'} <\gradeis$. A matching is \emph{stable} if it is not blocked by any student-college pair.
\end{definition}

This definition is the classical definition of stability, and is equivalent to the one provided in Section 2.3 based on cutoffs. To find a stable matching, one can extend the classic deferred acceptance algorithm by \cite{gale_college_1962} to the continuum model. This algorithm is described in Algorithm \ref{alg:GS}.

\begin{algorithm}
\caption{Deferred acceptance algorithm (DA)}\label{alg:GS}
\begin{algorithmic}
\State \textbf{First step:} All students apply to their favorite college, they are temporarily accepted. If the mass of students applying to college $\col$ is greater than its capacity $\capacity^{\col}$, then $\col$ only keeps the $\capacity^{\col}$ best 
\While {A positive mass of students are unmatched and have not yet been rejected from every college}
	\State Each student who has been rejected at the previous step proposes to her 		preferred college among those that have not rejected them yet
	\State Each college $\col$ keeps the best $\capacity^{\col}$ mass of students among those it had temporarily accepted and those who just applied, and rejects the others
\EndWhile 
\State \textbf{End:} If the mass of students that are either matched or rejected from every college is 1, the algorithm stops. However it could happen that it takes an infinite number of steps to converge.
\end{algorithmic}
\end{algorithm}

If the algorithm stops, the matching it outputs is stable; \cite*{abdulkadiroglu_expanding_2015} show that even when the number of steps is infinite, the algorithm converges to a stable matching. 

\begin{remark} \label{rmk: stable_match}
Note that stable matchings do not only result from centralized algorithms but are often the result of a decentralized process (see, e.g., \citealt{Rot90}).
\end{remark}

Figure \ref{fig:cutoff_app} illustrates the link between the cutoffs and the matching: students who prefer $\col^1$ are admitted there if and only if their priority score $\grade^1$ is higher than the cutoff $\cut^1$. Otherwise, they are admitted to $\col^2$ if their priority score $\grade_2$ is higher than $\cut^2$ and stay unmatched if it is not (left panel). The situation is symmetric for students who prefer college $\col^2$ (right panel).

\begin{figure}
    \FIGURE{
    \includegraphics[width = 0.95 \textwidth]{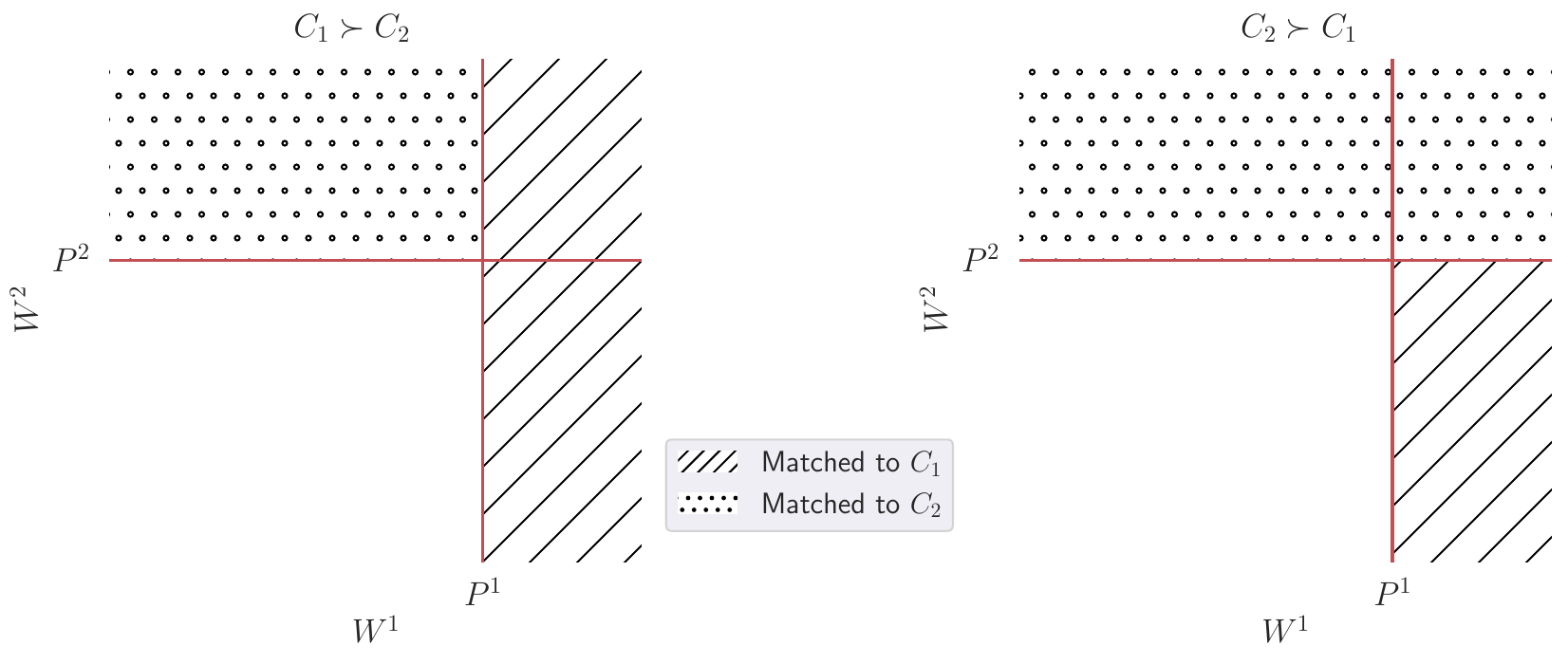}}{Cutoff description of stable matchings.\label{fig:cutoff_app}} {Students in the hashed area are matched to college $\col^1$, those in the dotted area to college $\col^2$, and those in the white area remain unmatched.}
\end{figure}

\subsection{Computing metrics} \label{app:metrics}

We provide expressions for the metrics introduced in Section 3.

\begin{lemma} \label{lemma:compV}
Let $j \in [\ngp]$, $k \in [m]$, $\preflist \in \preflistsm$, $\params \in \tset^\ngp$, and $\cutoffs(\params)$ be the cutoffs associated to the unique stable matching. We have:
\begin{align}
\Rjsk(\params) &=  \prob_{j, \paramj}\left(\grade^{\preflist(k)} \geq \cut^{\preflist(k)}(\params) \cap  \bigcap\limits_{l = 1}^{k-1} (\grade^{\preflist(l)} < \cut^{\preflist(l)}(\params))\right)
. \label{eq:VRk}
\end{align}
\end{lemma}
The notation $\prob_{j, \paramj}$ is used as shorthand for $\prob_{\grades \sim \pdfj}$.
Lemma \ref{lemma:compV} allows us to compare the satisfaction of different types of students and derive comparative statics with respect to differential correlation.

\proof{Proof.}
Consider student $s \in \gpj$ who has preferences $\preflist$. By \citet[][Lemma 1]{azevedo_supply_2016} (cf. Section 2.3), $\stud$ is admitted to $\coli$ which is their $k$th choice if and only if $\stud \in D^{i}(\cutoffs)$, i.e., if and only if they pass the cutoff at $\coli$ and do not pass it at the colleges their prefer to $\coli$. Since the grades of students from group $\gpj$ are drawn according to $\cdfj$, we obtain Equation \eqref{eq:VRk}.\hfill \Halmos
\endproof

\subsection{Regularity of cutoffs and metrics}\label{app:C1}

The comparative statics results presented in Section 4 often require the cutoffs to be smooth, and all the metrics defined based on cutoffs as well. We prove here that they are.

\begin{lemma} \label{lemma:C1}
    If Assumptions 1 and 2 hold (coherence and differentiability), $\cutoffs(\params)$ is a $\mathcal{C}^1$ function over $\mathring{\tset}^\ngp$, and so are $\Rjsk$, $\RjE$, $\eff$ and $\ineqjl$.
\end{lemma}

\proof{Proof.}
The market-clearing equation (1) can be written as 
\begin{equation} \label{eq:proof.V1-inc-MC}
    \left\{ \begin{array}{lcl} 
        &\sum\limits_{j \in [\ngp]} \propj \sum\limits_{k = 1}^m \sum\limits_{\substack{\preflist \in \preflistsm \\ \preflist(k)=1}} \prefsj \prob_{\paramj}\left(\bigcap\limits_{n < k} (\grade^{\preflist(n)} < \cut^{\preflist(n)}) \cap \grade^{\preflist(k)} \geq \cut^{\preflist(k)}\right)&  = \capacity^1 \\
    & \vdots & \\
    &\sum\limits_{j \in [\ngp]} \propj \sum\limits_{k = 1}^m \sum\limits_{\substack{\preflist \in \preflistsm \\ \preflist(k)=m}} \prefsj \prob_{\paramj}\left(\bigcap\limits_{n < k} (\grade^{\preflist(n)} < \cut^{\preflist(n)}) \cap \grade^{\preflist(k)} \geq \cut^{\preflist(k)}\right)& = \capacity^m. 
        \end{array} 
        \right.
\end{equation}
We fix all $\param_i$ except for $\paramj$ for some $j$, and we want to study how the solution $\cutoffs(\params)$ of the above equation varies as a function of $\paramj$. Define $T:\R^m \times \tset \to \R^m$, $(\cutoffs, \paramj) \mapsto (D^{\col^1}(\cutoffs(\params)) - \capacity^1,\dots,  D^{\col^m}(\cutoffs(\params)) - \capacity^m)$. Denote by $T^1, \dots, T^m$ the
components of $T$. Then for any $\paramj \in \tset$, the solution $\cutoffs$ of Equation \eqref{eq:proof.V1-inc-MC} is also the solution of the equation $T(\cutoffs, \paramj) = 0_{\R^m}$ (where $x_{\R^m}$ denotes the vector $(x,\ldots,x)$ in $\R^m$). In order to show that $\cut^1, \dots, \cut^m$ are all decreasing in $\paramj$, we apply the implicit function theorem. Let $\cutoffs \in \R^m$ and $\paramj \in \tset$ such that $T(\cutoffs, \paramj)=0_{\R^m}$. Function $T$  is of class $\mathcal{C}^1$ because $\copcdffamily$ is differentiable. We first verify that the partial Jacobian $J_{T, \cutoffs}(\cutoffs, \paramj)$ is invertible, where 
\begin{equation} \label{eq:Jacobian_mult}
    J_{T, \cutoffs}= 
    \begin{pmatrix}
    \frac{\partial T^1}{\partial \cut^1} & \dots & \frac{\partial T^1}{\partial \cut^m} \\
    \vdots & & \vdots \\
    \frac{\partial T^m}{\partial \cut^1} & \dots & \frac{\partial T^m}{\partial \cut^m}
    \end{pmatrix}.
\end{equation}
To prove this, we show that no (non-trivial) linear combination of the rows of $J_{T, \cutoffs}(\cutoffs, \paramj)$ can be equal to zero. We start by proving the following inequality:
\begin{equation} \label{eq:proof.V1-inc.sum-delta-h}
    \forall i \in \{1, \dots, m\},  \sum\limits_{n \in \{1, \dots, m\}} \frac{\partial T^n}{\partial \cuti} < 0
\end{equation}
Note that $T^n$ is a sum of probabilities of intersections containing terms of the type
$\gradei < \cuti$ for all $i \neq n$, and $\grade^n \geq \cut^n$. 
Therefore, for every $i, n\in [m], i\neq n$, $\frac{\partial T^i}{\partial \cuti} < 0$ and $\frac{\partial T^n}{\partial \cuti} > 0$. So proving the relation \eqref{eq:proof.V1-inc.sum-delta-h} amounts to proving that for every $i \in [m]$, $- \frac{\partial T^i}{\partial \cuti} > \sum\limits_{n \neq i} \frac{\partial T^n}{\partial \cuti}$. The first term can be written as:
\begin{align} \label{eq:proof.V1-inc.hiPi}
    - \frac{\partial T^i}{\partial \cuti} = & -  \sum\limits_{k = 1}^m \sum\limits_{\substack{\preflist \in \preflistsm \\ \preflist(k)=i}} \sum\limits_{\ell \in [\ngp]} \prop_\ell \prefsig_\ell \frac{\partial \prob_{\param_\ell}\left(\bigcap\limits_{p < k} (\grade^{\preflist(p)} < \cut^{\preflist(p)}) \cap \grade^{\preflist(k)} \geq \cut^{\preflist(k)}\right)}{\partial \cuti}. 
\end{align}
Notice that 
\begin{align}
    &\prob_{\param_\ell}\left(\bigcap\limits_{p < k} (\grade^{\preflist(p)} < \cut^{\preflist(p)}) \cap \grade^{\preflist(k)} \geq \cut^{\preflist(k)}\right) 
    =\prob_{\param_\ell}\left(\bigcap\limits_{p < k} (\grade^{\preflist(p)} < \cut^{\preflist(p)}) \right) - \prob_{\param_\ell}\left(\bigcap\limits_{p \leq k} (\grade^{\preflist(p)} < \cut^{\preflist(p)})\right)
\end{align}
where the first term is constant in $\cuti = \cut^{\preflist(k)}$. We then deduce that 
\begin{align}
    & \frac{\partial \prob_{\param_\ell}\left(\bigcap\limits_{p < k} (\grade^{\preflist(p)} < \cut^{\preflist(p)}) \cap \grade^{\preflist(k)} \geq \cut^{\preflist(k)}\right)}{\partial \cuti} 
    = - \frac{\partial \prob_{\param_\ell}\left(\bigcap\limits_{p \leq k} (\grade^{\preflist(p)} < \cut^{\preflist(p)})\right)}{\partial \cuti}.
\end{align}
Inserting the latter  in \eqref{eq:proof.V1-inc.hiPi} gives: 
\begin{align} \label{eq:proof.V1-inc.hiPi-2}
    - \frac{\partial T^i}{\partial \cuti} = &  \sum\limits_{k = 1}^m \sum\limits_{\substack{\preflist \in \preflistsm \\ \preflist(k)=i}}\sum\limits_{\ell \in [\ngp]} \prop_\ell \prefsig_\ell \frac{\partial \prob_{\param_\ell}\left(\bigcap\limits_{p \leq k} (\grade^{\preflist(p)} < \cut^{\preflist(p)})\right)}{\partial \cuti}.
\end{align}
Now consider the terms in $T^n$ for $n \neq i$:
\begin{align} 
    \frac{\partial T^n}{\partial \cuti} = &  \sum\limits_{k = 2}^m \sum\limits_{\substack{\preflist \in \preflistsm \\ \preflist(k)=n \\ \preflist^{-1}(i) < n}} \sum\limits_{\ell \in [\ngp]} \prop_\ell \prefsig_\ell \frac{\partial \prob_{\param_\ell}\left(\bigcap\limits_{p < k} (\grade^{\preflist(p)} < \cut^{\preflist(p)}) \cap \grade^{\preflist(k)} \geq \cut^{\preflist(k)}\right)}{\partial \cuti}.
\end{align}

The terms of $T^n$ where $\cuti$ does not appear are constant in $\cuti$ and do not appear here, which is encompassed in the condition $\preflist^{-1}(i) < n$ in the second sum and explains why the first sum starts at $k=2$. We want to prove that $\sum\limits_{n \neq i} \frac{\partial T^n}{\partial \cuti} < - \frac{\partial T^i}{\partial \cuti}$. We can do so by comparing term by term what is inside the derivative on each side (by linearity). Let $\preflist \in \preflistsm$, $\ell \in [\ngp]$, and consider the terms on each side of the inequality that have $\prefsig_\ell$ as a factor. On the right side, we have
\begin{align}
   &  \prob_{\param_\ell}\left(\bigcap\limits_{p \leq \preflist^{-1}(i)} (\grade^{\preflist(p)} < \cut^{\preflist(p)})\right).
\end{align}
On the left side, those terms exist only for $k > \preflist^{-1}(i)$:
\begin{align} \label{eq:proof.V1-inc.total_prob}
   & \sum\limits_{k > \preflist^{-1}(i)}  \prob_{\param_\ell}\left(\bigcap\limits_{n < k} (\grade^{\preflist(n)} < \cut^{\preflist(n)}) \cap \grade^{\preflist(k)} \geq \cut^{\preflist(k)}\right).
\end{align}
Consider the last term  of the sum in \eqref{eq:proof.V1-inc.total_prob}, and upper-bound it by removing the last intersection:
\begin{align}
    \prob_{\param_\ell}\left(\bigcap\limits_{n < m}(\grade^{\preflist(n)} < \cut^{\preflist(n)}) \cap  \grade^{\preflist(m)} \geq \cut^{\preflist(m)}\right) < \prob_{\param_\ell}\left(\bigcap\limits_{n < m}(\grade^{\preflist(n)} < \cut^{\preflist(n)}) \right).
\end{align} 
The penultimate term being 
\begin{align}
\prob_{\param_\ell}\left(\bigcap\limits_{n < m-1}(\grade^{\preflist(n)} < \cut^{\preflist(n)}) \cap  \grade^{\preflist(m-1)} \geq \cut^{\preflist(m-1)}\right),
\end{align}
we can add those two to obtain
\begin{align}
\prob_{\param_\ell}\left(\bigcap\limits_{n < m-1}(\grade^{\preflist(n)} < \cut^{\preflist(n)}) \right).
\end{align}
We continue packing the terms together, and we finally obtain
\begin{align}
  & \sum\limits_{k > \preflist^{-1}(i)}  \prob_{\param_\ell}\left(\bigcap\limits_{n < k} (\grade^{\preflist(n)} < \cut^{\preflist(n)}) \cap \grade^{\preflist(k)} \geq \cut^{\preflist(k)}\right)
   < \prob_{\param_\ell}\left(\bigcap\limits_{p \leq \preflist^{-1}(i)} (\grade^{\preflist(p)} < \cut^{\preflist(p)})\right),
\end{align}
which is the term associated to $\prefsig_\ell$ in $- \frac{\partial T^i}{\partial \cuti}$ (cf. Equation \eqref{eq:proof.V1-inc.hiPi-2}). We finally conclude that $\forall i \in \{1, \dots, m\},  \sum\limits_{n \in \{1, \dots, m\}} \frac{\partial T^n}{\partial \cuti} < 0$.

We can now use this result to prove that $J_{T, \cutoffs}(\cutoffs, \paramj)$ is invertible. Let us call $R^i \coloneqq (\frac{\partial T^i}{\partial \cut^k})_k$ the $i$-th row of $J_{T, \cutoffs}(\cutoffs, \paramj)$. Assume that there exist $\lambda^1, \dots, \lambda^m$, not all zero, such that $\sum \lambda^i R^i = 0_{\R^m}$. Let $i_0 \in \arg\max\limits_i \lambda^i$. We assume without loss of generality that $\lambda^{i_0} > 0$. Then on the $i_0$-th column, using the inequality from Equation \eqref{eq:proof.V1-inc.sum-delta-h}, we have 
\begin{align}
\sum\limits_i \lambda^i \frac{\partial T^i}{\partial \cut^{i_0}} \leq \lambda^{i_0} \sum\limits_i \frac{\partial T^i}{\partial \cut^{i_0}}
  < 0,
\end{align}
which contradicts $\sum \lambda^i R^i = 0_{\R^m}$. We conclude that no non-trivial linear combination of the rows of $J_{T, \cutoffs}(\cutoffs, \paramj)$ can be zero, therefore it is invertible.

The assumptions of the implicit function theorem are verified, therefore there exists a neighborhood $U \subseteq \R^m \times \tset$ of $(\cutoffs, \paramj)$, a neighborhood $V \subseteq \tset$ of $\paramj$, and a $\mathcal{C}^1$ function $\psi:V \to \R^m$ such that for all $(x, \param) \in \R^m \times \tset$,
\begin{equation*}
    ( \, (x, \param) \in U \mbox{ and } T(x, \param) = 0 \, ) \Leftrightarrow (\, \param \in V \mbox{ and } x = \psi (\param) \,) .
\end{equation*}
Applying this reasoning for every $\paramj \in \mathring{\tset}$, the function $\psi$ is uniquely defined over each element of an open cover of $\mathring{\tset}$, and therefore is well defined on $\mathring{\tset}$. Since this is true for all $j \in [\ngp]$, we finally conclude that $\cutoffs(\params)$ is a $\mathcal{C}^1$ function. All the other metrics are defined as the composition of the cutoffs with $\mathcal{C}^1$ functions and are therefore also of class $\mathcal{C}^1$.
\hfill \Halmos
\endproof

\newpage

\section{Additional results}\label{app:additional}

\subsection{The latent quality plus noise setting}\label{app:latentnoise}

We here discuss the classical model where an (unknown) latent quality is observed with an added noise term (cf. \citealt*{phelps72,emelianov_fair_2020,emelianov22,garg_dropping_2021}). This provides an example of how differential correlation can arise.

Assume that all students have a latent quality $W$, and their priority  score $\widehat{\grade}^i$ at college $\col^i$ is the sum of the latent quality and a noise term $\varepsilon$ drawn independently at each college, that is,
\begin{equation*}
\forall \stud \in \Stud \mbox{, for }i \in [m], \widehat{\grade}^{i}_s = \grade_s + \varepsilon^{i}_s.
\end{equation*}
Further, assume that the latent qualities of all students (independently of their group) are drawn from the same Gaussian distribution, and that the noises are also normally distributed and depend on the group: 
\begin{equation*}
\forall \stud \in \Stud \mbox{, for }i  \in [m], \grade_s \sim \mathcal{N}(0, \chi^2) \mbox{, } \varepsilon^{i}_s \sim \mathcal{N}(0, \sigma_{\gp(\stud)}^2).
\end{equation*}
The fact that the noise's variance is different for each group can be interpreted as colleges having different accuracies when evaluating students from different groups. Consider two examples: First, $\gpi$ could consist of students from well-known high schools, which colleges can evaluate well since they have a lot of applicants from there each year; and $\gpii$ could consist of students from unknown high schools, for which colleges do not have a lot of prior information. Second, each group $\gpj$ could consist of students from different demographic groups, defined by sensitive attributes such as gender, ethnicity, or social class.

Suppose that colleges know each student's group, and are aware of the difference that exists in noise variance across groups. Further, assume that colleges implement equal opportunity policies, that is, ceteris paribus the rank distribution of students must be the same for all groups.\footnote{Without the equal opportunity assumption, colleges would compute the expected true qualities based on the different variances for each group, see \cite*{emelianov22,garg_dropping_2021}.}
To do so while maximizing the expected quality of admitted candidates, it is optimal for a college to not change the order of  priority scores within groups, but to only fit each group's priority scores $\widehat{\grade}^i$ to the same, standardized distribution. The transformation that achieves this goal is the following

\begin{equation*}
    \forall \stud \in \Stud \mbox{, for }i \in [m] \mbox{, } \widetilde{\grade}_s^{i} = \frac{\widehat{\grade}_s^{i}}{\sqrt{\chi^2 + \sigma_{\gp(\stud)}^2}}.
\end{equation*}

With these new standardized priority scores, the marginal priority distribution of each group is $\mathcal{N}(0, 1)$ at each college. The priority vectors then follow a centered bivariate normal distribution with variance 1 and a correlation that is different between the two groups: formally, 
\begin{equation*} \label{eq:lat_qual_cor}
      \widetilde{\grades}_s \sim \mathcal{N} \left(0_{\mathbb{R}^m}, 
      \begin{pmatrix}
    1 & \cor_{\gp(\stud)} & \cdots & \cor_{\gp(\stud)} \\
    \cor_{\gp(\stud)} & \ddots &  & \vdots \\
    \vdots & & \ddots & \cor_{\gp(\stud)} \\
    \cor_{\gp(\stud)} & \cdots & \cor_{\gp(\stud)} & 1
\end{pmatrix}
      \right),
\end{equation*}
with
\begin{equation*} \label{eq: cor_noise}
   \cor_{\gp(\stud)} = \frac{\chi^2}{\chi^2 + \sigma_{\gp(\stud)}^2}.
\end{equation*}
This setting satisfied the assumptions of our general model, and thus our analysis applies. Notice that this is also the case for the priority scores $\widehat{\grades}$ before renormalization, the difference being that in this case the marginals are not identical. The parameter $\paramj$ for each group $\gpj$ can then be chosen as $-\sigma^2_{\gpj}$, or $\cor_{\gpj}$, or any increasing function of one of these quantities.

Beyond this example, note that any priority score vector can be decomposed into an unknown latent quality and the remainder, interpreted as noise. As long as the observed priority score vectors' distributions depend on a parameter so that they form a coherent family, our results apply.

\subsection{Excess capacity} \label{app:excess}

We explore a case where correlation does not play any role in the matching.

\begin{proposition} \label{prop:excess}
If for all pairs of colleges $\coli, \col^{i'}$, $\capi + \capacity^{i'} \geq 1$, then correlation has no effect on the matching, i.e., $\cutoffs$ is constant in $\params$, and every student gets either their first or second choice.
\end{proposition}

\proof{Proof.}

If for all $i, i' \in [m]$, $\capi + \capacity^{i'} \geq 1$, at most one college is full. Because if two colleges were full, the mass of students admitted to those two colleges would exceed the total mass of students. Assume college $\coli$ is full, then students whose favorite college is not $\coli$ get their first choice independently of their priority scores, and those who prefer $\coli$ get it if they pass the cutoff, otherwise they get their second choice. Therefore, the demand at each college depends only on preferences and marginal distributions, not the correlation, and the same holds for the cutoffs, and thus the matching.\hfill \Halmos
\endproof

\subsection{Two preferred colleges} \label{app:2_cols}

Here we formalize the result mentioned in Section 4.3.1, where two colleges are preferred to all others, which technically violates the assumptions of the model since we assumed that all preference lists are used by a positive mass of students.

\begin{proposition} \label{prop:2_col}
    Assume that two colleges, say, $\col^1$ and $\col^2$, are preferred to all other colleges by all students.\footnote{By that we mean that the mass of students using a preference list $\preflist$ such that $\{\preflist(1), \preflist(2) \} \neq \{1, 2\}$ is zero.} Further assume that the projection of $\copcdffamily$ on its two first variables is coherent. Then there exists a unique stable matching, $\cut^1$ and $\cut^2$ are decreasing, and all the results presented until now apply to the college admissions problem obtained by restricting $\colset$ to $\{\col^1, \col^2\}$.
\end{proposition}

\proof{Proof.}
Students only get matched to colleges $\coli, i \geq 3$ if they do not pass the cutoffs at $\col^1$ and $\col^2$. Therefore, we can compute the matching on the restricted set of colleges $\{\col^1, \col^2\}$, then remove the matched students and start again with the remaining colleges. As long as the projection of $\copcdffamily$ on the variables associated to $\col^1$ and $\col^2$ is coherent, we recover Theorem $3$, so their respective cutoffs are decreasing, and all subsequent results apply to the reduced market $\{\col^1, \col^2\}$. \hfill \Halmos
\endproof

\newpage

\section{Omitted proofs} \label{app:proof}

\subsection{Proof of Proposition 1}\label{app.proof.prop:1st-choice}

The result follows directly by applying Lemma \ref{lemma:compV} and noticing that for $k = 1$ only the marginal distribution of the favorite college plays a role, not the joint distribution. Therefore, if groups $\gpj$ and $\gpl$ have the same marginal at college $\coli$ ($\pdfCij = \pdfCil$), then for all $\preflist$ such that $\preflist(1) = i$, $\Rjsi = \Rlsi$.
\hfill \Halmos

\subsection{Proof of Proposition 2}\label{app.proof.lemma:unmatched}

It is sufficient to notice that $\RjsE = \mathbb{P}_{j, \paramj}(\bigcap\limits_{i = 1}^m \gradei < P^i)$ does not depend on the preferences to obtain the first part of the lemma. The second part follows from the fact that either there is excess capacity and everyone is matched, or all colleges are full and the mass of matched students is the sum of the capacities. \hfill \Halmos

\begin{remark}
The first part of Proposition 2 could also be derived from the strategy-proofness for students of the student-proposing deferred acceptance algorithm \citep{roth_equivalent_1985}. Indeed, the fact that students cannot improve their outcome by modifying the order of their preferences implies that them being unmatched or not does not depend on their preferences.
\end{remark}

\subsection{Proof of Proposition 3} \label{app.proof.prop:FOSD}

Let $\paramj = \paraml = \param \in \tset$, $k \in [m]$ and $\preflist \in \preflistsm$. The projection of $\copcdfj = \copcdf_\ell$ on coordinates $\{\preflist(1), \dots, \preflist(k)\}$ induces a copula $\Bar{H}$.

By Sklar's theorem, 
\begin{align}
\mathbb{P}_{j,\param}\left(\bigcap\limits_{i = 1}^k \grade^{\preflist(i)} < \cut^{\preflist(i)} \right) = \Bar{H}\left(\cdf_j^{\preflist(1)}(\cut^{\preflist(1)}), \dots, \cdf_j^{\preflist(k)}(\cut^{\preflist(k)})\right).
\end{align}
Moreover, $\Bar{H}$ is increasing in all coordinates. Therefore, since $\cdfCij < \cdfCil$ for all $i \in [m]$, 
\begin{align}
\Bar{H}\left(\cdf_j^{\preflist(1)}(\cut^{\preflist(1)}), \dots, \cdf_j^{\preflist(k)}(\cut^{\preflist(k)})\right) < \Bar{H}\left(\cdf_\ell^{\preflist(1)}(\cut^{\preflist(1)}), \dots, \cdf_\ell^{\preflist(k)}(\cut^{\preflist(k)})\right).
\end{align}
Since 
\begin{align}
\Rjsk = 1 - \mathbb{P}_{j,\param}\left(\bigcap\limits_{i = 1}^k \grade^{\preflist(i)} < \cut^{\preflist(i)} \right),
\end{align}
we conclude that $\Rjsk > \Rlsk$. Choosing $k = m$, we also get $\RjE < \RlE$ and therefore $\ineqjl > 0$.\hfill \Halmos

\subsection{Proof of Theorem 2}\label{app.proof.cor:unmatched}

By Assumption 3, all $\cuti$ are decreasing in all $\paramj$, thus for $\ell \neq j \in [\ngp]$:
\begin{align*}
       \frac{d\RjE}{d \paraml} & = \frac{d  \prob_{j, \paramj}\left(\bigcap\limits_{i \in [m]}\gradei < \cuti\right)}{d \param_\ell} \\
       &= \left(\frac{d  \prob_{j, \paramj}\left(\bigcap\limits_{i \in [m]}\gradei < \cuti\right)}{d\cut^1} ~, \dots, ~ \frac{d  \prob_{j, \paramj}\left(\bigcap\limits_{i \in [m]}\gradei < \cuti\right)}{d\cut^m} \right) \cdot \left(\frac{d \cut^1}{d \param_\ell} ~, \dots, ~ \frac{d \cut^m}{d \param_\ell} \right)^T \\
       & <  0. 
\end{align*}
Since the total capacity is constant, the mass of unmatched students must also be constant. Therefore, we have
\begin{equation}
    \propj  \RjE + \sum\limits_{\ell \neq j} \propl  \RlE = 1 - \sum\limits_{i \in [m]}\capi.
\end{equation}
By differentiating this equation we get
\begin{align*}
   & \prop_i  \frac{d \RjE}{d \paramj} + \sum\limits_{\ell \neq j} \propl  \frac{d \RlE}{d \paramj} = 0 \\ 
    \Leftrightarrow & \frac{d \RjE}{d \paramj} = - \frac{1}{\propj} \sum\limits_{\ell \neq j} \propl  \frac{d \RlE}{d \paramj} \\
    \Rightarrow & \frac{d \RjE}{d \paramj} > 0
\end{align*}
which proves the first part of the theorem. Moreover, since $\ineqjl = \vert \RjE - \RlE \vert$, assume without loss of generality that $\RjE > \RlE$, then $\ineqjl = \RjE - \RlE$, and we can use the first part of the theorem to deduce the variations of $\ineqjl$. \hfill \Halmos

\subsection{Proof of Proposition 4} \label{app.proof.prop:efficiency}

$\eff(\params)$ is a convex combination of the $\Rjsi(\params)$ that are increasing in all coordinates $\paramj$. Moreover it is continuous, and we assumed $\tset$ to be an interval, so the image set of $\eff(\params)$ is an interval, say $[\eff^{\min}, \eff^{\max}]$.
Fix $\hat{E} \in (\eff^{\min}, \eff^{\max})$, and consider the solutions of the equation $\eff(\params) = \hat{E}$. By continuity, this equation has a solution. The implicit function theorem applied to express some $\paramj$ (the choice of $j$ does not matter) as a function $\phi$ of all the other coordinates $\paraml$ shows that the solutions of $\eff(\params) = \hat{E}$ is a connected subset of $\tset^\ngp$, and also an hypersurface because the function $\phi$ is monotonous in all $\paraml$ (this comes from the fact that $\eff(\params)$ is itself monotonous). This proves the first part of the proposition.

Let us choose two groups $\gpj, \gpl$, and fix all $\param_k$ for $k \neq j, \ell$. We  apply the implicit function theorem to express $\paramj$ as a function of $\paraml$, which shows that there exists an interval $U \coloneqq [\underline{\param}, \Bar{\param}] \subseteq \tset$ and a differentiable function $\phi: U \to \tset$ such that ($\paraml \in U$ and $\paramj = \phi(\paraml)) \Leftrightarrow (\eff(\params) = \hat{E}$ and $ \params \in U \times \phi(U))$. Since $\eff(\params)$ is increasing in all arguments, $\phi$ is necessarily decreasing. Along the line $\paramj = \phi(\paraml)$, then $\frac{d \RlE}{d \paraml} = \frac{\partial \RlE}{\partial \param_i} + \frac{\partial \RlE}{\partial \paramj} \phi'(\paraml)$, which is positive by Corollary 1, so $(\paraml, \paramj) = (\underline{\param}, \phi(\underline{\param}))$ minimizes $\RlE$, and $(\paraml, \paramj) = (\Bar{\param}, \phi(\Bar{\param}))$ maximizes it. The same reasoning shows that those two points respectively maximize and minimize $\RjE$. Finally, since $\RlE$ is increasing and $\RjE$ decreasing, $\ineqjl= \vert \RlE - \RjE \vert$ has a unique global minimum on the line $\paramj = \phi(\paraml)$.
\hfill \Halmos

\subsection{Proof of Proposition 5} \label{app.proof.prop:TB}

We start by building a distribution family that can represent both STB and MTB for two different values of the parameter. Let $\copcdffamily$ be a coherent copula family such that $\param = 0$ gives independent variables and $\param = 1$ gives equal variables, and let $\coppdft$ be the derivative of $\copcdft$ for all $\param$. We denote by $\kappa^i_p = \mass(\class^i_p)$ the mass of students inside class $p$ of college $\coli$. For all $i \in [m]$, let $a^i_0 = 0, a^i_1 = \kappa^i_1, a^i_2 = \kappa^i_1 + \kappa^i_2, \dots , a^i_{\tau^i} = 1$, such that they form a partition of $[0, 1]$ with the $p$-th segment having length $\kappa^i_p$. Finally, for any $\textbf{p} = (p^1, \dots, p^m) \in \prod\limits_{i \in [m]} [\tau^i]$, let $\kappa_\textbf{p} = \eta(\prod\limits_{i \in [m]}\class^i_{p^i})$ be the mass of students belonging to class $p^1$ at $\col^1$, class $p^2$ at $\col^2$, and so on.

Let $\Tilde{h}_\param: [0, 1]^m \to \mathbb{R}$ be defined as:
\begin{equation*}
    \Tilde{h}_\param (x) = \kappa_{\textbf{p}} h_\param\left(\left(\frac{x_i - a^i_{p^i-1}}{\kappa^i_{p^i}}\right)_{i \in [m]}\right),
\end{equation*}

\begin{figure}[ht]
\FIGURE{
   \includegraphics[width = 0.6 \textwidth]{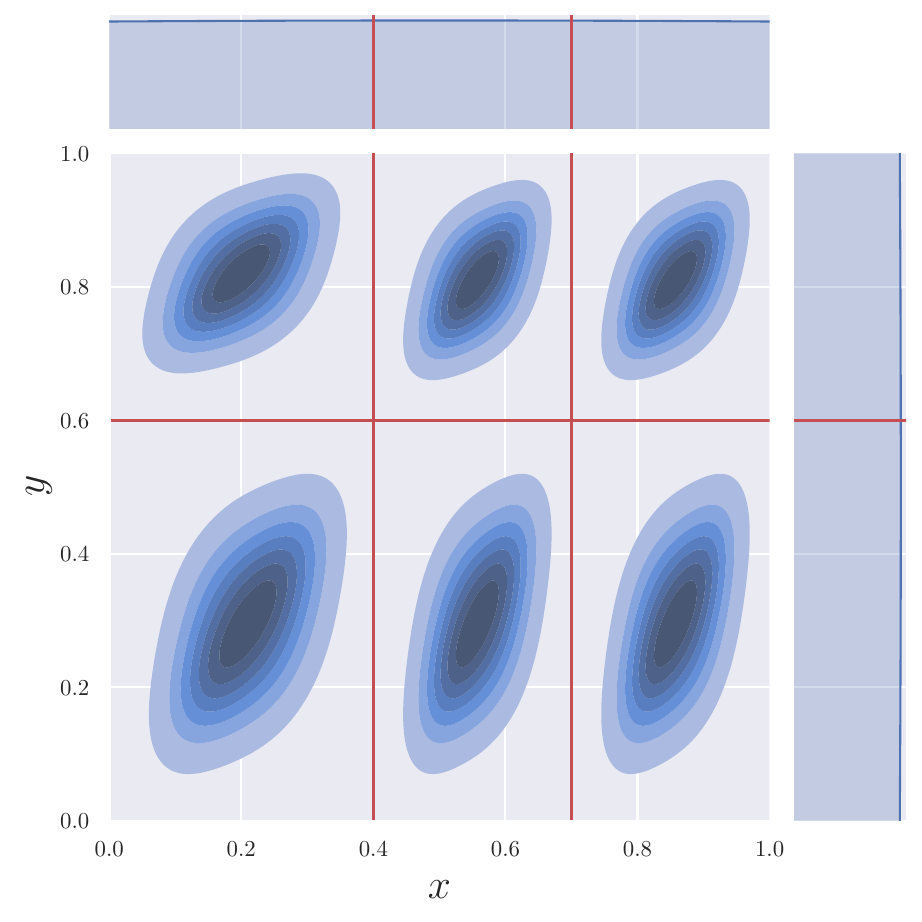}}
    {Illustration of the distribution $\Tilde{H}_\param$.\label{fig:tie-breaking}} {Parameters: two colleges, three priority classes at $\col^1$ (30\% of applicants in the first class, 30\% in the second, 40\% in the third), two priority classes at $\col^2$ (40\% in the first class, 60\% in the second). Base copula: Gaussian copula, with correlation chosen as the covariance (equal to $0.5$ here).}
    
\end{figure}

with $\mathbf{p}$ such that $\forall i\in [m], a^i_{p^i-1} \leq x \leq a^i_{p^i}$. Let $\Tilde{H}_\param$ be the primitive of $\Tilde{h}_\param$ that cancels out on $0_{\mathbb{R}^m}$. Defined this way, $\Tilde{H}_\param$ is a copula since it is non-negative, increasing in all coordinates, marginals are uniform and do not depend on $\param$, and $\Tilde{H}_\param(1_{\mathbb{R}^m}) = 1$. Moreover, each hyper-rectangle $\prod\limits_{i \in [m]}\class^i_{p^i}$ contains a ``copy'' of $H_\param$ adjusted to its dimensions, that has mass $\kappa_\textbf{p}$. $\Tilde{H}_\param$ is consistent with priority classes: if student $\stud$ is in a higher priority class at college $\coli$ than student $\stud'$, then $\stud$ will have a higher priority  score with probability 1. This distribution is depicted in Figure \ref{fig:tie-breaking}.

We can verify that this definition recovers MTB and STB: if $\param = 0$, if two students are in the same priority class for a college, they have the same ex-ante probability of getting a seat there, and if they also are in the same priority class for another college, the result of this second tie-breaker is independent from the first one. When $\param = 1$, if two students are in the same priority class for a college, they have the same ex-ante probability of getting a seat there, but if they also are in the same priority class for another college, the winner of the tie-breaking is the same as in the first college since priority scores inside the rectangle are perfectly correlated. Therefore MTB corresponds to $\param = 0$ and STB to $\param = 1$.

Let us now prove the two parts of the proposition:
\begin{enumerate} 
    \item  We have: \begin{itemize}
        \item  The family $(\Tilde{H}_\param)_{\param \in \tset}$ is differentiable because $(H_\param)_{\param \in \tset}$ is. It is also coherent (except for the $x$ such that $x_i = a^i_p$ for some $i$ and some $p$, i.e., the sides of the hyper-rectangles, in which case the cdf is constant and not increasing). Therefore by applying Theorem 1, $\Rjsk, \Rlsi$ and $\eff$ are either increasing or constant.
        \item  Moreover, the case where it could be constant can only happen if there are several priority classes, so if there is only one they are  increasing.
        \item  Consider the case of multiple priority classes. Suppose that $\exists \params \in \tset$ such that $\cuti(\params) \neq a^i_p$ for all $i$ and all $p$. We can apply Theorem 1, and deduce that the $\Rjsk$, $\Rjsi$ and $\eff$ are increasing in $\paraml$ on the whole interval $\tset$. If there exists no such $\params$, it implies that $\cutoffs$ is constant in $\params$ and so are all the metrics. However, as any perturbation of either $\props, \prefs$, or $\capacities$ would change the cutoffs and resolve the issue, the set of problematic values of $(\props, \prefs, \capacities)$ has Lebesgue measure 0.
    \end{itemize}
    \item Finally, Theorem 2 can be applied with the same adjustments, which proves the second part.
\end{enumerate}  \hfill \Halmos

\subsection{Proof of Theorem 3} \label{app.proof.thm:cutoff_inc}

Let $j \in [\ngp]$ and $\params \in \tset^2$. By the coherence assumption, the mass of students below fixed cutoffs increases in $\paramj$, however, since the mass of unmatched students is constant at least one of the cutoffs has to be decreasing. Without loss of generality, assume that $\cut^1$ is decreasing, and suppose that $\cut^2$ is non-decreasing. College $\col^2$'s component of the market-clearing equation (1) gives
\begin{equation} \label{eq:thm_3}
    \left(\sum\limits_{\ell \in [\ngp]}\prefl^{(2 1)} \right) \mathbb{P}(\grade^2 \geq \cut^2) + \left(\sum\limits_{\ell \in [\ngp]}\prefl^{(1 2)}  \mathbb{P}_{\ell, \paraml}(\grade^1 < \cut^1, \grade^2 \geq \cut^2) \right) = \capacity^2
\end{equation}
 Let us analyze the variations of each term. $\mathbb{P}(\grade^2 \geq \cut^2)$ is decreasing in $\cut^2$, its partial derivative w.r.t. $\paramj$ is zero, and $\cut^2$ is assumed to be non-decreasing in $\paramj$ so $\mathbb{P}(\grade^2 \geq \cut^2)$ is overall decreasing in $\paramj$. For $\ell \neq j$, $\mathbb{P}_{\ell, \paraml}(\grade^1 < \cut^1, \grade^2 \geq \cut^2)$ is increasing in $\cut^1$, decreasing in $\cut^2$, and its partial derivative w.r.t. $\paramj$ is zero, so it is overall decreasing in $\paramj$. Finally, $\mathbb{P}_{j, \paramj}(\grade^1 < \cut^1, \grade^2 \geq \cut^2)$ is increasing in $\cut^1$, decreasing in $\cut^2$, and its partial derivative w.r.t. $\paramj$ is negative by the coherence assumption (because $\mathbb{P}_{j, \paramj}(\grade^1 < \cut^1, \grade^2 \geq \cut^2) = \mathbb{P}(\grade^1 < \cut^1) - \mathbb{P}_{j, \paramj}(\grade^1 < \cut^1, \grade^2 < \cut^2)$). Therefore, all terms are decreasing in $\paramj$, but the right-hand side of Equation \eqref{eq:thm_3} is a constant. We deduce by contradiction that $\cut^2$ is decreasing, which concludes the proof. 
 
 \hfill \Halmos

\section{Numerical experiments} \label{app:numerical}

In this section, we show in more detail the results of the numerical experiments from Section 4.3.2. We compute cutoffs and rank metrics for various values of the parameters, for three and four colleges. For three colleges we consider three different preference vectors (i.e., distributions over the possible preference lists of students). For four colleges, we only consider uniform preferences, because accounting for 24 different preferences lists each used by a different proportion of students is  computationally too costly. For the same reason, we stop our exploration at four colleges, since the number of preference lists is equal to $m!$. 

\subsection{Three colleges} \label{app:num_3}

We compute the cutoffs as functions of group $\gp_2$'s correlation $\param_2$, for 324 different sets of values of the other parameters: 
\begin{itemize}
    \item \textbf{Group Proportions ($\prop_1:\prop_2$):} 10:90, 30:70, 50:50
    \item \textbf{Total Capacity ($\sum_{i\in [3]}\capacity^i$):} 1/3, 1/2, 2/3
    \item \textbf{College Capacity Allocation ($\frac{100}{\sum_{i\in [3]}\capacity^i}(\capacity^1, \capacity^2, \capacity^3$)):} $(10, 10, 80)$, $(33.3, 33.3, 33.3)$, $(10, 45, 45)$
    \item \textbf{Correlation of Group 1 ($\param_1$):} 0, 1/3, 2/3, 0.99
    \item \textbf{Preference Vector ($\prefs$):} $\prefs^{\mathrm{I}}=$Uniform, $\prefs^{\mathrm{II}}=$(1/32; 1/8; 1/32; 1/4; 1/16; 1/2), $\prefs^{\mathrm{III}}=$(1/2; 1/16; 1/4; 1/32; 1/8; 1/32).
\end{itemize}
The preferences vectors are to be read along the following order of permutations of colleges: $(123, 132, 213, 231, 312, 321)$.

\subsubsection{Cutoffs}
Out of the 324 points in the grid, some give cutoff functions with very similar shapes. Specifically, group size, total capacity and correlation of group $\gp_1$ change the value of cutoffs but have very little impact on the shape of the functions; only students preferences and the allocation of the total capacity among colleges makes a noticeable difference.

 Figure \ref{fig:sim_cut_3} therefore only displays nine  points, corresponding to the three different preference vectors and three different capacity allocations; the choice of the other parameters has little importance and is as follows: {group proportions:} 50:50, {total capacity:} 2/3, {correlation of group $\gp_1$:} 1/3.

Note that all cutoffs decrease over the whole interval $[0, 1]$ for the middle and right columns, corresponding to capacity vectors $(0.22, 0.22, 0.22)$ and $(0.07, 0.3, 0.3)$. When the capacity is $(0.07, 0.07, 0.53)$ (in the left column), however, we find counterexamples. In Figures \ref{fig:sim_cut_3_a} and \ref{fig:sim_cut_3_g}, that correspond respectively to uniform preferences and preference vector $(1/2; 1/16; 1/4; 1/32; 1/8; 1/32)$, the cutoff $\cut^3$  increases slightly as the  correlation approaches 1. We can conjecture from this observation that counterexamples are found when preferences and capacity are greatly misaligned, i.e., when some college has low demand and very high capacity.

\begin{figure}[ht]
    \FIGURE{\shortstack{
    \subcaptionbox{$\capacities =(0.07, 0.07, 0.53)$, $\prefs = \prefs^{\mathrm{I}}$\label{fig:sim_cut_3_a}}  {\includegraphics[width=0.32\linewidth]{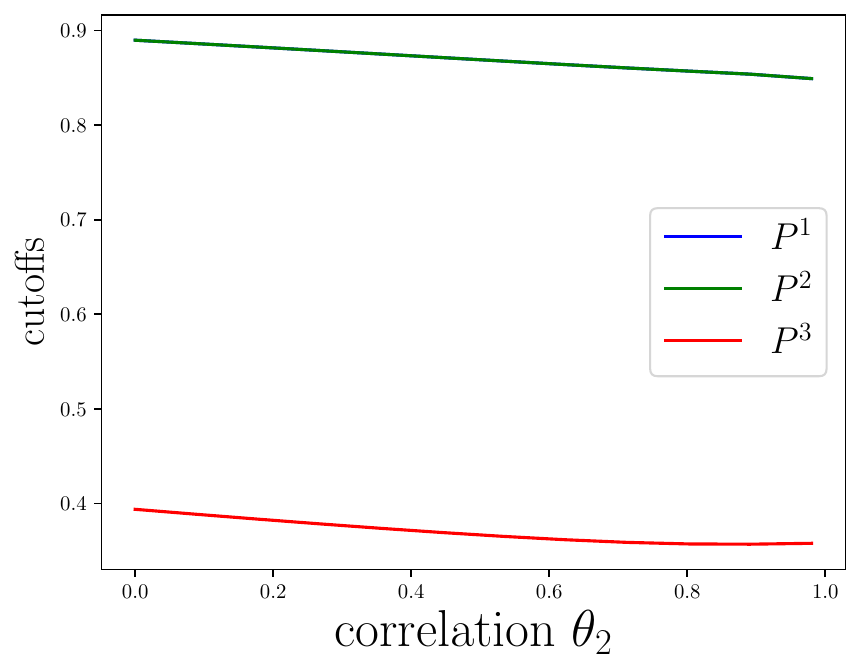}}
    \subcaptionbox{$\capacities =(0.22, 0.22, 0.22)$, $\prefs = \prefs^{\mathrm{I}}$\label{fig:sim_cut_3_b}}{\includegraphics[width=0.32\linewidth]{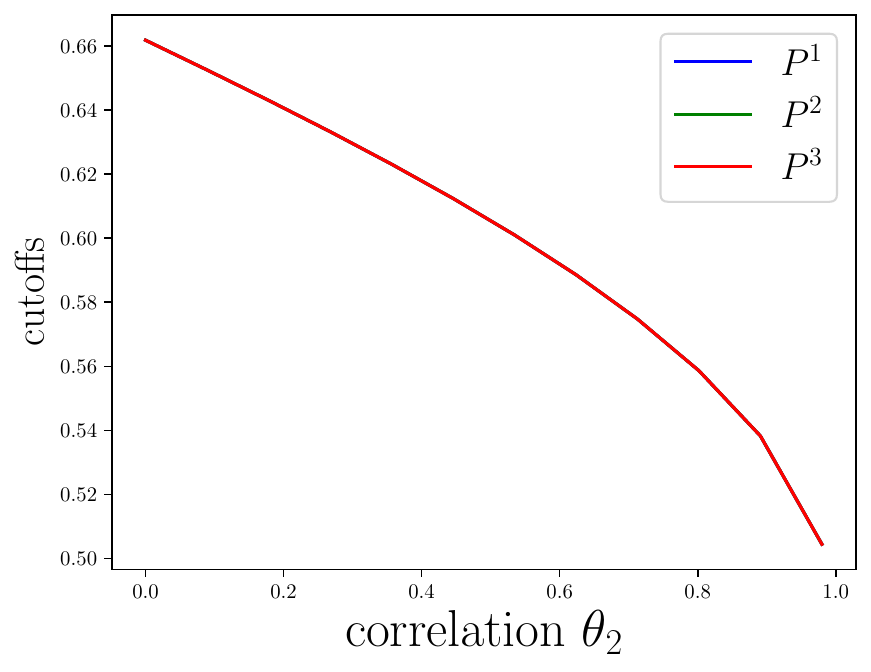}}
    \subcaptionbox{$\capacities =(0.07, 0.3, 0.3)$, $\prefs = \prefs^{\mathrm{I}}$\label{fig:sim_cut_3_c}} {\includegraphics[width=0.32\linewidth]{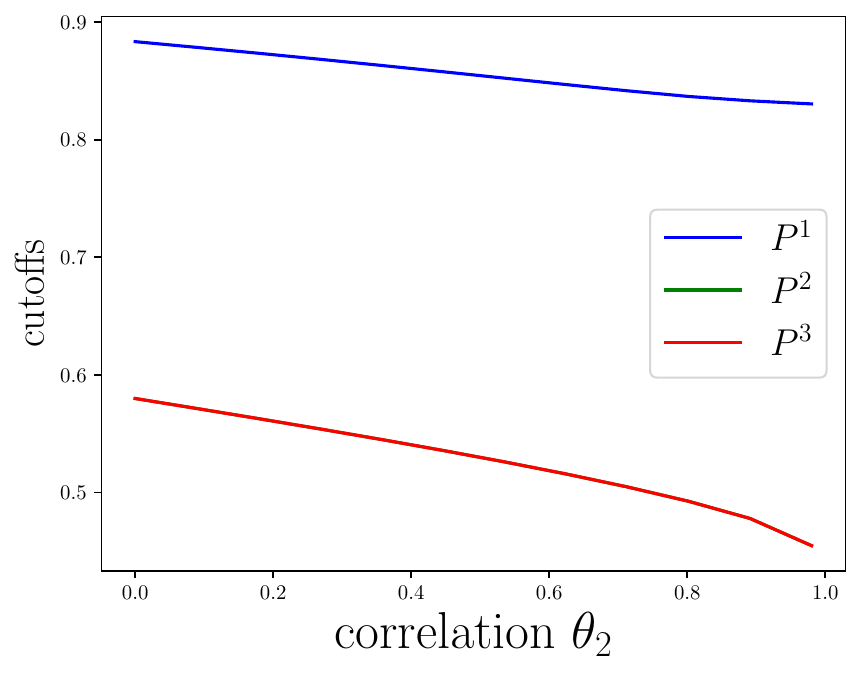}} \\
    \subcaptionbox{$\capacities =(0.07, 0.07, 0.53)$, $\prefs = \prefs^{\mathrm{II}}$ \label{fig:sim_cut_3_d}}{\includegraphics[width=0.32\linewidth]{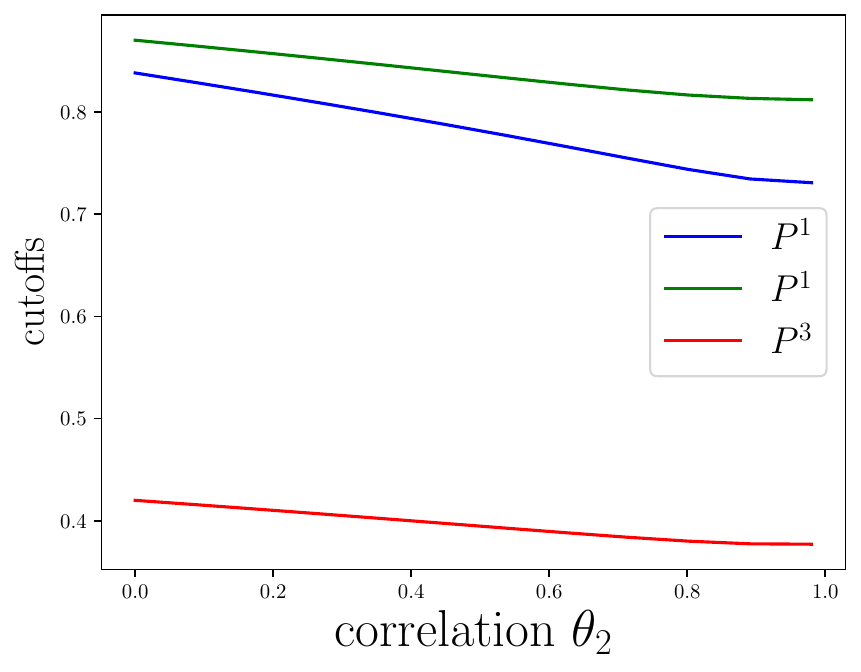}}
    \subcaptionbox{$\capacities =(0.22, 0.22, 0.22)$, $\prefs = \prefs^{\mathrm{II}}$\label{fig:sim_cut_3_e}}{\includegraphics[width=0.32\linewidth]{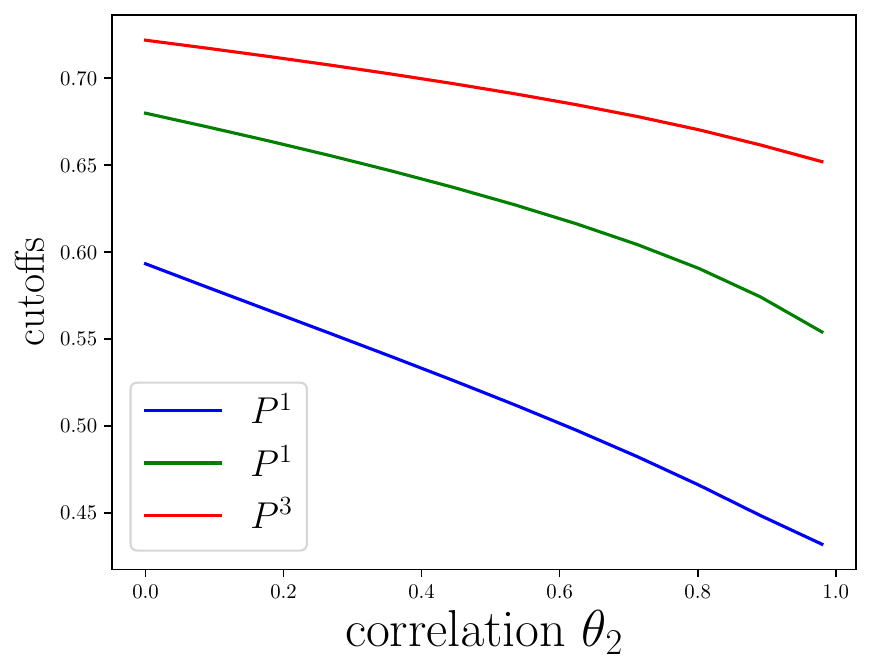}}
    \subcaptionbox{$\capacities =(0.07, 0.3, 0.3)$, $\prefs = \prefs^{\mathrm{II}}$ \label{fig:sim_cut_3_f}}{\includegraphics[width=0.32\linewidth]{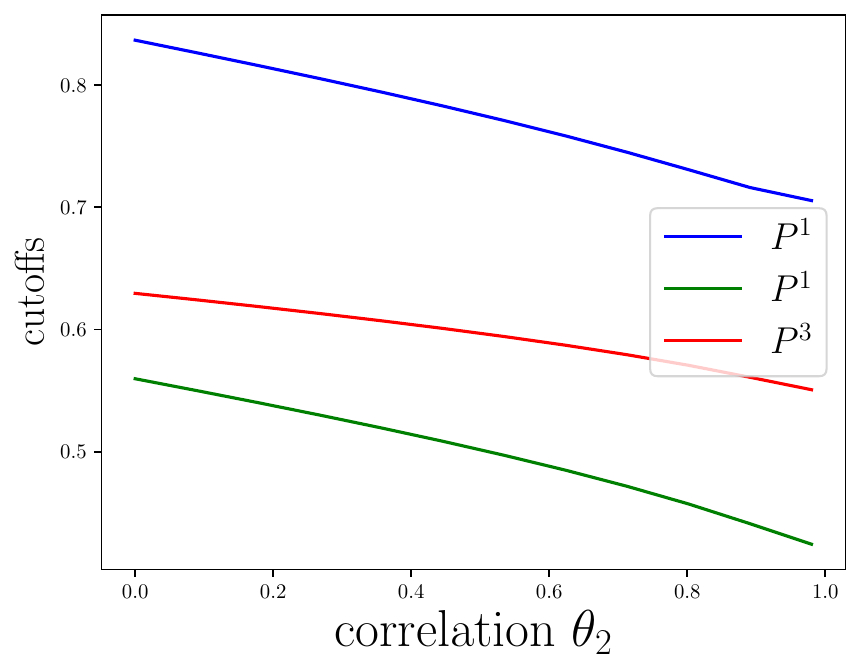}} \\
    \subcaptionbox{$\capacities =(0.07, 0.07, 0.53)$, $\prefs = \prefs^{\mathrm{III}}$ \label{fig:sim_cut_3_g}}{\includegraphics[width=0.32\linewidth]{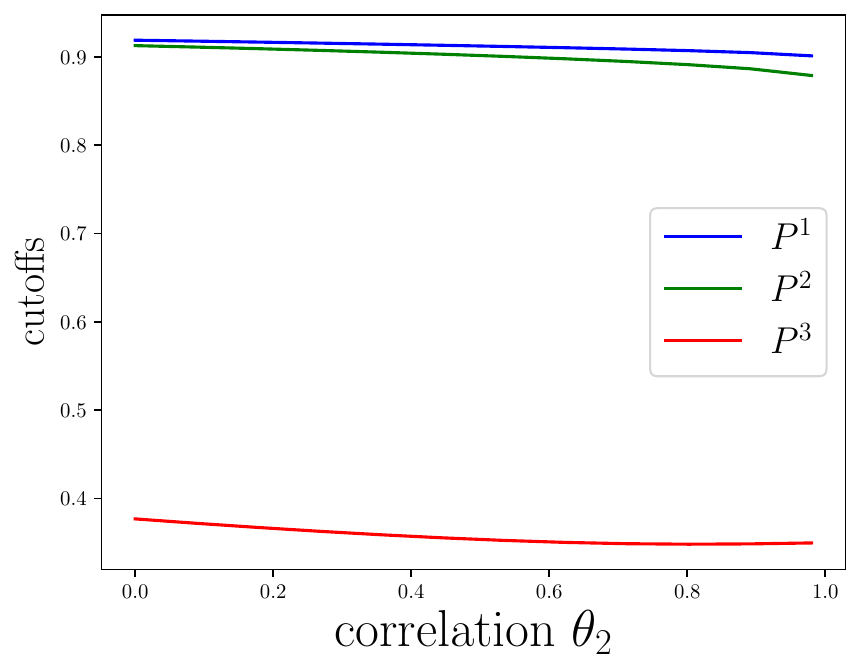}}
    \subcaptionbox{$\capacities =(0.22, 0.22, 0.22)$, $\prefs = \prefs^{\mathrm{III}}$ \label{fig:sim_cut_3_h}}{\includegraphics[width=0.32\linewidth]{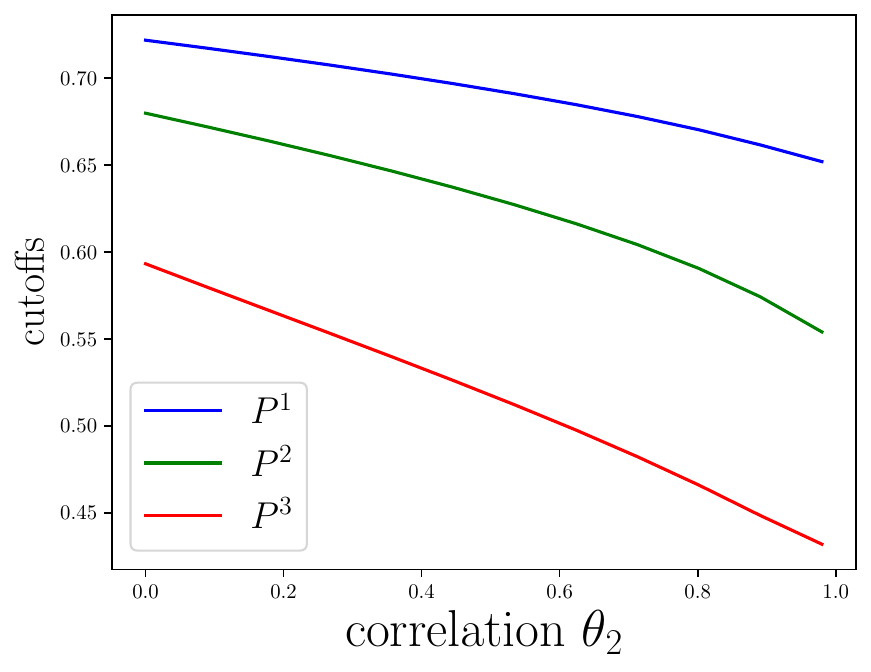}}
    \subcaptionbox{$\capacities =(0.07, 0.3, 0.3)$, $\prefs = \prefs^{\mathrm{III}}$ \label{fig:sim_cut_3_i}}{\includegraphics[width=0.32\linewidth]{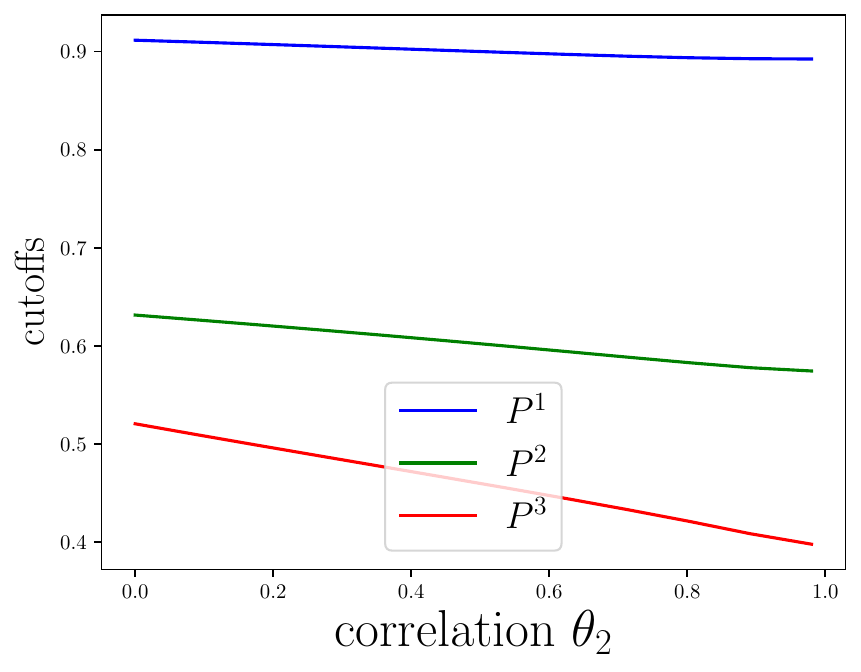}}}
    }
    {Variations of the cutoffs for three colleges. \label{fig:sim_cut_3}}
    {Note that in some plots the cutoffs overlap. Preference vectors: $\prefs^{\mathrm{I}} = (1/6, 1/6, 1/6, 1/6, 1/6, 1/6)$, $\prefs^{\mathrm{II}} = (1/32; 1/8; 1/32; 1/4; 1/16; 1/2)$, $\prefs^{\mathrm{III}} =(1/2; 1/16; 1/4; 1/32; 1/8; 1/32)$.}
\end{figure}

 \subsubsection{Ranks}

We compute the rank metrics $R_1^{k, \preflist}$ with $\preflist = (312)$ for all points of the grid to study the validity of Theorem 1. We choose the preferences $\preflist$ as it allows to highlight potential counterexamples of Theorem 1, as it has college $\col^3$ as the first choice and in our examples it is $\col^3$'s cutoff that is sometimes increasing. The results are displayed in Figure \ref{fig:sim_rk_3}. For the values of the parameters where cutoffs are decreasing (middle and right columns, as well as Figure \ref{fig:sim_rk_3_d}), Theorem 1 applies and we find that all $R_1^{k, \preflist}$ are increasing as predicted. For the cases where the cutoffs do not always decrease, i.e., in Figures \ref{fig:sim_rk_3_a} and \ref{fig:sim_rk_3_g}, the proportion of students who get their first choice $R_1^{1, \preflist}$ decreases exactly where cutoffs increase. This is predicted because the proof of Theorem 1 implies that $R_1^{1, \preflist}$'s variations are the opposite of those of the first choice's cutoff, here $\cut^3$. Interestingly, for preference vector $(1/2; 1/16; 1/4; 1/32; 1/8; 1/32)$ (Figure \ref{fig:sim_rk_3_g}), $R_1^{2, \preflist}$  also decreases slightly for values of $\param_2$ very close to 1, which is not the case for uniform preferences (Figure \ref{fig:sim_rk_3_a}). This tends to further support the conjecture that what drives counterexamples is the misalignment between capacities and preferences.

\begin{figure}[ht]
    \FIGURE{\shortstack{
    \subcaptionbox{$\capacities =(0.07, 0.07, 0.53)$, $\prefs = \prefs^{\mathrm{I}}$\label{fig:sim_rk_3_a}}  {\includegraphics[width=0.32\linewidth]{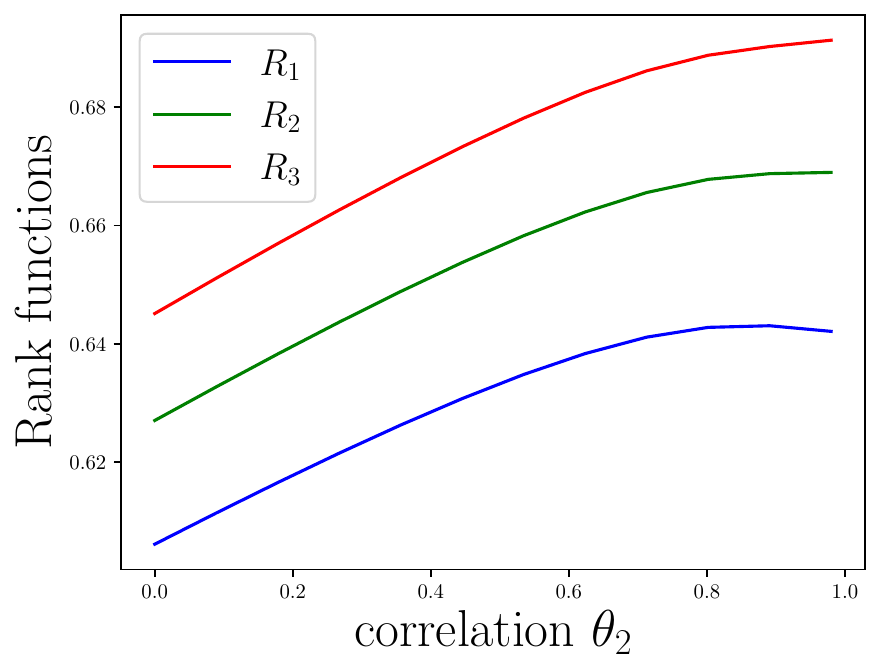}}
    \subcaptionbox{$\capacities =(0.22, 0.22, 0.22)$, $\prefs = \prefs^{\mathrm{I}}$\label{fig:sim_rk_3_b}}{\includegraphics[width=0.32\linewidth]{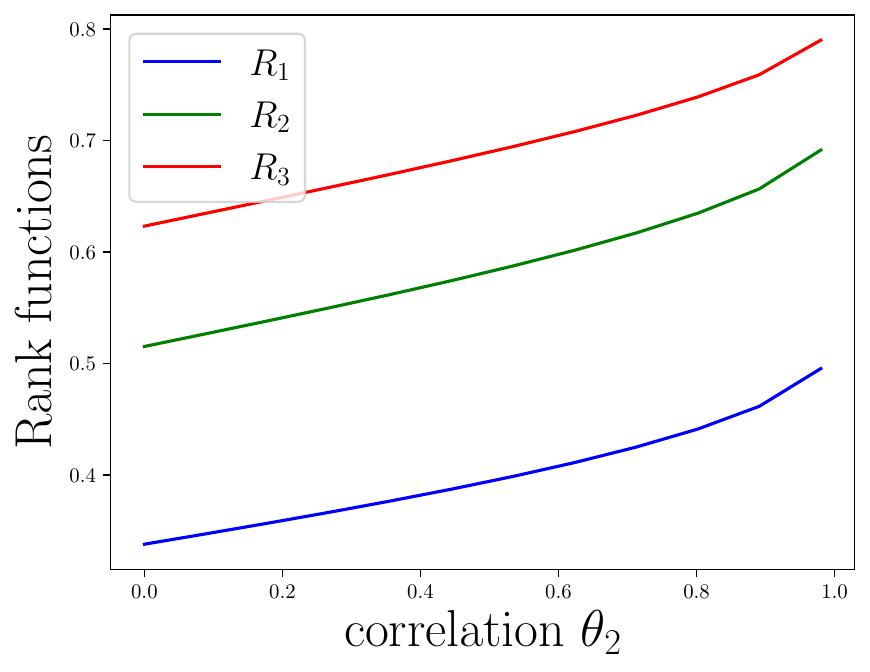}}
    \subcaptionbox{$\capacities =(0.07, 0.3, 0.3)$, $\prefs = \prefs^{\mathrm{I}}$\label{fig:sim_rk_3_c}} {\includegraphics[width=0.32\linewidth]{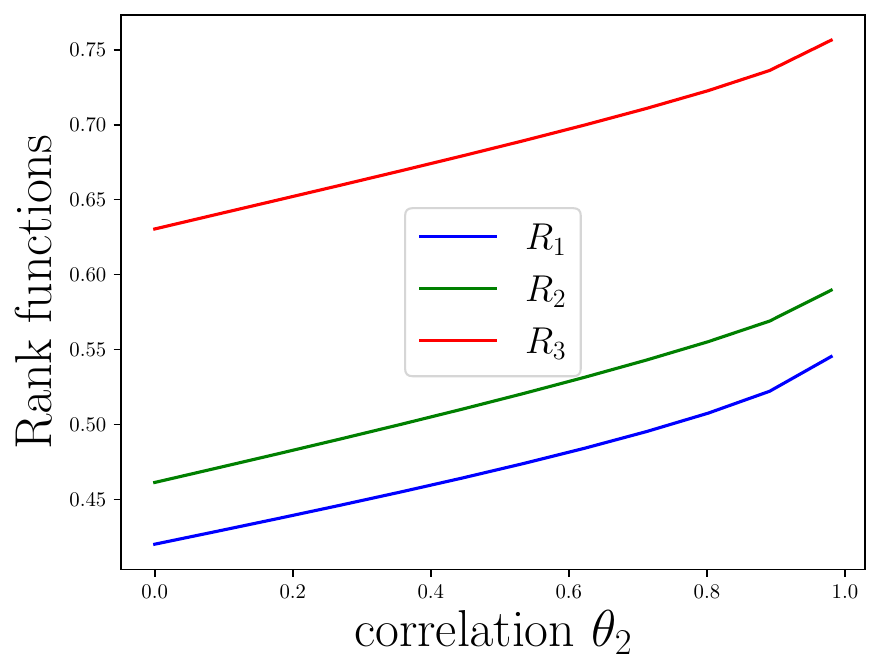}} \\
    \subcaptionbox{$\capacities =(0.07, 0.07, 0.53)$, $\prefs = \prefs^{\mathrm{II}}$ \label{fig:sim_rk_3_d}}{\includegraphics[width=0.32\linewidth]{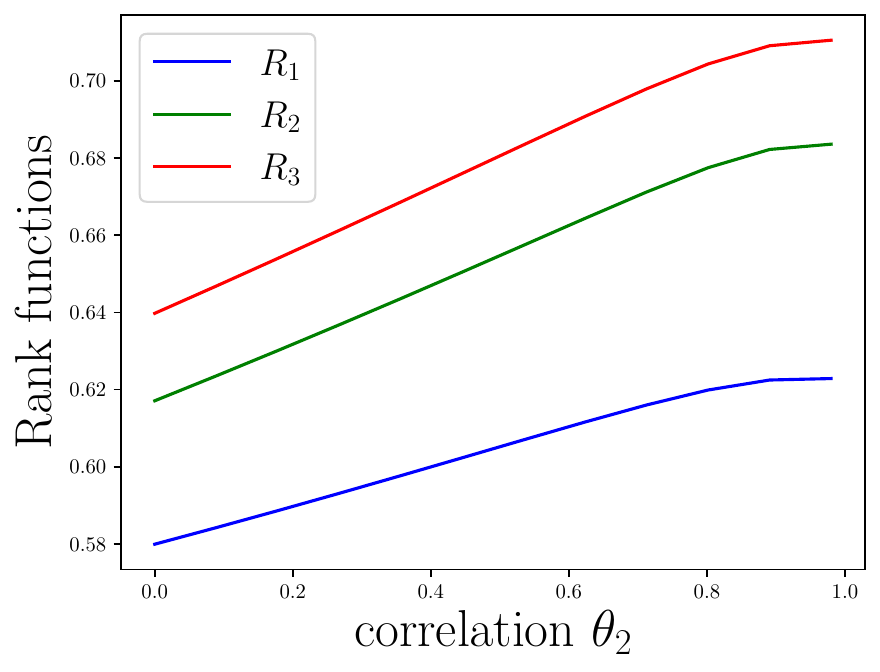}}
    \subcaptionbox{$\capacities =(0.22, 0.22, 0.22)$, $\prefs = \prefs^{\mathrm{II}}$\label{fig:sim_rk_3_e}}{\includegraphics[width=0.32\linewidth]{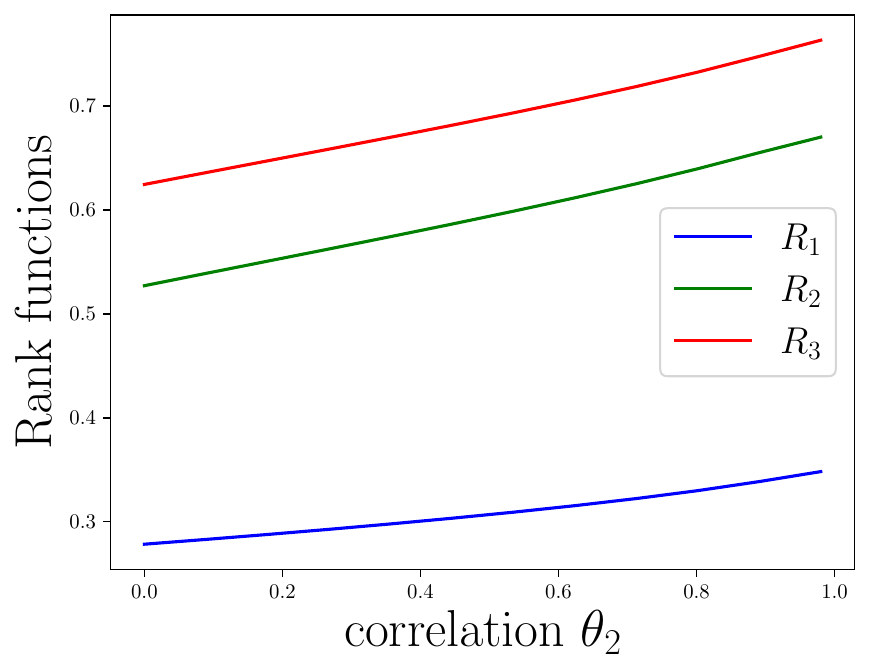}}
    \subcaptionbox{$\capacities =(0.07, 0.3, 0.3)$, $\prefs = \prefs^{\mathrm{II}}$ \label{fig:sim_rk_3_f}}{\includegraphics[width=0.32\linewidth]{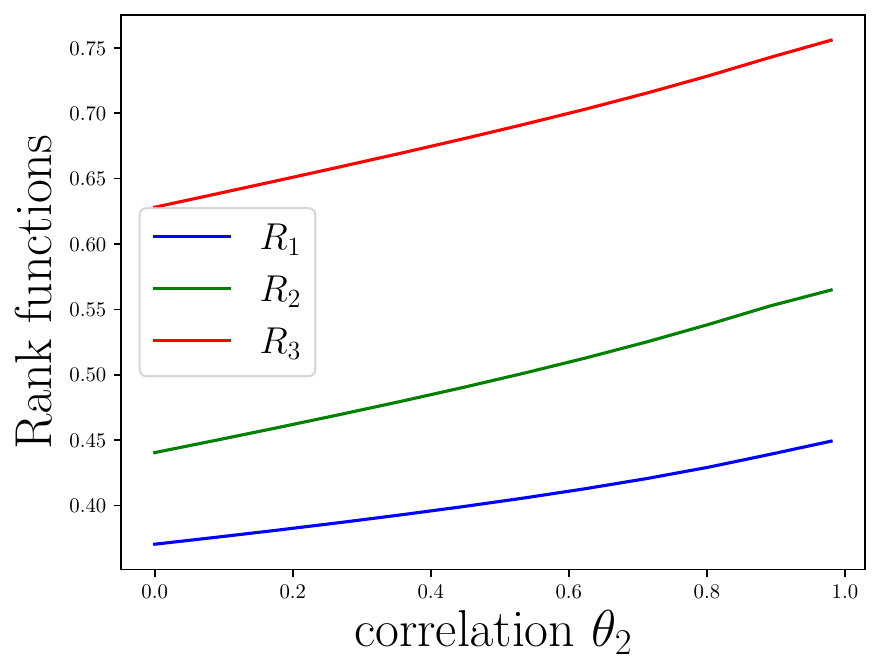}} \\
    \subcaptionbox{$\capacities =(0.07, 0.07, 0.53)$, $\prefs = \prefs^{\mathrm{III}}$ \label{fig:sim_rk_3_g}}{\includegraphics[width=0.32\linewidth]{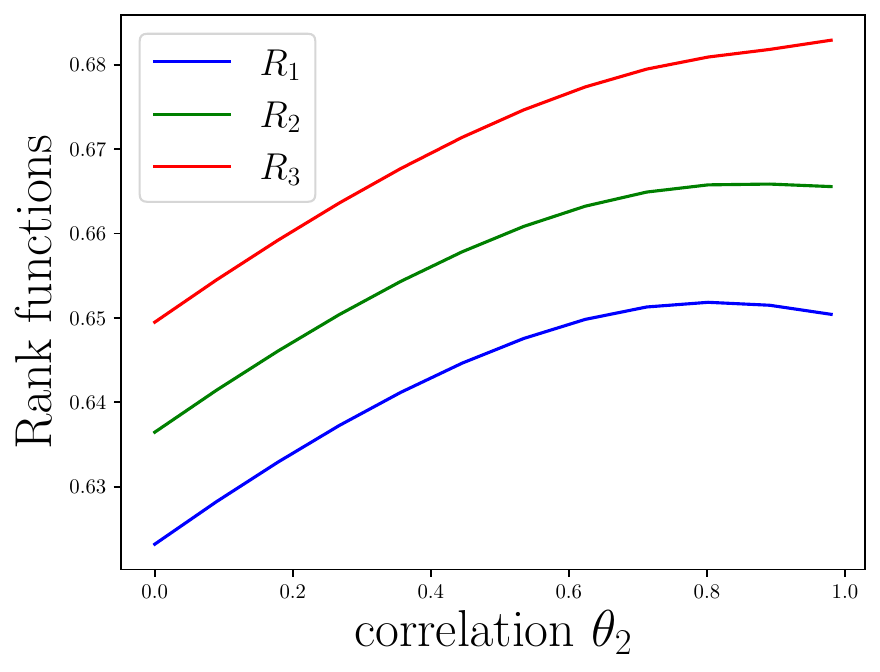}}
    \subcaptionbox{$\capacities =(0.22, 0.22, 0.22)$, $\prefs = \prefs^{\mathrm{III}}$ \label{fig:sim_rk_3_h}}{\includegraphics[width=0.32\linewidth]{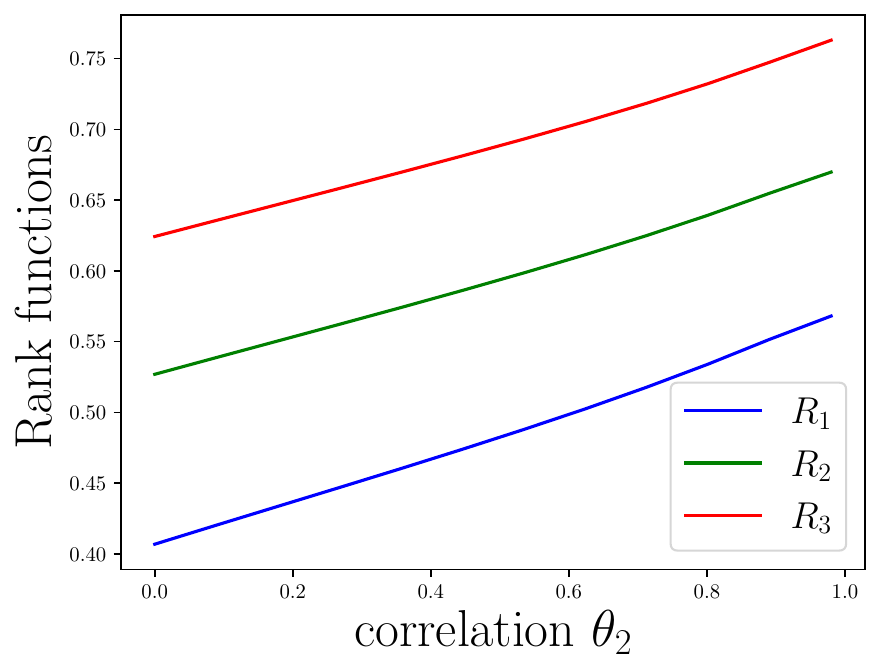}}
    \subcaptionbox{$\capacities =(0.07, 0.3, 0.3)$, $\prefs = \prefs^{\mathrm{III}}$ \label{fig:sim_rk_3_i}}{\includegraphics[width=0.32\linewidth]{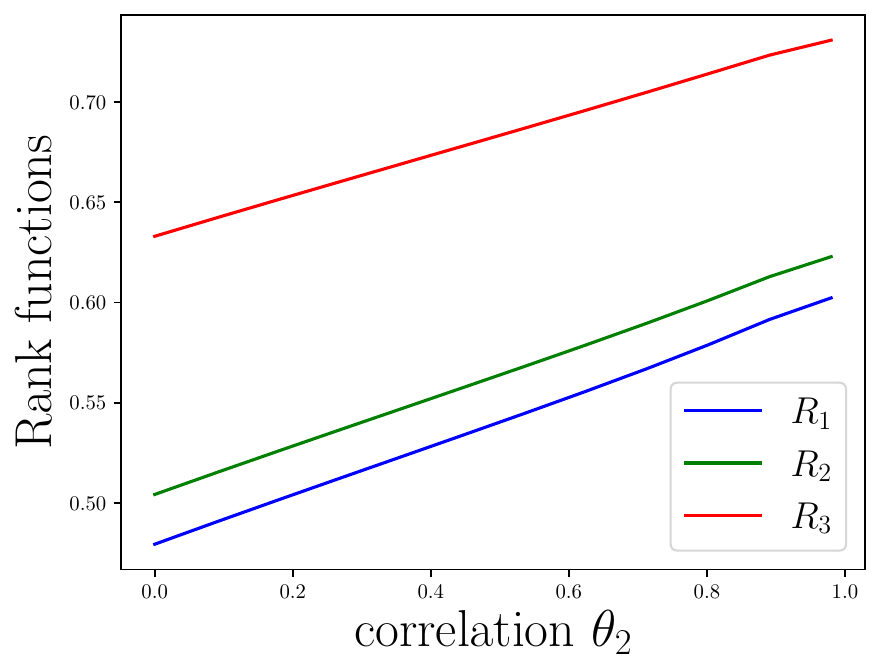}}}
    }
    {Variations of rank metrics for three colleges. \label{fig:sim_rk_3}}
    {Preference vectors: $\prefs^{\mathrm{I}} = (1/6, 1/6, 1/6, 1/6, 1/6, 1/6)$, $\prefs^{\mathrm{II}} = (1/32; 1/8; 1/32; 1/4; 1/16; 1/2)$, $\prefs^{\mathrm{III}} =(1/2; 1/16; 1/4; 1/32; 1/8; 1/32)$.}
\end{figure}

\subsection{Four colleges} \label{app:num_4}
We compute the cutoffs as functions of group $\gp_2$'s correlation $\param_2$, for 36 different sets of values of the other parameters:
\begin{itemize}
    \item \textbf{Group Proportions  ($\prop_1:\prop_2$):} 10:90, 50:50
    \item \textbf{Total Capacity ($\sum_{i\in [4]}\capacity^i$):} 1/3, 1/2, 9/10
    \item \textbf{College Capacity Allocation ($\frac{100}{\sum_{i\in [4]}\capacity^i}(\capacity^1, \capacity^2, \capacity^3, \capacity^4)$):} $(10, 20, 30, 40)$, $(6.25, 12.5, 18.75, 62.5)$
    \item \textbf{Correlation of Group 1 ($\param_1$):} 0, 1/2, 0.99
    \item \textbf{Preference Vector ($\prefs$):} Uniform.
\end{itemize}

\subsubsection{Cutoffs}
Out of the 36 points in the grid, we find four (around 11\%) where one cutoff (namely, $\cut^4$) is not always decreasing. Some sets of parameters give cutoff functions with very similar shapes. Specifically, as for the three colleges case, group size and correlation of group $\gp_1$ change the value of cutoffs but have very little impact on the shape of the functions; as before, the allocation of the total capacity among colleges makes a noticeable difference. Total capacity, while not having an equally sizeable impact, still sometimes plays a role in making cutoffs non-decreasing, which is why we also display the results for different values of total capacity. 
Figure \ref{fig:sim_cut_4} displays six points, corresponding to the three different total capacities and two different capacity allocations; the choice of the other parameters is of little importance and is as follows: group proportions 50:50, correlation of group $\gp_1$: 1/2.

Note that  all cutoffs decrease over the whole interval $[0, 1]$ on the top row, corresponding to capacity allocation 10:20:30:40. Cutoffs are also decreasing on the left figure of the bottom row, Figure \ref{fig:sim_cut_4_d}, corresponding to capacity allocation 6.25:12.5:18.75:62.5 and total capacity  1/3. For total capacities 2/3 and 0.9, however, we find that cutoffs are not always decreasing. More specifically, in Figures \ref{fig:sim_cut_4_e} and \ref{fig:sim_cut_4_f}, the cutoff $\cut^4$ increases slightly when correlation approaches 1, similarly to what we observed in the three colleges case. With those parameters, college $\col^4$ has a high capacity (respectively 0.42 and 0.56) compared to its demand (the proportion of students ranking it first is 0.25), which further confirms the intuition we had in the three colleges case that counterexamples arise in colleges that have a capacity significantly larger than their demand.

\begin{figure}[ht]
    \FIGURE{\shortstack{
    \subcaptionbox{$\capacities =(0.03, 0.07, 0.1, 0.13)$\label{fig:sim_cut_4_a}}  {\includegraphics[width=0.32\linewidth]{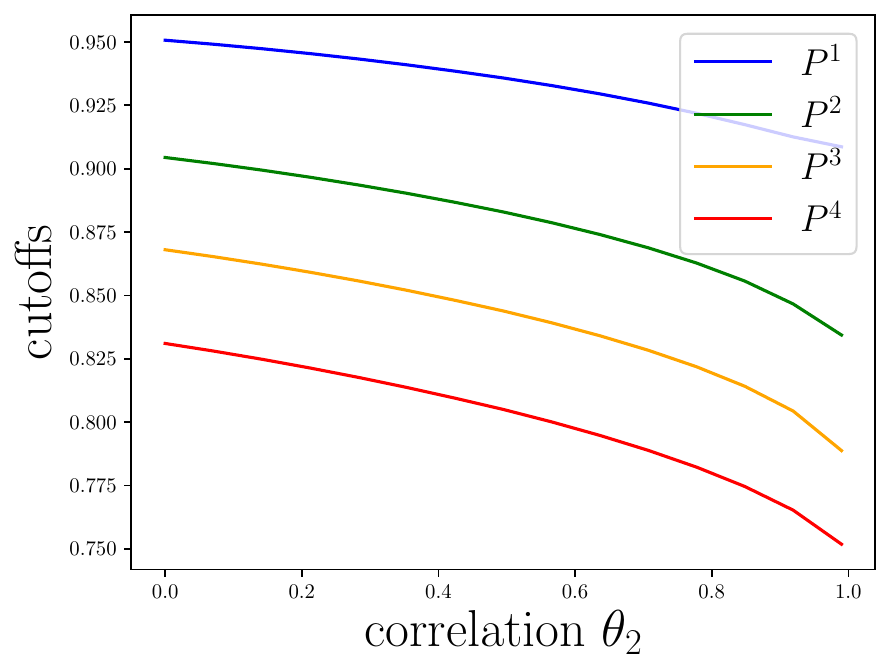}}
    \subcaptionbox{$\capacities =(0.07, 0.13, 0.2, 0.27)$\label{fig:sim_cut_4_b}}{\includegraphics[width=0.32\linewidth]{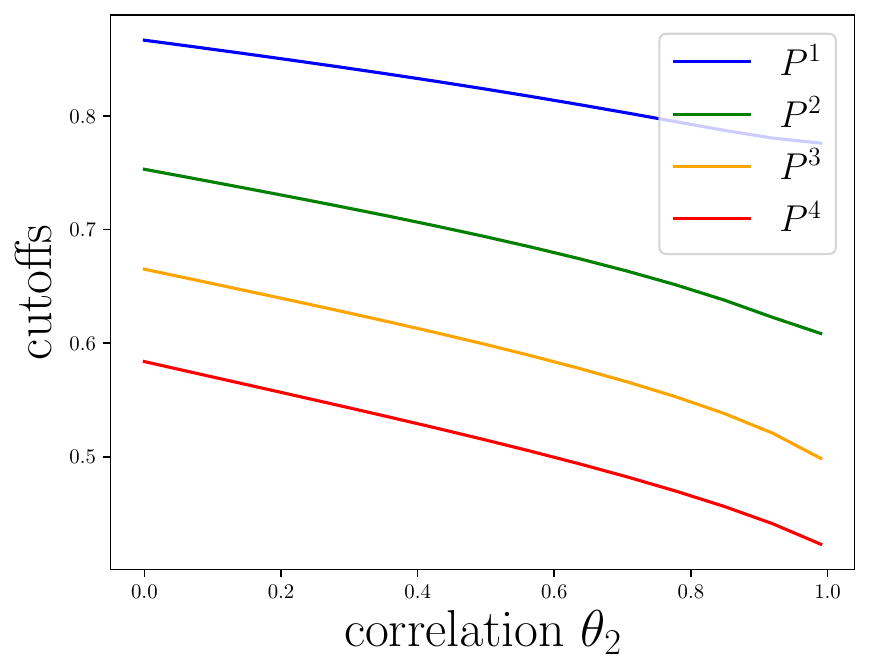}}
    \subcaptionbox{$\capacities =(0.09, 0.18, 0.27, 0.36)$\label{fig:sim_cut_4_c}} {\includegraphics[width=0.32\linewidth]{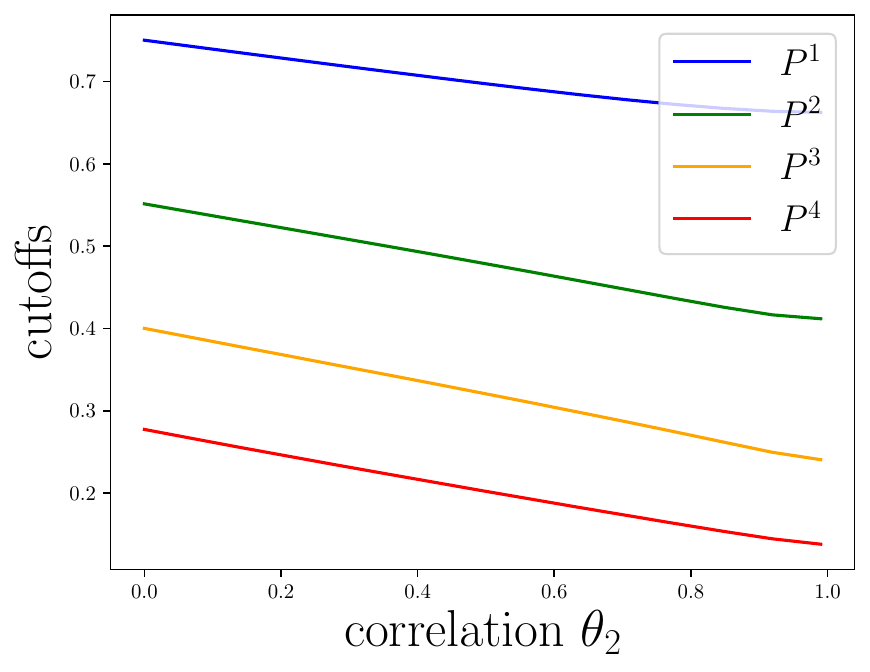}} \\
    \subcaptionbox{$\capacities =(0.02, 0.04, 0.06, 0.21)$ \label{fig:sim_cut_4_d}}{\includegraphics[width=0.32\linewidth]{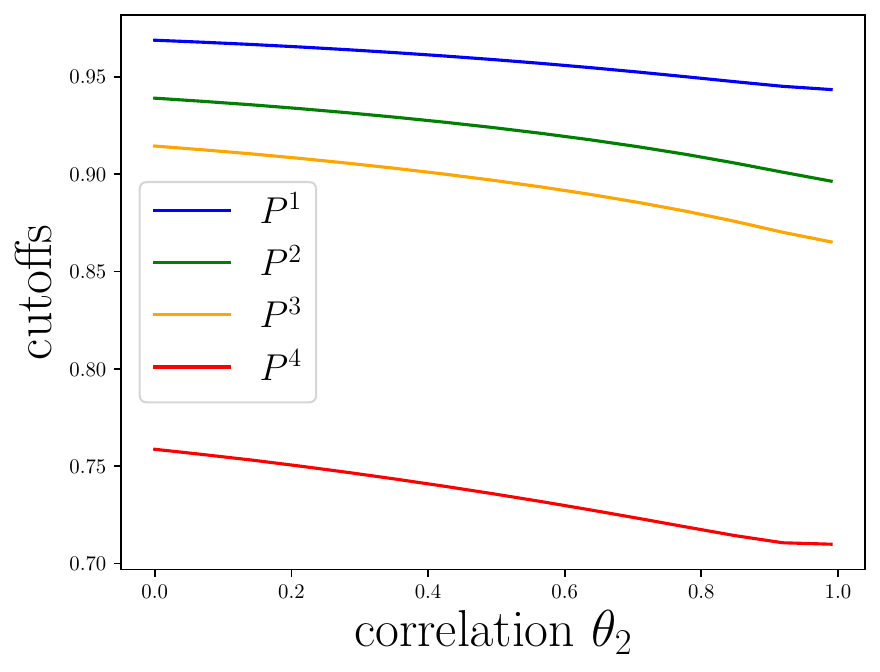}}
    \subcaptionbox{$\capacities =(0.04, 0.08, 0.12, 0.42)$\label{fig:sim_cut_4_e}}{\includegraphics[width=0.32\linewidth]{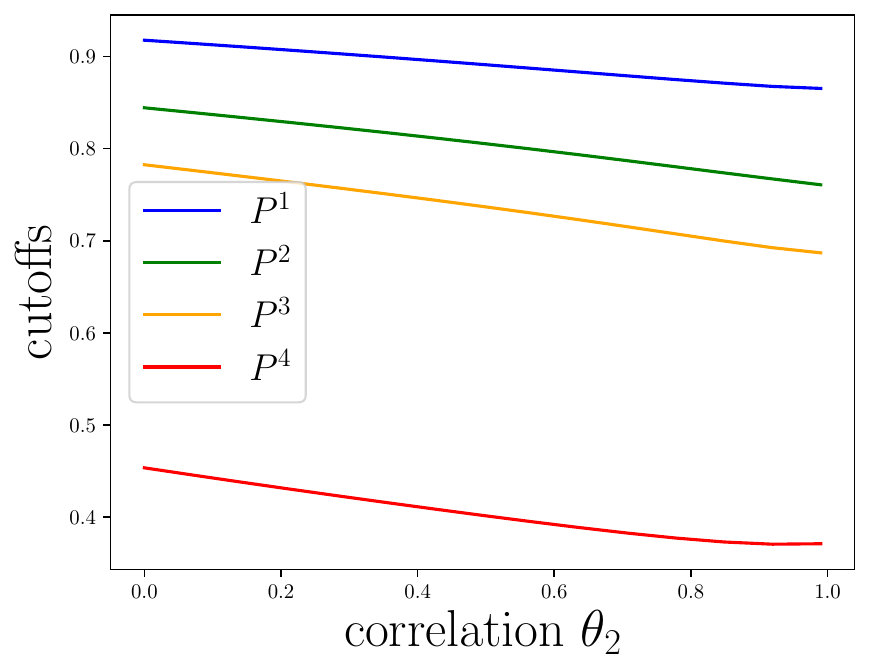}}
    \subcaptionbox{$\capacities =(0.06, 0.11, 0.17, 0.56)$ \label{fig:sim_cut_4_f}}{\includegraphics[width=0.32\linewidth]{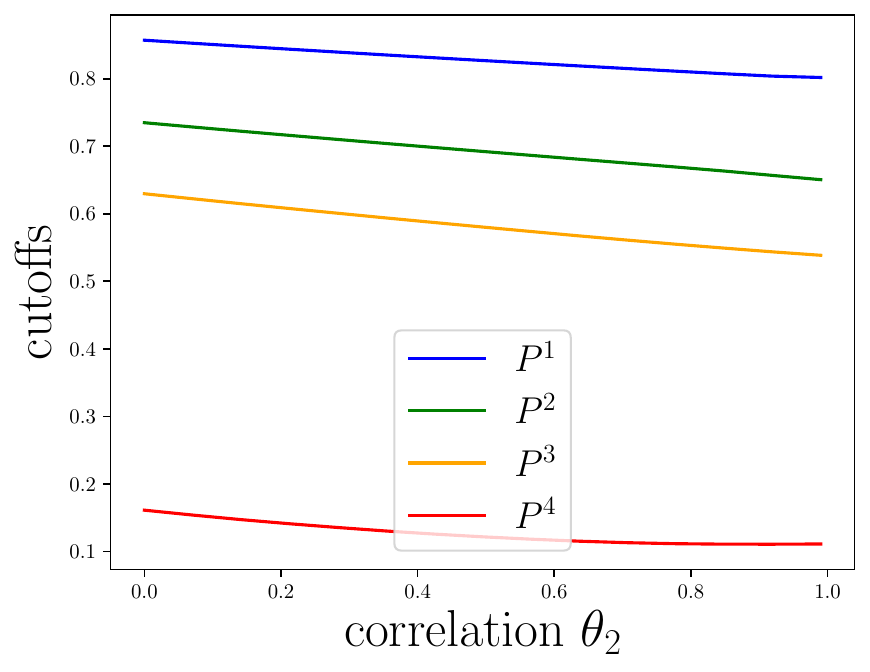}}}
    }
    {Variations of the cutoffs for four colleges. \label{fig:sim_cut_4}}
    {First row: capacity allocation 10:20:30:40, second row: 6.25:12.5:18.75:62.5. Left column: total capacity 1/3, middle: 2/3, right:0.9.}
\end{figure}

\subsubsection{Ranks}

We perform the same analysis that we performed for three colleges.
We compute the rank metrics $R_1^{k, \preflist}$ with $\preflist = (4321)$ for all points of the grid, to look for counterexamples to Theorem 1. The results are displayed in Figure \ref{fig:sim_rk_4}. For the values of the parameters where cutoffs are decreasing (top row, as well as Figure \ref{fig:sim_rk_4_d}), Theorem 1 applies and we find that all $R_1^{k, \preflist}$ are increasing as predicted. For the cases where cutoffs do not always decrease, i.e., in Figures \ref{fig:sim_rk_4_e} and \ref{fig:sim_rk_4_f}, the proportion of students who get their first choice $R_1^{1, \preflist}$ decreases exactly where cutoffs increase as predicted. Contrary to the three colleges case, here the proportion of students getting one of their two top choices $R_1^{2, \preflist}$ is never decreasing. This may be explained by the fact that $\cut^3$, in the three colleges case, was increasing at a higher rate than $\cut^4$ is in the four colleges case. Making $R_1^{2, \preflist}$ decrease, thus probably requires a sufficiently high rate of increase.

\begin{figure}[h!]
    \FIGURE{\shortstack{
    \subcaptionbox{$\capacities =(0.03, 0.07, 0.1, 0.13)$\label{fig:sim_rk_4_a}}  {\includegraphics[width=0.32\linewidth]{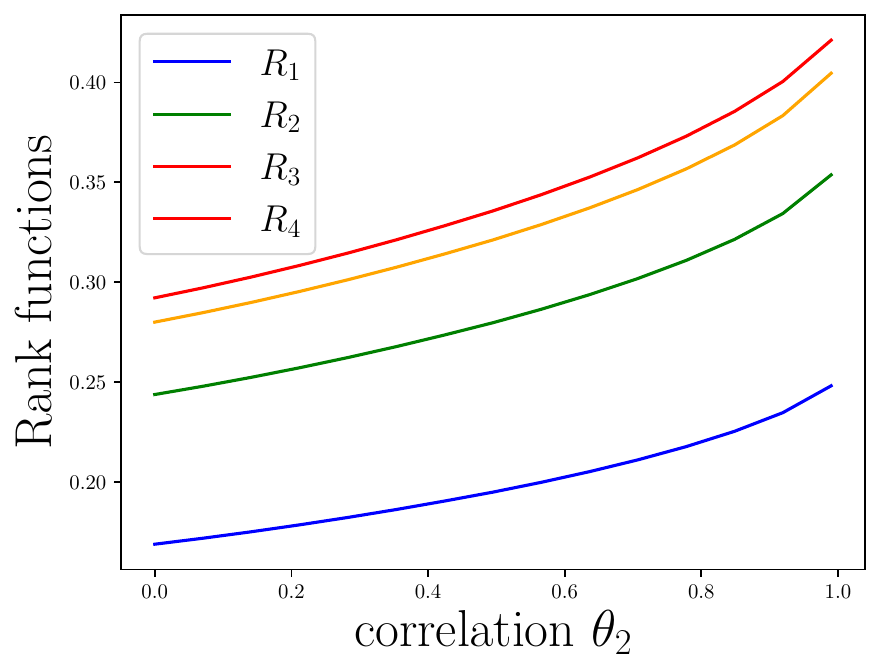}}
    \subcaptionbox{$\capacities =(0.07, 0.13, 0.2, 0.27)$\label{fig:sim_rk_4_b}}{\includegraphics[width=0.32\linewidth]{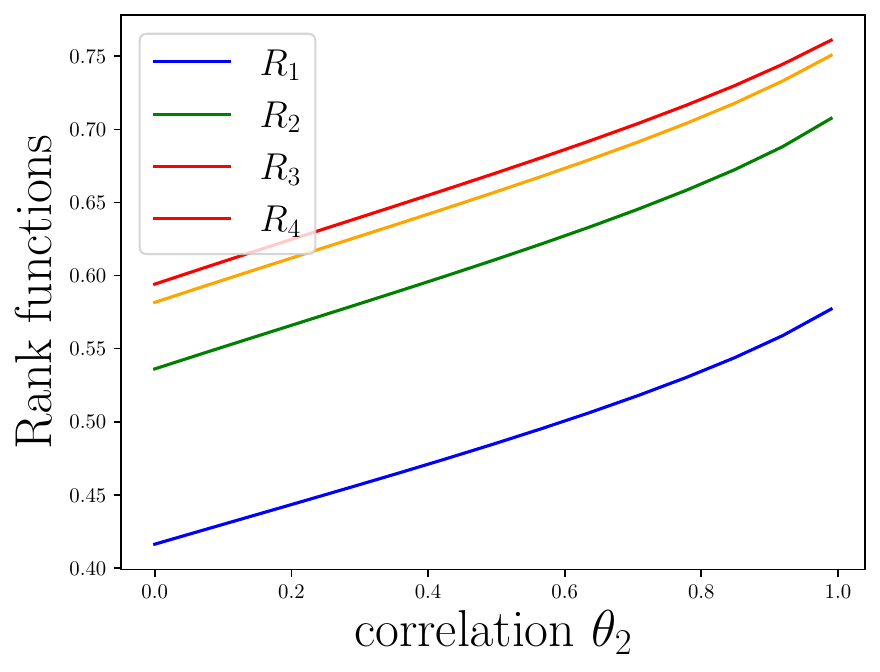}}
    \subcaptionbox{$\capacities =(0.09, 0.18, 0.27, 0.36)$\label{fig:sim_rk_4_c}} {\includegraphics[width=0.32\linewidth]{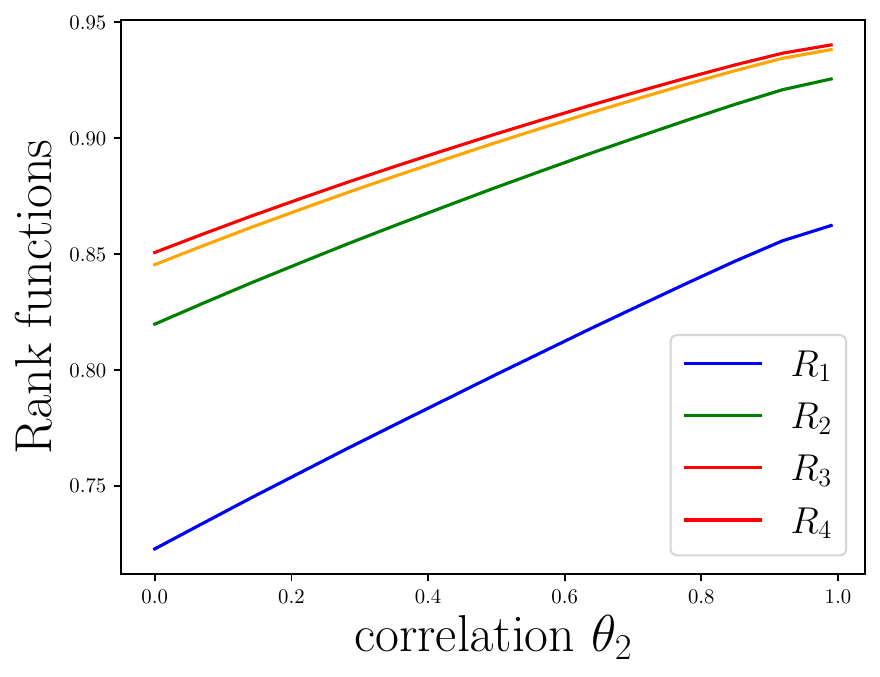}} \\
    \subcaptionbox{$\capacities =(0.02, 0.04, 0.06, 0.21)$ \label{fig:sim_rk_4_d}}{\includegraphics[width=0.32\linewidth]{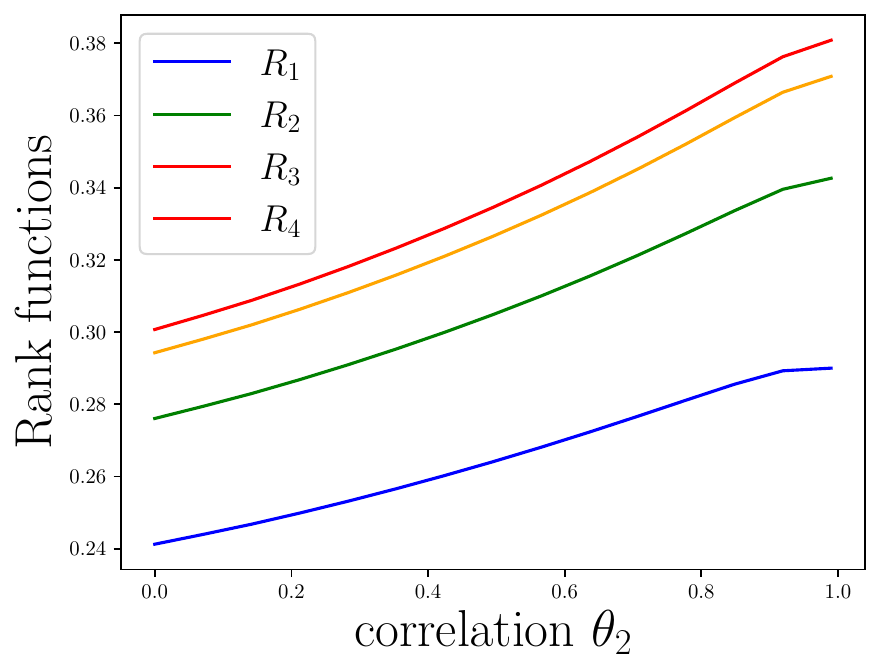}}
    \subcaptionbox{$\capacities =(0.04, 0.08, 0.12, 0.42)$\label{fig:sim_rk_4_e}}{\includegraphics[width=0.32\linewidth]{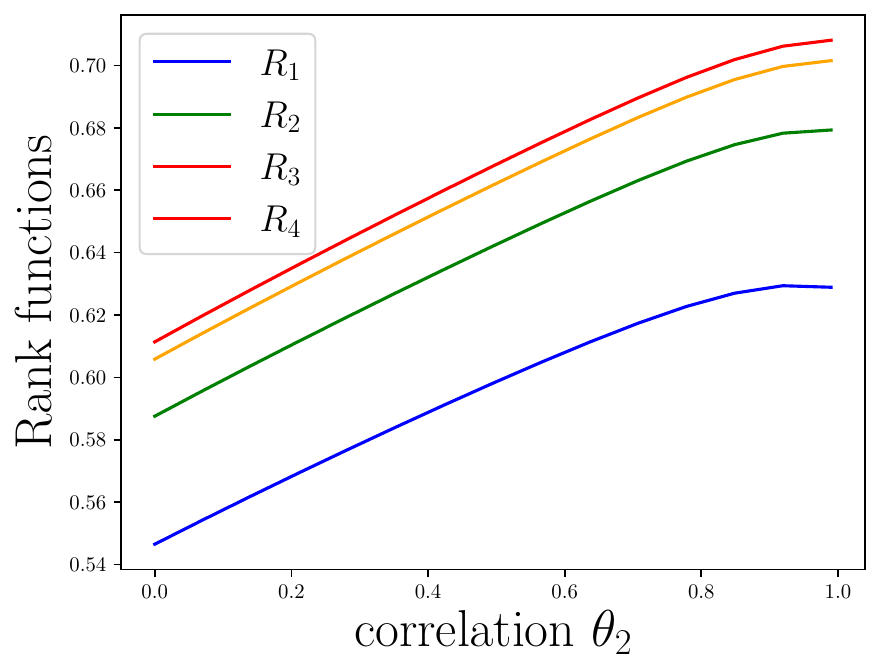}}
    \subcaptionbox{$\capacities =(0.06, 0.11, 0.17, 0.56)$ \label{fig:sim_rk_4_f}}{\includegraphics[width=0.32\linewidth]{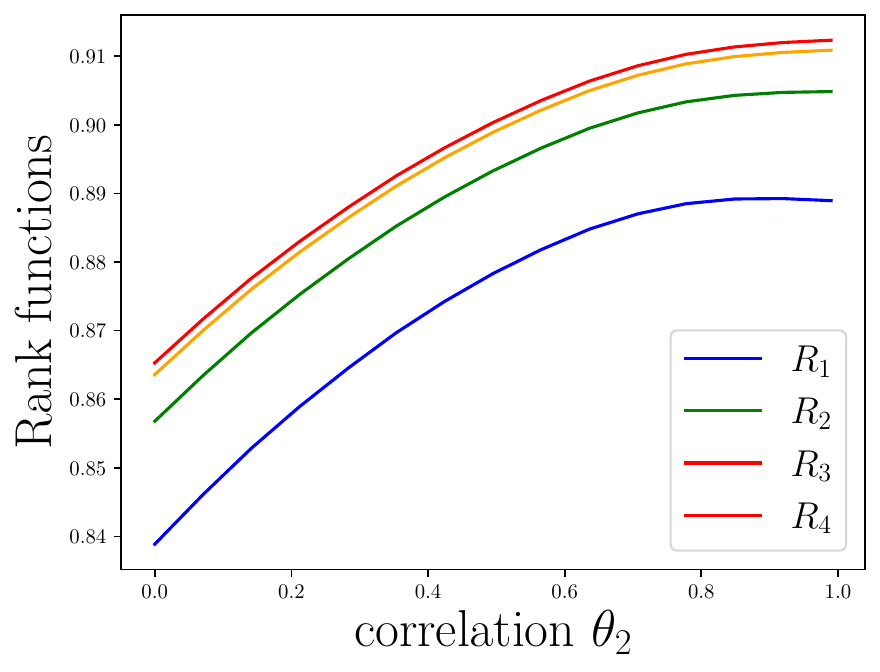}}}
    }
    {Variations of the rank metrics for four colleges. \label{fig:sim_rk_4}}
    {First row: capacity allocation 10:20:30:40, second row: 6.25:12.5:18.75:62.5. Left column: total capacity 1/3, middle: 2/3, right: 0.9.}
\end{figure}

\end{APPENDICES}

\end{document}